\documentclass[twocolumn]{aastex631}

\usepackage{amsmath}
\usepackage{xspace}
\usepackage{xcolor}
\usepackage{tikz-imagelabels}

\imagelabelset{
    image label font = \normalfont\small,
    image label distance = 0.5mm,
    image label back = white,
    image label text = black,
    coordinate label font = \bfseries\scriptsize,
    coordinate label distance = 2mm,
    coordinate label back = white,
    coordinate label text = black,
    annotation font = \normalfont\small,
}

\newcommand{\logOH}{$12 + \log\left(\frac{\text{O}}{\text{H}}\right)$\xspace}
\newcommand{\HI}{\ion{H}{1}\xspace}
\newcommand{\HII}{\ion{H}{2}\xspace}
\newcommand{\OII}{[\ion{O}{2}]\xspace}
\newcommand{\NII}{[\ion{N}{2}]\xspace}
\newcommand{\OIII}{[\ion{O}{3}]\xspace}
\newcommand{\Htwo}{H$_2$\xspace}
\newcommand{\MHI}{$M_\text{H{\sc i}}$\xspace}
\newcommand{\RHI}{$R_\text{H{\sc i}}$\xspace}
\newcommand{\MHItot}{$M_{\text{H{\sc i},tot}}$\xspace}
\newcommand{\MHIout}{$M_{\text{H{\sc i},out}}$\xspace}
\newcommand{\MHtwo}{$M_{\text{H}_{2}}$\xspace}
\newcommand{\logMHtwo}{$\log(M_{\text{H}_2})$\xspace}
\newcommand{\MHe}{$M_\text{He}$\xspace}

\newcommand{\Halpha}{H$\alpha$\xspace} 
\newcommand{\Hbeta}{H$\beta$\xspace} 

\newcommand{\Mdust}{$M_\text{dust}$\xspace}
\newcommand{\MdZ}{$M_\text{dZ}$\xspace}
\newcommand{\Mgas}{$M_\text{gas}$\xspace}
\newcommand{\Mvis}{$M_\text{vis}$\xspace}
\newcommand{\Mtot}{$M_\text{tot}$\xspace}

\shorttitle{Visible Mass in Disk Galaxies}
\shortauthors{Ravi, Douglass \& Demina}

\begin{document}

\title{A Full Accounting of the Visible Mass in SDSS MaNGA Disk Galaxies}

\correspondingauthor{Nitya Ravi}
\email{nravi3@ur.rochester.edu}

\author[0000-0002-4248-2840]{Nitya Ravi}
\affiliation{Department of Physics \& Astronomy, University of Rochester, 500 Joseph C. Wilson Blvd., Rochester, NY  14627}
 
\author[0000-0002-9540-546X]{Kelly A. Douglass}
\affiliation{Department of Physics \& Astronomy, University of Rochester, 500 Joseph C. Wilson Blvd., Rochester, NY  14627}

\author[0000-0002-7852-167X]{Regina Demina}
\affiliation{Department of Physics \& Astronomy, University of Rochester, 500 Joseph C. Wilson Blvd., Rochester, NY  14627}

\begin{abstract}
  We present a study of the ratio of visible mass to total mass in spiral galaxies to better understand the relative amount of dark matter present in galaxies of different masses and evolutionary stages.  
  Using the velocities of the \Halpha emission line measured in spectroscopic observations from the Sloan Digital Sky Survey (SDSS) MaNGA Data Release 17 (DR17), we evaluate the rotational velocity of over 5500 disk galaxies at their 90\% elliptical Petrosian radii, $R_{90}$.  We compare this to the velocity expected from the total visible mass, which we compute from the stellar, \HI, \Htwo, and heavy metals and dust masses. \Htwo mass measurements are available for only a small subset of galaxies observed in SDSS~MaNGA~DR17, so we derive a parameterization of the \Htwo mass as a function of absolute magnitude in the $r$~band using galaxies observed as part of SDSS~DR7. With these parameterizations, we calculate the fraction of visible mass within $R_{90}$ that corresponds to the observed velocity. Based on statistically analyzing the likelihood of this fraction, we conclude that the null hypothesis (no dark matter) cannot be excluded at a confidence level better than 95\% within the visible extent of the disk galaxies.  We also find that when all mass components are included, the ratio of visible-to-total mass within the visible extent of star-forming disk galaxies increases with galaxy luminosity. 
  
\end{abstract}

\section{Introduction}

Current cosmological models indicate that the dominant component of matter in the Universe is dark matter \citep{Planck20}, which manifests itself primarily through gravity. Dark matter is expected to have minimal to no interaction with the electromagnetic force, therefore emitting little to no light. It is also unlikely to participate in the strong interaction, since otherwise it would be embedded in nuclei. It is currently unclear whether or not dark matter engages in the weak interactions \citep[see][and references therein]{Porter11}.

Phenomena such as gravitational lensing around galaxy clusters \citep[see][and references therin]{Bartelmann10} and galaxy kinematics \citep[e.g.,][]{Freeman70,Bosma78,Carignan85,Salucci19} contribute to the observational evidence for dark matter across most scales in the Universe. Constraints from big bang nucleosynthesis \citep{Yao06} and detailed measurements of the imprint of baryon acoustic oscillations on the anisotropy of the cosmic microwave background \citep{Komatsu11} strongly suggest that dark matter is of a nonbaryonic nature. Simulations based on cold dark matter models are able to reproduce the current distribution of galaxies \citep[e.g.,][]{Springel05}, indicating that dark matter is likely composed of heavy, weakly interacting particles. However, ground-based experiments have failed to observe any effects associated with the passage of such particles through normal matter \citep{Boveia18}. Moreover, results from the Large Hadron Collider exclude most models that offer plausible candidates for dark matter \citep[for the latest results, see][]{CMS:2022qva, ATLAS:2022ygn, ATLAS23, Tumasyan23}. Hence, solving the puzzle of dark matter is one of the leading problems currently faced by the scientific community. 

Modern large-scale galaxy surveys offer high-quality data that allow us to reevaluate the astronomical evidence for the existence of dark matter. One of the original sources of such evidence was galactic rotation curves \citep{Rubin:1970zza, Rubin:1980zd, Rubin:1982kyu, Rubin:1985ze}. These studies were based on samples with low statistics, containing only about 20 galaxies. The expected rotational velocities of galaxies were estimated based only on stellar mass and did not include gas or dust. Since the 1980s, rotation curve analysis has been performed on larger galaxy samples to study various galaxy properties. \cite{Mathewson92} analyzed long-slit spectroscopy, where velocities were measured along the semi-major axes of galaxies, to construct the rotation curves of over 900 galaxies. \cite{Persic96} analyzed the rotation curves of the same sample and found that the stellar disk did not contain sufficient matter to produce the observed rotation curve.

Other studies that construct rotation curves from long-slit spectroscopy \citep[e.g.,][]{Catinella06, DiTeodoro21} support the observation that rotation curves ubiquitously flatten at the outer radii of galaxies and find that the stellar mass scales with the inferred mass of the dark halos. More recently, studies have fit rotation curves to stellar and gas velocity fields using integral field spectroscopy \citep[e.g.,][]{deBlok08, TorresFlores11, Kalinova17, Schmidt23} for tens to hundreds of galaxies to estimate the galaxies' dynamical masses and model dark matter halo profiles. \cite{Douglass19} and \cite{Yoon21} each study the rotation curves of almost 2000 Sloan Digital Sky Survey (SDSS) Mapping Nearby Galaxies at Apache Point Observatory (MaNGA) galaxies using either gas or stellar kinematics.  Because of the large variations in galaxy properties throughout these samples, one of the biggest short-comings of these prior studies has been their limited statistical power.

In this paper, we reevaluate the amount of dark matter needed to explain the observed rotational velocities and revisit the statistical significance of the null hypothesis using the high statistics afforded by SDSS MaNGA \citep{MaNGA}. The null, i.e., ``no dark matter," hypothesis is that the rotation of a disk galaxy can be explained by its visible mass --- galaxies do not have a dark matter halo \citep[e.g.,][]{Sellwood01, vanDokkum18}.

We analyze the rotation curves of over 5500 galaxies in SDSS MaNGA Data Release 17 \citep[DR17;][]{SDSS17} to study the dark matter content of spiral galaxies. We construct models of rotation curves for each galaxy using the \Halpha emission-line velocities measured across a galaxy's surface. Based on the rotational velocity, we infer the value of the total (gravitational) mass and compare it to the visible mass. A similar analysis was conducted on 1988 galaxies in SDSS MaNGA DR15 \citep{SDSS15} in \cite{Douglass22}, where visible mass was defined as the sum of the stellar and atomic hydrogen masses. The ratios of visible to total mass for these galaxies were studied by splitting the sample into three subsamples based on color-magnitude classification and analyzing the mass ratios' dependence on the luminosity, gas-phase metallicity, and color-magnitude classification.

We improve on these earlier studies \citep[e.g.,][]{TorresFlores11, DiPaolo19, Douglass22} by defining the visible mass as the sum of the stellar, neutral atomic hydrogen, molecular hydrogen, helium, heavy metal, and dust masses.  We present the ratio of visible to total mass as a function of galaxy luminosity. For each galaxy in our sample, we construct a statistical model that accounts for the statistical and systematic uncertainties on the measured rotational velocity, as well as the uncertainties on each of the visible mass components. Using this statistical model, we evaluate the level of consistency of the observed rotational velocities with the null, i.e., ``no dark matter,'' hypothesis.  

The paper is structured as follows. In Section~\ref{sec:data}, we discuss the data selection process. In Section~\ref{sec:model}, we describe the modeling of the rotation curves and stellar mass distributions. In Section~\ref{sec:mass}, we detail the estimation of the mass components. In Section~\ref{sec:stat}, we describe the statistical model. We present the results  in Section~\ref{sec:results}, and we conclude in Section~\ref{sec:conclusion}.

\section{SDSS MaNGA DR17 and Galaxy Selection}\label{sec:data}

We use the \Halpha emission-line velocity maps from SDSS MaNGA DR17 \citep{SDSS17} to model the rotation curves of spiral galaxies. The SDSS MaNGA survey used integral field spectroscopy to measure spectra at different points throughout a galaxy by placing an integral field unit (IFU) on each galaxy. The IFU is a bundle of spectroscopic fibers arranged in a hexagonal shape containing between 19 and 127 fibers and covering 12.5" to 32.5" in diameter \citep{Law15}. The light received by the fibers was fed to two spectrographs with wavelength ranges 3600--6000{\AA} and 6000--10300{\AA}, respectively, with a resolution of $\lambda/\Delta \lambda \sim 2000$ \citep{Drory15}.

SDSS MaNGA DR17 is the final data release for the MaNGA survey and contains more than 10,000 nearby galaxies in the northern sky. The target selection process prioritized maintaining a flat distribution in luminosity \citep{Wake17}, so the survey consists of three subsamples: the primary sample, with the IFU covering out to $1.5R_e$, where $R_e$ is the half-light radius of a galaxy; the secondary sample, covering out to $2.5R_e$; and the color-enhanced sample, which supplements the primary sample with high-mass blue galaxies and low-mass red galaxies. In order to check for possible systematic bias, we present the results of our analysis for the entire data set and each of these individual subsamples, referred to as MaNGA sample 1, 2, and 3, respectively.

We extract each galaxy's rotation curve using the \Halpha velocity map and $g$-band-weighted mean flux map as processed by the MaNGA Data Analysis Pipeline \citep[DAP;][]{Westfall19}. The stellar mass rotation curve is extracted from the stellar mass density maps processed by Pipe3D \citep{Sanchez16, Sanchez18}. Absolute magnitudes are obtained from version 1.0.1 of the NASA-Sloan Atlas \citep{Blanton11}.  Distances are in units of Kpc~$h^{-1}$, where $h$ is the reduced Hubble constant defined by $H_0 = 100~h$ km s$^{-1}$ Mpc$^{-1}$.

\subsubsection{SDSS DR7}

SDSS DR7 \citep{SDSS7} observed approximately one quarter of the northern sky in both photometry and spectroscopy.  A dedicated 2.5~m telescope at the Apache Point Observatory in New Mexico with a wide-field imager and a pair of double fiber-fed spectrometers was used to conduct the multiband imaging and spectroscopic survey. Photometric data was taken in the five SDSS filters: $u$, $g$, $r$, $i$, and $z$ \citep{Fukugita96, Gunn98}. Using 320 fibers placed into fiber plug plates with a minimum fiber separation of 55", follow-up spectroscopic analysis was performed on galaxy targets with Petrosian $r$-band magnitudes $m_r \leq 17.77$ and $r$-band Petrosian half light radii $\mu_{50} \leq 24.5$~mag~arcsec$^{-2}$ \citep{Lupton01,Strauss02}. For SDSS DR7, the spectrometers covered a wavelength range of 3800--9200{\AA} with a resolution of $\lambda/\Delta \lambda \sim 1800$ \citep{Smee13}.  

We make use of the photometric data for MaNGA galaxies available from the Korea Institute for Advanced Study Value-Added Galaxy Catalog \citep[KIAS-VAGC;][]{Choi10}. The KIAS-VAGC is based on SDSS DR7 and the New York University Value-Added Galaxy Catalog \citep[NYU-VAGC;][]{Blanton05}. The NYU-VAGC contains multiple crossmatched galaxy catalogs including SDSS   and independently reduced data from SDSS. We use the $u - r$ color, $\Delta(g - i)$ color gradient, and inverse concentration index from the KIAS-VAGC, which are calculated using the NYU-VAGC data. Colors are calculated using fluxes within the $r$-band Petrosian radius. $\Delta(g - i)$ is the difference in the $(g - i)$ color between the region within $0.5R_{Pet}$ and the annulus between $0.5R_{Pet}$ and $R_{Pet}$, where $R_{Pet}$ is the $i$-band Petrosian radius. A galaxy with a negative color gradient is bluer on the outside, whereas a galaxy with a positive color gradient is redder on the outside. The inverse concentration index, $c_{inv}$, is measured as $c_{inv} = R_{50, i}/R_{90,i}$, where $R_{50,i}$ and $R_{90,i}$ are the $i$-band 50\% and 90\% Petrosian radii, respectively. We use the global emission line fluxes from the Portsmouth group galaxy properties catalog \citep{Thomas13} to calculate the gas-phase metallicity.  

\subsubsection{\HI Observations}

\HI mass estimates are obtained from the \HI--MaNGA DR3 \citep{Stark21}. \HI--MaNGA is a follow-up survey of MaNGA galaxies conducted on the Robert C. Byrd Green Bank Telescope (GBT) in Green Bank, West Virginia. MaNGA galaxies with redshifts z $< 0.05$ are observed in the GBT $L$ band (1.15--1.73 GHz). The third data release of \HI--MaNGA has HI observations of 3358 MaNGA galaxies from GBT and also includes a crossmatch between the Arecibo Legacy Fast ALFA (ALFALFA) survey \citep{Haynes18} performed at the Arecibo Observatory in Arecibo, Puerto Rico, and MaNGA DR17 targets. ALFALFA was a blind 21 cm survey with observations of 31,500 \HI sources within z $< 0.06$ using the Arecibo $L$-band Feed Array (1.355--1.435 GHz). \HI--MaNGA DR3 includes 3,274 of these sources, which are crossmatched with MaNGA and have a redshift z $< 0.05$.

\subsubsection{CO Observations}

\Htwo masses are inferred from measurements of the CO(1-0) line emission from two surveys: the MaNGA Arizona Radio Observatory (ARO) Survey of Targets (MASCOT) first data release \citep{Wylezalek22} and the xCOLD GASS survey \citep{Saintonge17}. The MASCOT survey performs observations of MaNGA galaxies at the ARO using the 12~m ARO antenna with a 3 mm receiver with frequency range 84--116 GHz. MASCOT has observations of the CO(1-0) emission line for 187 galaxies selected from MaNGA DR15 with stellar masses greater than $10^{9.5}M_\odot$. The xCOLD GASS survey conducted CO(1-0) observations of SDSS galaxies on the IRAM 30-meter telescope in Spain. xCOLD GASS targets were selected from SDSS DR7 with redshifts $0.01 < \text{z} < 0.05$ and stellar masses greater than $10^9 M_\odot$. xCOLD GASS has observations of 532 SDSS~DR7 galaxies. We crossmatch 41 galaxies from xCOLD GASS with MaNGA DR17, 2 of which also have observations in MASCOT. Excluding CO nondetections, we have a total of 204 galaxies with CO observations in MaNGA DR17 from MASCOT and xCOLD GASS combined.


\subsection{Color-Magnitude Classification}

\begin{figure}
    \includegraphics[width=0.45\textwidth]{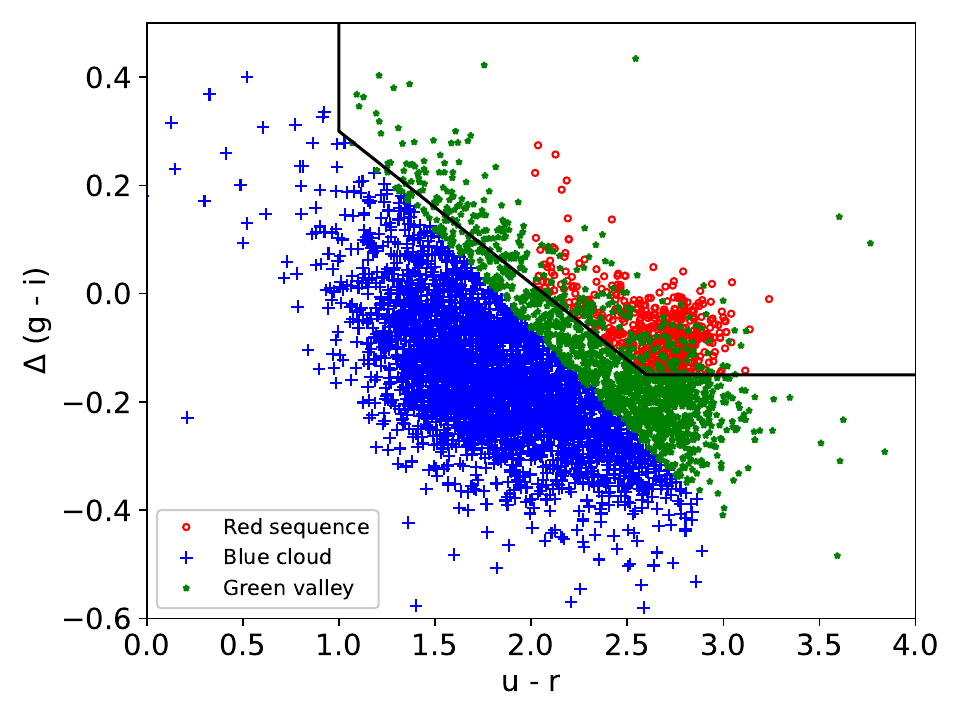}
    \caption{$\Delta (g - i)$ color gradient vs. $u - r$ color for our 
    sample of SDSS MaNGA galaxies with stellar mass estimates, marked by their 
    CMD classification: open red circles for the red sequence, 
    green stars for the green valley, and blue crosses for the blue cloud.  The 
    black boundary is the separation between early- and late-type galaxies as 
    defined by \cite{Choi10}.
    \label{fig:CMD_class}}
\end{figure}

As shown in \cite{Douglass22}, a galaxy's ratio of total to stellar mass 
depends on the galaxy's evolutionary stage.  We therefore separate the galaxies 
into three populations---blue cloud, green valley, and red sequence---in the 
color-magnitude diagram (CMD) to better understand these relationships.  Galaxies 
in the blue cloud are typically fainter and more blue, while galaxies in the red
sequence are brighter and more red.  It is believed that galaxies transitioning 
between the blue cloud and red sequence occupy the green valley \citep{Martin07}.

We use the same method to classify the galaxies into one of these three 
populations as used in \cite{Douglass22}, where the classification is based on the 
inverse concentration index, $c_\text{inv}$, color, $u - r$, and color gradient, 
$\Delta (g - i)$.  As shown in Figure~\ref{fig:CMD_class}, galaxies that are part of 
the red sequence are those that generally fall above and to the right of the 
depicted boundary originally defined by \cite{Park05} (normal early-type galaxies), while galaxies that are part of the 
blue cloud are those that generally fall below and to the left of the boundary 
(late-type galaxies).  Galaxies that are part of the green valley are either those 
above the boundary but with $u - r < 2$ (blue early-type galaxies) or a high 
$c_\text{inv}$, or those below the boundary with $\theta < 20^\circ$, where
\begin{equation}
    \theta = \tan^{-1} \left( \frac{-\Delta (g - i) + 0.3}{(u - r) - 1} \right).
\end{equation}
See \cite{Douglass22} for a more detailed description of the CMD classification.

In this study, we analyze the rotation curves of galaxies. Thus, we require our objects to be dominated by rotational motion 
(described in Section~\ref{sec:model} below). Disk galaxies have velocity dispersions which are small compared to their rotational velocities, so the total dynamical mass of a disk galaxy can be calculated assuming that the centripetal acceleration is due to the gravitational force \citep{Sofue17}. As a result, we expect few 
galaxies in our sample to be in the red sequence.  After visual inspection, we 
find that the red sequence galaxies that are in our final sample appear to be 
either red-disk galaxies with little to no star formation (likely lenticulars) 
or elliptical galaxies that are still supported by rotation.

\section{Modeling the rotation curves and stellar mass distribution}\label{sec:model}

\subsection{Modeling the Velocity Map}\label{sec:Vmodel}

\begin{figure}
    \centering
    \includegraphics[width=0.49\textwidth]{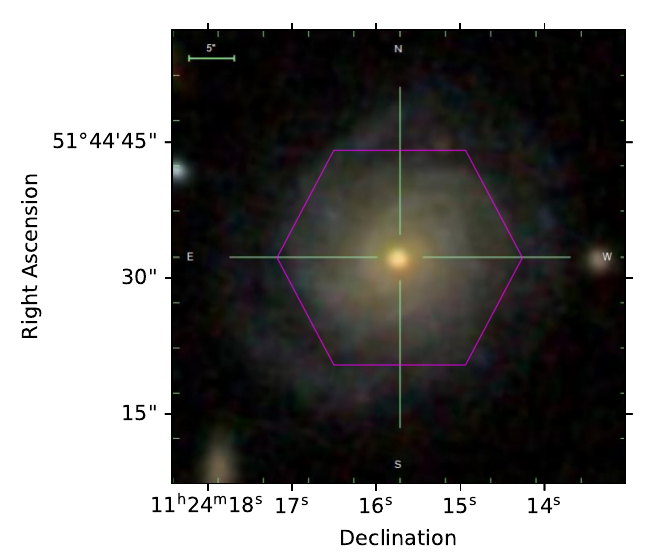}
    \caption{IFU (magenta hexagon) overlaid on RGB composite image of MaNGA galaxy 8997--9102 \citep[made with the SDSS Marvin python package by][]{SDSS-Marvin}. The IFU does not cover the entire visible extent of the galaxy, as is common for MaNGA observations.}
    \label{fig:IFU}
\end{figure}

\begin{figure*}
    \centering
    \includegraphics[height=0.2\textheight]{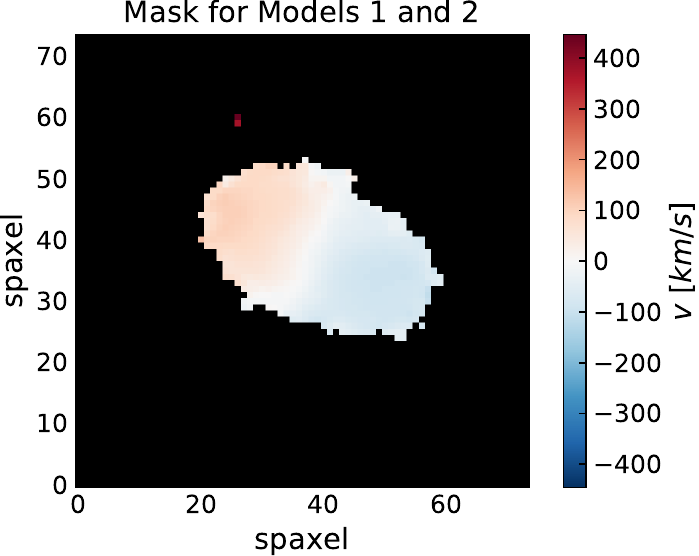} 
    \includegraphics[height=0.2\textheight]{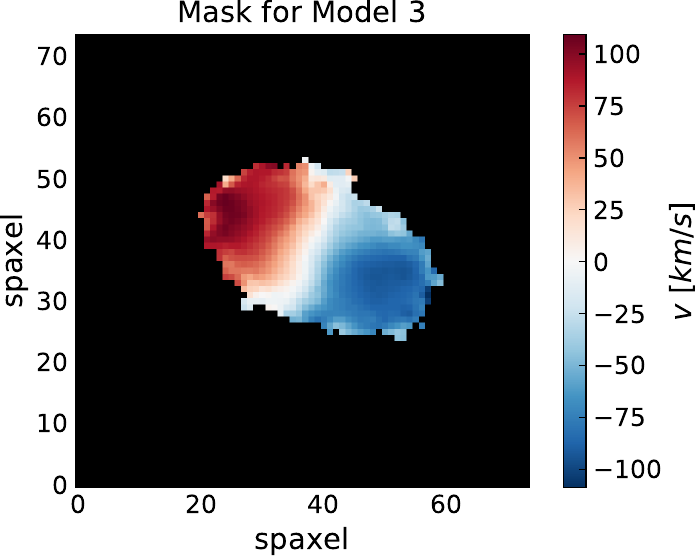}
    \includegraphics[height=0.2\textheight]{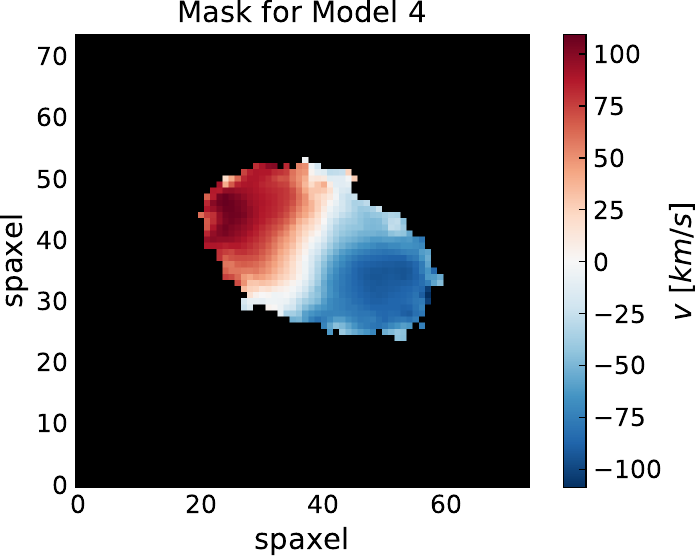}\\
    \vspace{1em}
    \includegraphics[width=0.45\textwidth]{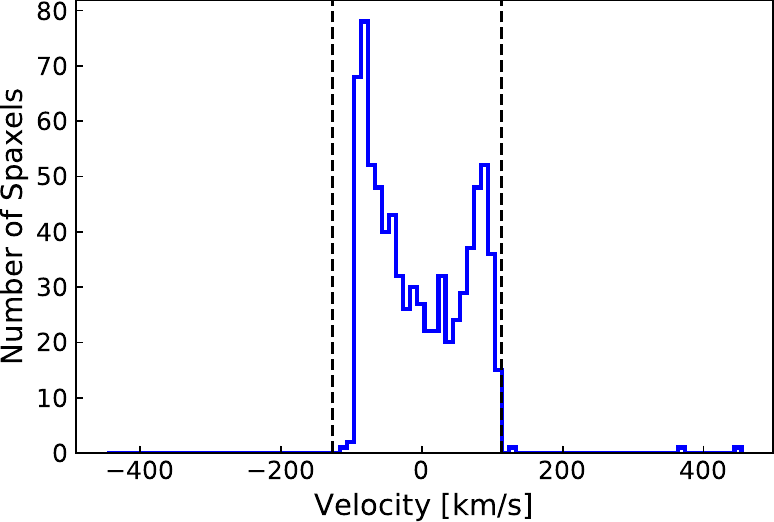}
    \includegraphics[width=0.45\textwidth]{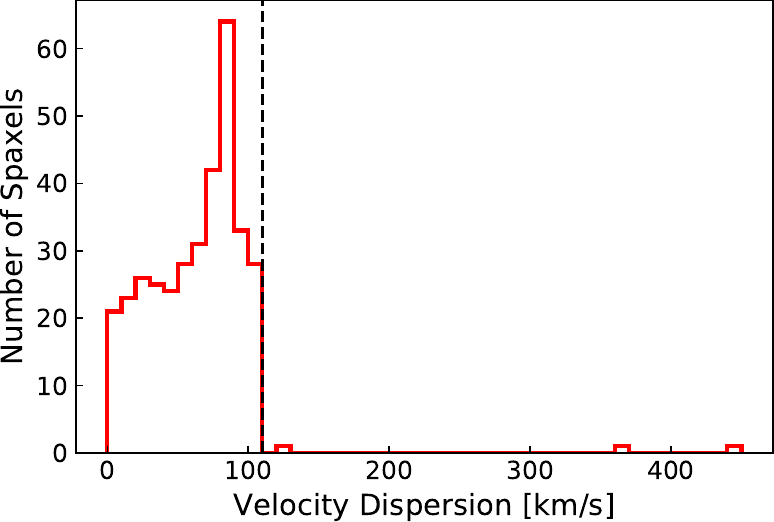}
    \caption{Masks for the different velocity map models for example galaxy 10001--12701. The masks for models 1 and 2 is shown to the top left, the mask for model 3 is shown in the center, and the mask for model 4 is shown to the top right. The histogram on the bottom left shows the distribution of unmasked spaxel velocities used in models 1 and 2. Model 3 masks spaxels outside of the vertical dashed lines. The histogram to the bottom right shows the distribution of unmasked spaxel velocity dispersions used in models 1 and 2. Model 4 masks spaxels to the right of the vertical dashed line.  Note that masking the outlying spaxels in the velocity distribution reduces the dynamic range of the velocity gradient (indicated by the colormap) to that expected for a rotating disk galaxy.}
    \label{fig:mask}
\end{figure*}

\begin{figure*}
    \centering
    \includegraphics[height=0.16\textheight]{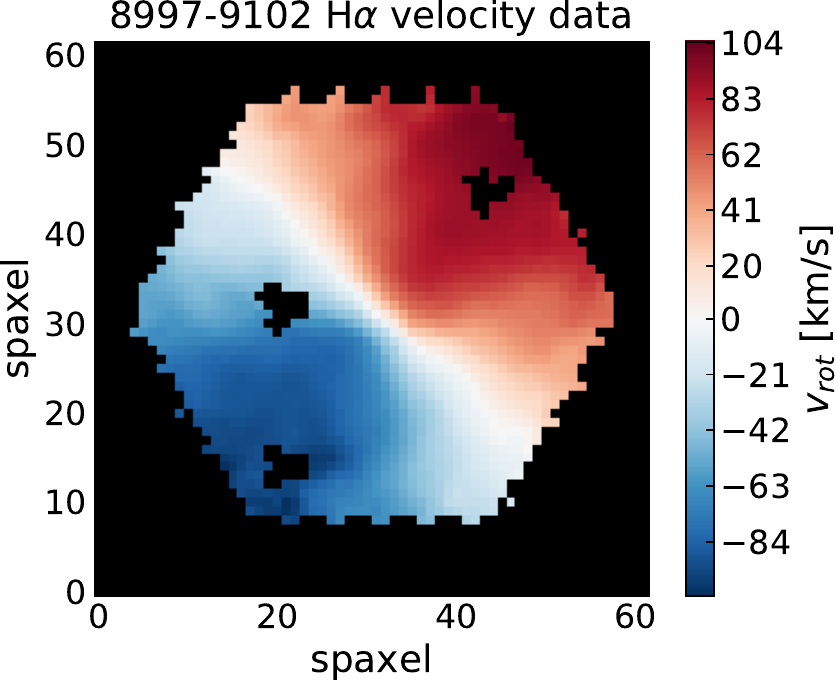}
    \includegraphics[height=0.16\textheight]{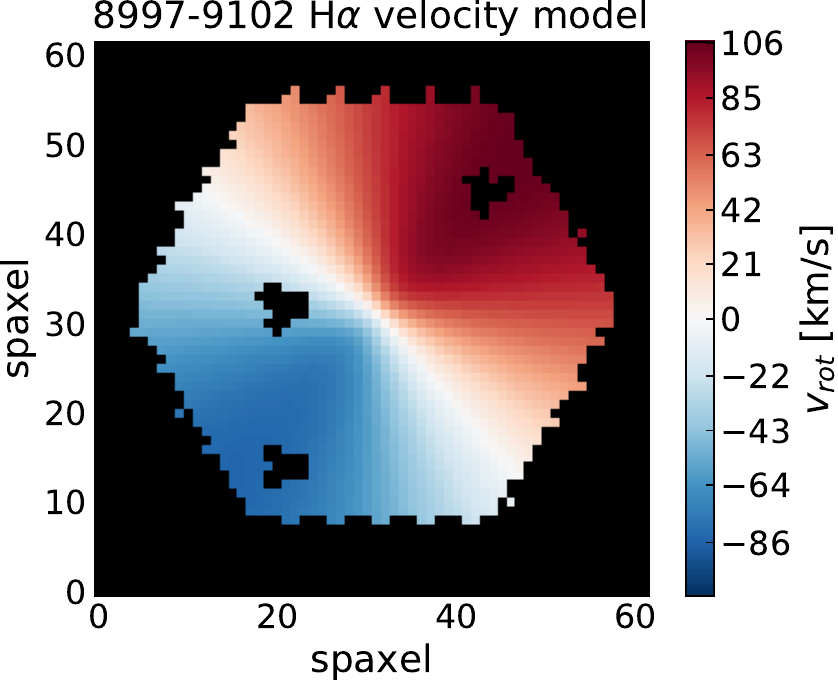}
    \includegraphics[height=0.16\textheight]{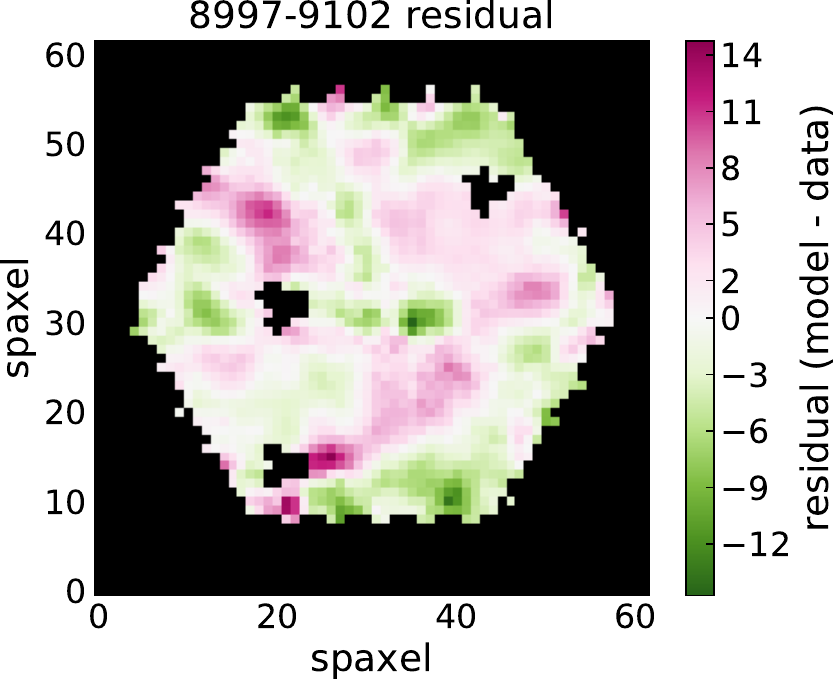}
    \includegraphics[height=0.165\textheight]{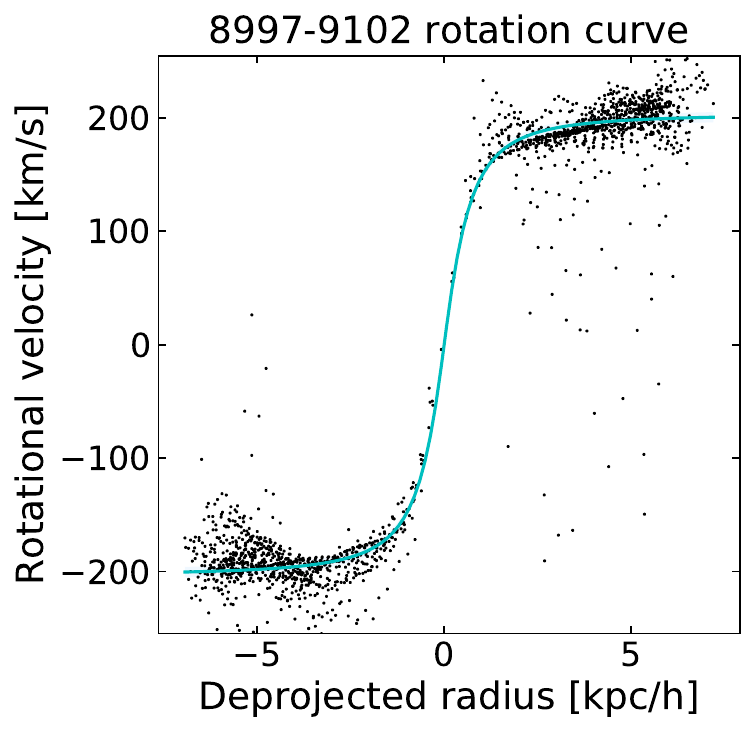}
    \caption{Example \Halpha velocity map from the MaNGA DAP (first column), our best fit model to the velocity map (second column), the residual between the velocity map and our best-fit model (third column), and the deprojected rotation curve for the galaxy (fourth column).}
    \label{fig:model_map}
\end{figure*}

We estimate a galaxy's total dynamical mass using its \Halpha velocity map obtained from the SDSS MaNGA DAP. We restrict our analysis to spaxels with a data quality bit of 0, as provided by the SDSS MaNGA DAP. Spaxels with a nonzero bit value indicate data that experienced issues in observations or in the data analysis process \citep{Westfall19}. In addition, we only include spaxels with signal-to-noise ratio $\geq 5$ in the H$\alpha$ flux to ensure that only spaxels with a significant detection, and therefore a reliable redshift, are considered in the analysis.

We also require that all galaxies have a smooth velocity gradient with a maximum ``smoothness score'' of 2.0, as described in \cite{Douglass19}. We further restrict the analysis to galaxies with a T-Type $> 0$ (late-type galaxies) as classified by the MaNGA Morphology Deep Learning DR17 Value Added Catalog \citep{DominguezSanchez22}.

Similar to both \cite{Douglass19} and \cite{Douglass22}, the velocity map of each galaxy is fit to the rotation curve parameterization defined in \cite{BarreraBallesteros18}: 
\begin{equation}\label{eqn:rot_curve}
    V(r) = \frac{V_{\text{max}} r}{(R^\alpha_{\text{turn}} + r^\alpha)^{1/\alpha}},
\end{equation}
where $V(r)$ is the rotational velocity at a distance $r$ from the center of the galaxy. The free parameters are $V_{\text{max}}$, the magnitude of the velocity at which the rotation curve plateaus; $R_{\text{turn}}$, the radius at which the rotation curve changes from increasing to flat; and $\alpha$, which describes the sharpness of the curve. The extent of the MaNGA \Halpha velocity maps and the radii to which we can measure rotational velocities are limited by the visible extent of the galaxy. Rotation curves are only fit out to the maximum radius, $R_\text{max}$, covered by the IFU, the extent of which is shown for an example galaxy in Figure~\ref{fig:IFU}.

Each galaxy's systemic velocity, kinematic center, inclination angle, and position angle are also free parameters in this fit, resulting in a total of eight free parameters. 
When determining the best-fit model for each galaxy, we make use of $\chi^2 = \Sigma ((\text{data} - \text{model})/\text{uncertainty})^2$ and $\chi^2_\nu$, where we normalize $\chi^2$ by the difference between the number of unmasked spaxels in the velocity map and the number of free parameters in the fit. We define four best-fit models, as follows:
\begin{description}
  \item[Model 1] the model with the smallest $\chi^2$.
  
  \item[Model 2] the model with the smallest residual, $\Sigma (\text{data}-\text{model})^2$.
  
  \item[Model 3] to help remove foreground artifacts from the analysis, we define upper- and lower-velocity bounds by binning all unmasked spaxels with a bin width of 10~km~s$^{-1}$. The velocity bounds are defined as the nearest empty bin on either side of the bin with the most spaxels, as shown in the histogram to the bottom left of Figure~\ref{fig:mask}. Spaxels with values outside of this velocity range are masked; see the top center of Figure~\ref{fig:mask} for an example of the resulting mask. We then select the model with the smallest $\chi^2$.
  
  \item[Model 4] to help remove spaxels that are potentially contaminated by emission from active galactic nuclei, which are defined as bins with an unusually high velocity dispersion, we define an upper limit on the velocity dispersion by binning the velocity dispersion of the unmasked spaxels with a bin width of 10~km~s$^{-1}$. The velocity dispersion upper bound is defined as the nearest empty bin to the lowest velocity dispersion bin containing spaxels, as shown in the histogram to the bottom right of Figure~\ref{fig:mask}. Spaxels with velocities above this upper limit are masked; see the top right plot in Figure~\ref{fig:mask} for an example of the resulting mask.  We then select the model with the smallest $\chi^2$.
\end{description}

Of these four models, we eliminate those with $\alpha = 100$, where $\alpha$ is the parameter from Equation~(\ref{eqn:rot_curve}). This eliminates models where the fitting algorithm failed, as 100 is the upper limit of $\alpha$ while fitting. Of the remaining models, we select the one with the lowest $\chi^2_\nu$ as the best-fit model for each galaxy. An example \Halpha velocity map and best-fit model map is shown in Figure~\ref{fig:model_map}.

\subsection{Modeling the Stellar Mass}\label{sec:Vstar}

\begin{figure}
    \centering
    \includegraphics[width=0.45\textwidth]{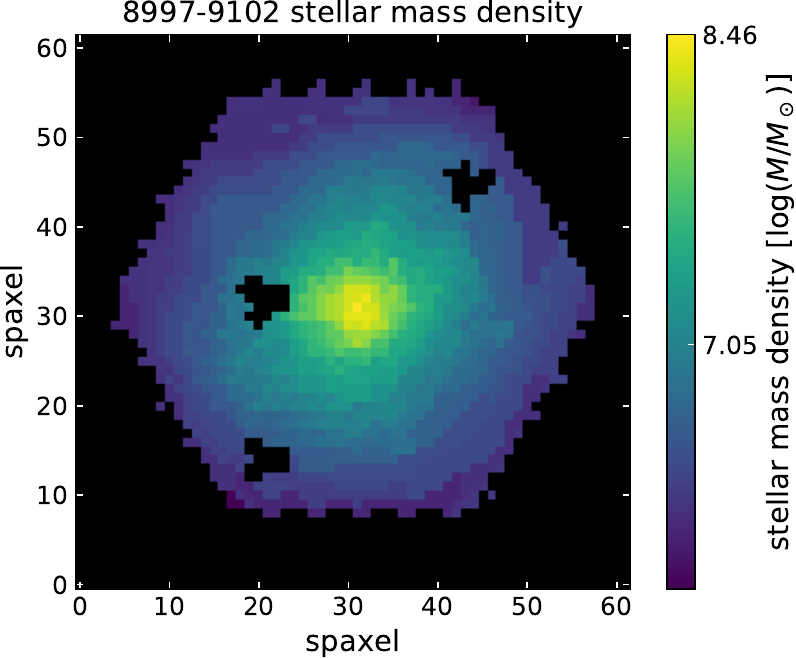}
    \includegraphics[width=0.49\textwidth]{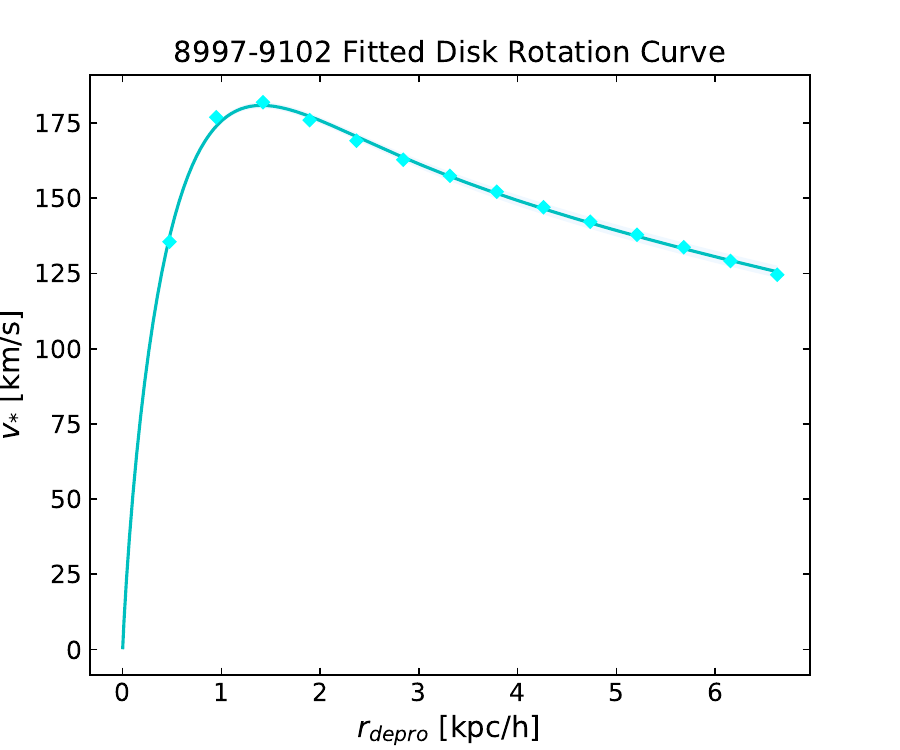}
    \caption{Example stellar mass density map extracted from MaNGA Pipe3D (top) and our best fit to the rotation curve extracted from this map (bottom).}
    \label{fig:smass_fit}
\end{figure}

We estimate each galaxy's stellar mass by fitting a rotation curve due to the stellar component of the galaxy using the stellar mass density maps available through the Pipe3D MaNGA analysis pipeline \citep{Sanchez16, Sanchez18}.  An example stellar mass density map is shown at the top of Figure~\ref{fig:smass_fit}. Using the best-fit model H$\alpha$ velocity map values for the galaxy's kinematic center, inclination angle, and position angle described above, we define concentric ellipses that correspond to different orbital radii in the galaxy, with the radius of each ellipse increasing by 2 spaxels. We compute the stellar mass as a discretized function of radius, $M_*(r)$, by summing the stellar mass density per spaxel over all the spaxels within each ellipse. 

We assume that the stellar mass is the primary component of the galaxy's disk and model the stellar mass as the sum of a central bulge and exponential disk. The rotational velocity due to the bulge and disk is summed in quadrature to get the rotational velocity due to the stellar mass:
\begin{equation}\label{eqn:Vstar_total}
    V_* (r)^2 = V_b(r)^2 + V_d(r)^2,
\end{equation}
where $V_*(r)$ is the rotational velocity due to the stellar mass, $V_b(r)$ is the rotational velocity due to the bulge component, and $V_d(r)$ is the rotational velocity due to the disk component.

The bulge is modeled as an exponential sphere \citep{Sofue17} with rotational velocity
\begin{equation}
    V_b(r)^2 = \frac{GM_0}{R_b}\, F \left(\frac{r}{R_b} \right),
\end{equation}
where $G = 6.67408\times 10^{-11}$~m$^3$ kg$^{-1}$ s$^{-2}$ is the Newtonian gravitational constant, $F(x) = 1 - e^{-x}(1 + x + 0.5x^2)$ and $M_0 = 8\, \pi\, R_b^3\,\rho_b$.  The free parameters in this fit are the scale radius of the bulge, $R_b$, and the central density of the bulge, $\rho_b$.

The rotational velocity due to the exponential disk \citep[a thin disk without perturbation;][]{Freeman70} is
\begin{equation}\label{eqn:V_disk}
    V_d(r)^2 = 4\pi G\Sigma_d R_d y^2 [I_0 (y) K_0 (y) - I_1 (y) K_1 (y)],
\end{equation}
where $\Sigma_d$ is the central surface mass density of the disk, $R_d$ is the scale radius of the disk, $y = r/2R_d$, and $I_i$ and $K_i$ are the modified Bessel functions \citep{Sofue13}.  The free parameters in this fit are $\Sigma_d$ and $R_d$.

\section{Estimating the mass components}\label{sec:mass}

\subsection{Total Mass, $M_\text{tot}$}\label{sec:Mtot}

\begin{figure}
    \centering
    \includegraphics[width=0.49\textwidth]{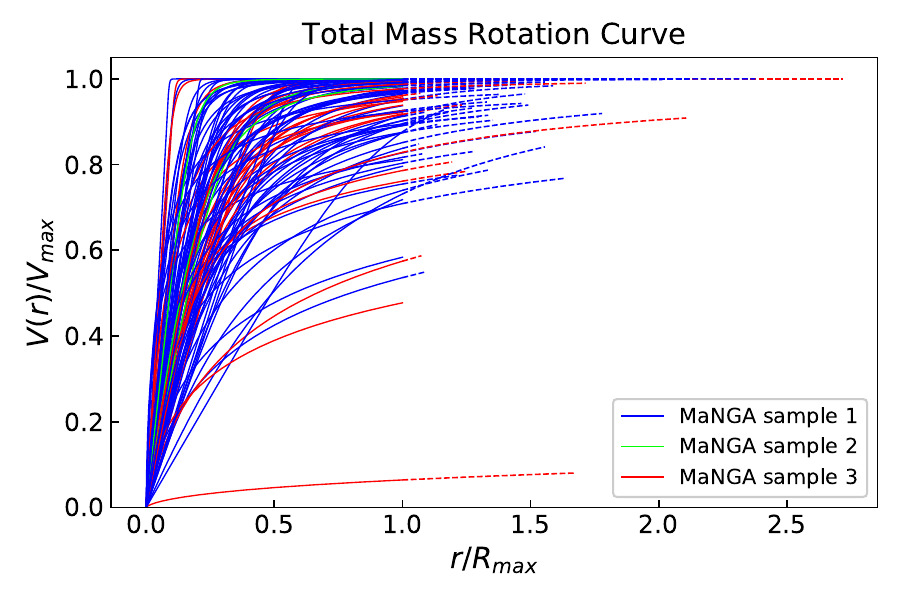}
    \includegraphics[width=0.49\textwidth]{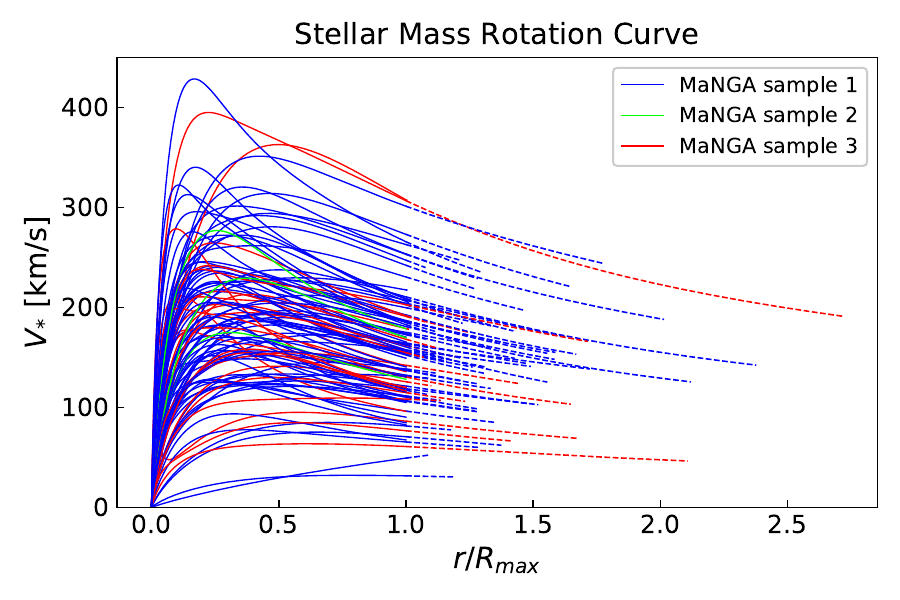}
    \caption{Rotation curves of the 107 MANGA galaxies with \HI and \Htwo masses from the \Halpha velocity field (top) and the stellar mass component (bottom). The solid lines extend to the maximum observed distance for each galaxy, $R_\text{max}$, and the dashed lines show the extrapolation of the model to $R_{90}$. The colors correspond to the different MaNGA samples.}
    \label{fig:rot_curves}
\end{figure}

We calculate the galaxy's total dynamical mass within the 90\% elliptical Petrosian radius, $R_{90}$, using the rotational velocity at this radius as determined from the best-fit rotation curve, found as described in Section~\ref{sec:Vmodel}.  We can calculate the mass of a galaxy within some radius $r$ from the center of the galaxy under the assumption that the galaxy's rotational motion is dominated by Newtonian orbital mechanics. Assuming axial symmetry, the velocity of a particle at distance $r$ from the center of the galaxy is a function of the mass within that radius, $M(r)$. Assuming that the orbital motion is circular in spiral galaxies, the centripetal acceleration of an orbiting particle is due to the gravitational force: 
\begin{equation}\label{eqn:M_within_r}
    M(r) = \frac{V(r)^2 r}{G}.
\end{equation}
Here, $V(r)$ is the rotational velocity a distance $r$ from the center of the galaxy. In order to study the same region of each galaxy, we estimate the mass within $R_{90}$, $M(R_{90}) = \text{\Mtot}$, by calculating $V(R_{90})$ from Equation~(\ref{eqn:rot_curve}). When $R_\text{max} < R_{90}$, we extrapolate our parameterization of the fitted rotation curve, Equation~(\ref{eqn:rot_curve}), out to $R_{90}$. On average, $R_{90}$ exceeds $R_{\text{max}}$ by about 10\%. Figure~\ref{fig:rot_curves} shows a subset of our rotation curves normalized by $R_\text{max}$, where the curves extrapolated out to $R_{90}$ for the galaxies with $R_{90} > R_\text{max}$ are shown as dashed extensions beyond $r/R_\text{max} = 1$. The stellar and \HI masses are also evaluated within $R_{90}$. Only global measurements are available for the remaining mass components, but these are expected to be concentrated within the visible disk and thus are also within $R_{90}$.

While the majority of the stellar mass is encompassed by $R_{90}$, gas and dark matter profiles are known to extend much farther than that \citep[e.g.,][]{Ostriker74, Begeman89, Kamphuis92, Pohlen10}.  Extrapolating the rotation curves to higher radii would significantly increase the uncertainty on the rotational velocity, so we focus our study on the mass content within the visible extent of the galaxy.

To calculate the total dynamical mass within $R_{90}$, we require:
\begin{enumerate}
    \item $\alpha \leq 99$;
    \item Velocity maps with less than 95\% of their spaxels masked;
    \item 10~km~s$^{-1}$ $< V(R_{90}) < 1000$~km~s$^{-1}$; and
    \item $\sigma_{V_{\text{max}}} /  V_{\text{max}} \leq 2$, where $\sigma_{V_\text{max}}$ is the uncertainty in the best-fit value of $V_\text{max}$.
\end{enumerate}
The first, third, and fourth conditions eliminate unsuccessful fits that result in nonphysical models. The first condition eliminates fits where $\alpha$ approaches the maximum allowed value which indicates an unsuccessful fit. The third condition eliminates galaxies where the inclination angle in the fit is incorrect and approaches the boundary values for the parameter, resulting in a very high or very low $V(R_{90})$. The fourth condition removes models with large uncertainties in $V_\text{max}$, also indicative of an unsuccessful fit. The second condition removes models for velocity maps where too many spaxels are masked and they therefore have too few data points to result in an trustworthy model.

\subsection{Stellar Mass, $M_*$}\label{sec:Star_mass}

To estimate the total stellar mass of each galaxy, $M_*$, within $R_{90}$, we use the parameters from the best-fit disk and bulge rotation curve as described in Section~\ref{sec:Vstar}.  The total mass of the bulge and disk at some radius $r$ is 
\begin{equation}
    M_*(r) = M_b(r) + M_d(r),
    \label{eqn:M_disk_tot}
\end{equation}
where the mass of the bulge component is 
\begin{equation}
    M_b(r) = M_0\, F \left( \frac{r}{R_b} \right)
\end{equation}
and the mass of the disk component is 
\begin{align}
    M_d (r) &= 2\pi \Sigma_d \int_0^r r e^{-r/R_d} \, dr \\
            &= 2\pi \Sigma_d R_d \left[ R_d - e^{-r/R_d} (r + R_d) \right]. \label{eqn:M_disk}
\end{align}

So that we study the stellar mass within the same region of each galaxy as the total mass, we evaluate Equation~(\ref{eqn:M_disk_tot}) at $R_{90}$. We apply a stellar mass cut and remove galaxies with $M_*(R_{90}) < 10^9 M_\odot$ from our analysis so that we can perform the \HI mass scaling described below (Section~\ref{sec:HI_mass}). After applying the quality cuts described in Section~\ref{sec:Mtot} and this stellar mass cut, our final sample consists of 5503 galaxies with best-fit rotation curves.

\subsection{Atomic Hydrogen, \HI}\label{sec:HI_mass}
 
We use the \HI mass from the \HI--MaNGA DR3 survey to quantify the neutral atomic gas content within each galaxy. As listed in Table~\ref{tab:v90_fit}, \HI mass estimates are available for 2588 galaxies in our sample.

We estimate the \HI mass within $R_{90}$ from the total \HI mass following the procedure in \cite{Wang20} for galaxies with $M_* > 10^9 M_\odot$. Using the total \HI mass, we calculate $R_{\text{H{\sc i}}}$, the radius where the \HI density is 1~M$_\odot$~pc$^{-2}$ \citep{Wang16}:
\begin{multline}
    \log(2R_{\text{H{\sc i}}}) = (0.506 \pm 0.003)\log\text{\MHItot}\\ 
    - (3.293 \pm 0.009).
\end{multline}
Here, \MHItot is the total \HI mass obtained from \HI--MaNGA, and $R_{\text{H{\sc i}}}$ is in units of kiloparsecs. We assume that within $R_{\text{H{\sc i}}}$, the \HI density follows the median profile from \cite{Wang16} and outside of $R_{\text{H{\sc i}}}$, it follows an exponential profile with scale radius 0.2$R_{\text{H{\sc i}}}$:
\begin{multline} \label{eqn:HI_profile}
    \Sigma_{\text{H{\sc i}}}(r) =
    \begin{cases}
        10^{0.73-1.3(r/R_{\text{H{\sc i}}}-0.23)^2} & r \leq R_{\text{H{\sc i}}}\\
        e^5 \exp(-r/0.2R_{\text{H{\sc i}}}) & r > R_{\text{H{\sc i}}}.
    \end{cases}
\end{multline}
$\Sigma_{\text{H{\sc i}}}$ is the \HI surface density at radius $r$. We consider 1.5$R_{\text{H{\sc i}}}$ to be the edge of the \HI disk, following \cite{Wang20}. The \HI mass outside some radius $r$ can then be calculated by integrating over the \HI surface density from $r$ to 1.5$R_{\text{H{\sc i}}}$:
\begin{equation} \label{eqn:HI_out}
    \text{\MHIout}(r) = \int_r^{1.5\text{\RHI}}\Sigma_{\text{H{\sc i}}}(r)2\pi r \, dr.
\end{equation}
If $R_{90}$ is greater than 1.5$R_{\text{H{\sc i}}}$, then we define the \HI mass within $R_{90}$, \MHI, as the total \HI mass, \MHItot. If $R_{90}$ is less then 1.5$R_{\text{H{\sc i}}}$, then we calculate the \HI mass between $R_{90}$ and 1.5$R_{\text{H{\sc i}}}$ using Equation~(\ref{eqn:HI_out}) and subtract this value from the total \HI mass to define \MHI for the galaxy.

\subsection{Molecular Hydrogen, \Htwo}\label{sec:H2_mass}

\begin{figure}
    \centering
    \includegraphics[width=0.49\textwidth]{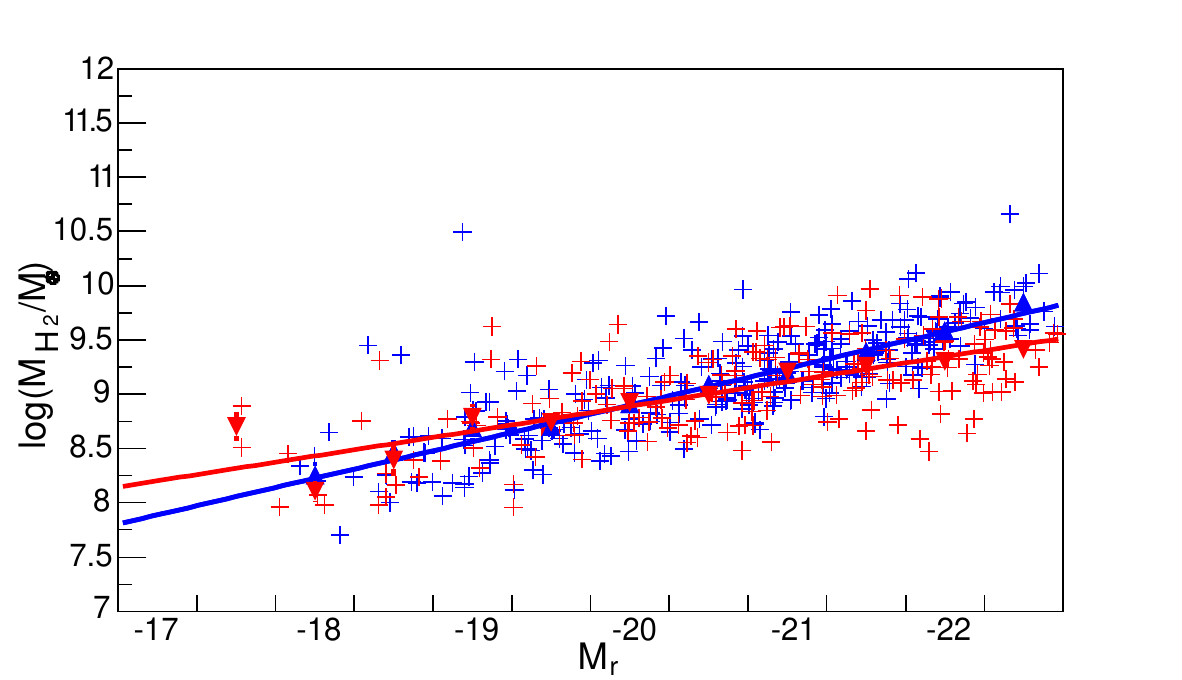}    
    \includegraphics[width=0.49\textwidth]{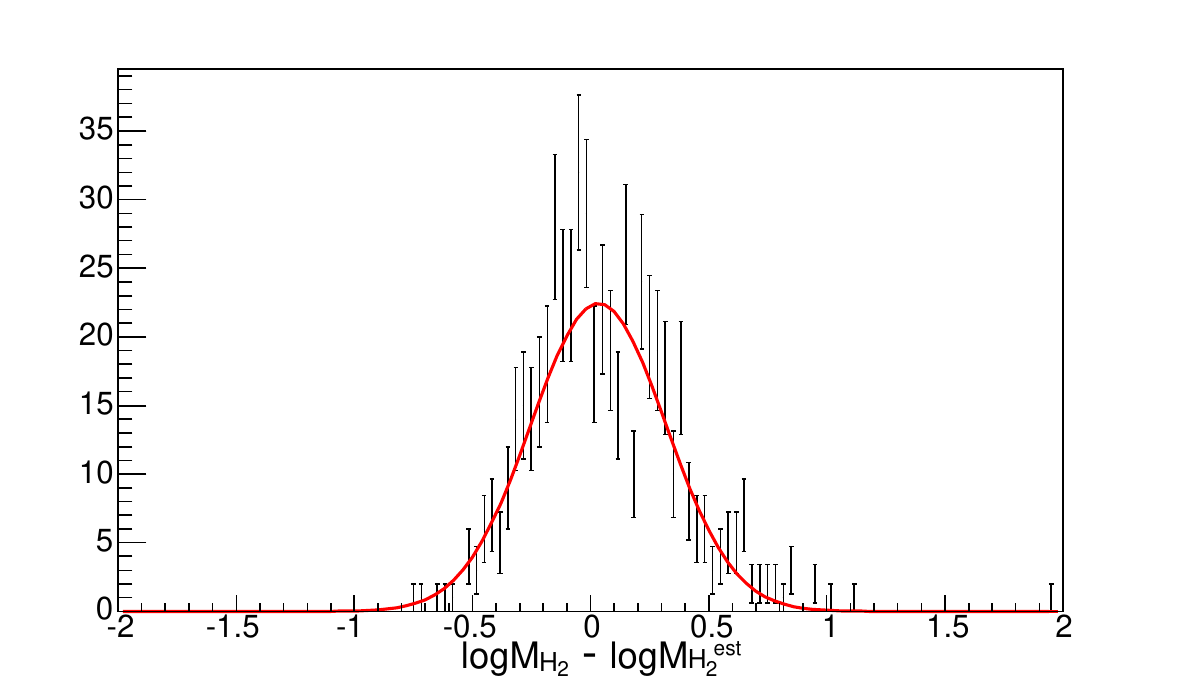}    
    \caption{Top: The dependence of \logMHtwo on $M_r$ for 531 galaxies in SDSS DR7 with \Htwo masses available through CO surveys. The blue crosses represent blue-cloud galaxies, while the red crosses are green-valley and red-sequence galaxies. The points are the mean of the \logMHtwo distribution in each bin in $M_r$. The lines are linear fits to the points: \logMHtwo $= aM_r + b$, with the coefficients shown in Table \ref{tab:H2_coeff}. Bottom: Resolution on \logMHtwo---the difference between the \Htwo mass evaluated based on CO mass and the parameterization from the top plot. The red line is a fit to a Gaussian with $\sigma = 0.27$.}
    \label{fig:MH2_rabsmag}
\end{figure}

Molecular hydrogen, \Htwo, is a low-mass, symmetric molecule without a dipole moment, and therefore it does not produce a significant amount of radiation, making it notoriously difficult to detect. Hence, to evaluate the molecular hydrogen content in a galaxy, it is customary to parameterize it with respect to some other observable. The most commonly used method is to use another molecular gas, particularly CO. We obtain mass estimates of \Htwo parameterized by the CO(1-0) line emission from the MASCOT and xCOLD~GASS surveys for 107 galaxies that also have total mass, stellar mass, and \HI mass estimates (as described above).

We have CO observations for only a small fraction of our galaxies, so we use CO observations of SDSS DR7 galaxies to derive a parameterization of the \Htwo mass as a function of galaxy luminosity in the $r$~band, $M_r$. A galaxy's \Htwo mass has been shown to be strongly correlated with its star formation rate \citep[e.g.,][]{Robertson08}. We choose to parameterize the \Htwo mass with luminosity since this quantity is related to star formation rate \citep[e.g.,][]{Hirashita03} and luminosity is directly measured whereas star formation rate is a derived quantity. Shown in Figure~\ref{fig:MH2_rabsmag}, we use $\chi^2$ minimization to find the coefficients that describe the linear relationship between $\log (\text{\MHtwo}/M_\odot)$ and $M_r$:
\begin{equation} \label{eqn:H2}
    \log(\text{\MHtwo}/M_\odot) = a\,M_r + b,
\end{equation}
where \MHtwo is the mass of molecular hydrogen. The values of $a$ and $b$ are listed in Table \ref{tab:H2_coeff} and depend on the color-magnitude classification.  We use this parameterization to estimate \MHtwo when CO observations are not available for galaxies in our sample. 

We assume that molecular hydrogen is concentrated within the optical disk of galaxies, so we use the global \Htwo mass of each galaxy in this analysis as the mass of \Htwo within $R_{90}$.

\begin{deluxetable}{l|CC}
    \tablewidth{0pt}
    \tablecolumns{7}
    \tablecaption{\MHtwo Mass Parameterization Coefficients \label{tab:H2_coeff}}
    \tablehead{ \colhead{CMD classification} & \colhead{$a$} & \colhead{$b$}}
    \startdata
        Blue cloud & -0.40 \pm 0.02 & 1.12 \pm 0.36\\
        Green valley and red sequence & -0.27 \pm 0.02 & 3.62 \pm 0.37
    \enddata
    \tablecomments{Coefficients for $\log(\text{\MHtwo}/M_\odot )$ parameterized as a function of $M_r$ as shown in Equation~(\ref{eqn:H2}).}
\end{deluxetable}
%

\subsection{Total Gas Mass, $M_\text{gas}$}\label{sec:gas_mass}

In this study, we define the total gas mass, $M_{\text{gas}}$, as the sum of the \HI mass, \Htwo mass, and helium mass:
\begin{equation}\label{eqn:Mgas}
    M_\text{gas} = \text{\MHI} + \text{\MHtwo} + \text{\MHe}.
\end{equation}
We approximate the helium mass, \MHe, by assuming a mass fraction of 25\%:
\begin{equation}
    \text{\MHe} = \left( \frac{0.25}{1 - 0.25} \right) (\text{\MHI} + \text{\MHtwo}).
\end{equation}
This is the amount of helium measured in the intergalactic medium and agrees well with the prediction from big bang nucleosynthesis \citep{Cooke18}. In this equation, $\text{\MHI}$ is the dominant component and is scaled to $R_{90}$, and $\text{\MHtwo}$ is assumed to be contained within $R_{90}$. Hence, the estimate for $\text{\MHe}$, and as a result the estimated total gas mass, can also be considered to be within $R_{90}$.

\subsection{Heavy Metals and Dust Mass, \Mdust}\label{sec:dust_mass}

The heavy metals and dust mass, \Mdust, is approximated from a galaxy's gas-phase metallicity. We compute the gas-phase metallicity, \logOH, following the R-calibration method described in \cite{Pilyugin16} using the flux of the \OII $\lambda \lambda$3727,3729 doublet and the \NII $\lambda$6548, \NII $\lambda$6584, \OIII $\lambda$4959, and \OIII $\lambda$5007 emission lines.  The fluxes are extinction-corrected using the Balmer decrement, assuming a flux ratio \Halpha/\Hbeta $= 2.86$ \citep{Osterbrock06}. We compute the metallicity as
\begin{multline}\label{eqn:Z}
    \text{\logOH} = a_1 + a_2 \log \left( \frac{R_3}{R_2}\right) + a_3 \log N_2\\
    + \left(a_4 + a_5 \log \left( \frac{R_3}{R_2}\right) + a_6 \log N_2 \right)\\
    \times \log R_2,
\end{multline}
where 
\begin{align}
    R_2 &= \frac{\text{\OII} \lambda \lambda 3727,3729}{\text{\Hbeta}},\\
    N_2 &= \frac{\text{\NII} \lambda 6548 + \text{\NII} \lambda 6584}{\text{\Hbeta}},\\
    R_3 &= \frac{\text{\OIII} \lambda 4959 + \text{\OIII} \lambda 5007}{\text{\Hbeta}}
\end{align}
are ratios of the specified emission-line fluxes.  The values of the coefficients in Equation~(\ref{eqn:Z}) depend on the value of $\log N_2$ and are listed in Table~\ref{tab:Z_coeff}.

We assume a constant dust-to-metals ratio corresponding to the metallicity calibration, \MdZ / \Mdust = 0.206
for galaxies with a gas-phase metallicity greater than 8.2 \citep{deVis19}.  \MdZ is the dust mass of each galaxy. The total mass of heavy metals and dust is then 
\begin{equation}\label{eqn:Mdust}
    M_{\text{dust}} = 1.259\, f_Z\, M_g,
\end{equation}
where $f_Z$ is the mass fraction of metals,
\begin{equation}
    f_Z = 27.36 \left( \frac{\text{O}}{\text{H}} \right),
\end{equation} 
and $M_g$ is the gas mass of the galaxy as defined in \cite{deVis19}:
\begin{equation}
    M_g = \xi \text{\MHI} \left( 1 + \frac{\text{\MHtwo}}{\text{\MHI}} \right),
\end{equation}
where
\begin{equation}
    \xi = \left( 1 - ( 0.2485 + 1.41 f_Z ) - f_Z \right)^{-1}.
\end{equation}

\begin{deluxetable}{L|CCCCCC}
    \tablewidth{0pt}
    \tablecolumns{7}
    \tablecaption{Gas-phase Metallicity Coefficients \label{tab:Z_coeff}}
    \tablehead{ \colhead{$\log N_2$} & \colhead{$a_1$} & \colhead{$a_2$} & \colhead{$a_3$} & \colhead{$a_4$} & \colhead{$a_5$} & \colhead{$a_6$} }
    \startdata
        \geq -0.6 & 8.589 & 0.022 & 0.399 & 0.137 & 0.164  & 0.589\\
        < -0.6    & 7.932 & 0.944 & 0.695 & 0.970 & -0.291 & -0.019
    \enddata
    \tablecomments{Coefficients for the gas-phase metallicity calculation shown in Equation~(\ref{eqn:Z}), from \cite{Pilyugin16}.
    }
\end{deluxetable}

\subsection{The Total Visible Mass, \Mvis}

\begin{figure}
    \includegraphics[width=0.46\textwidth]{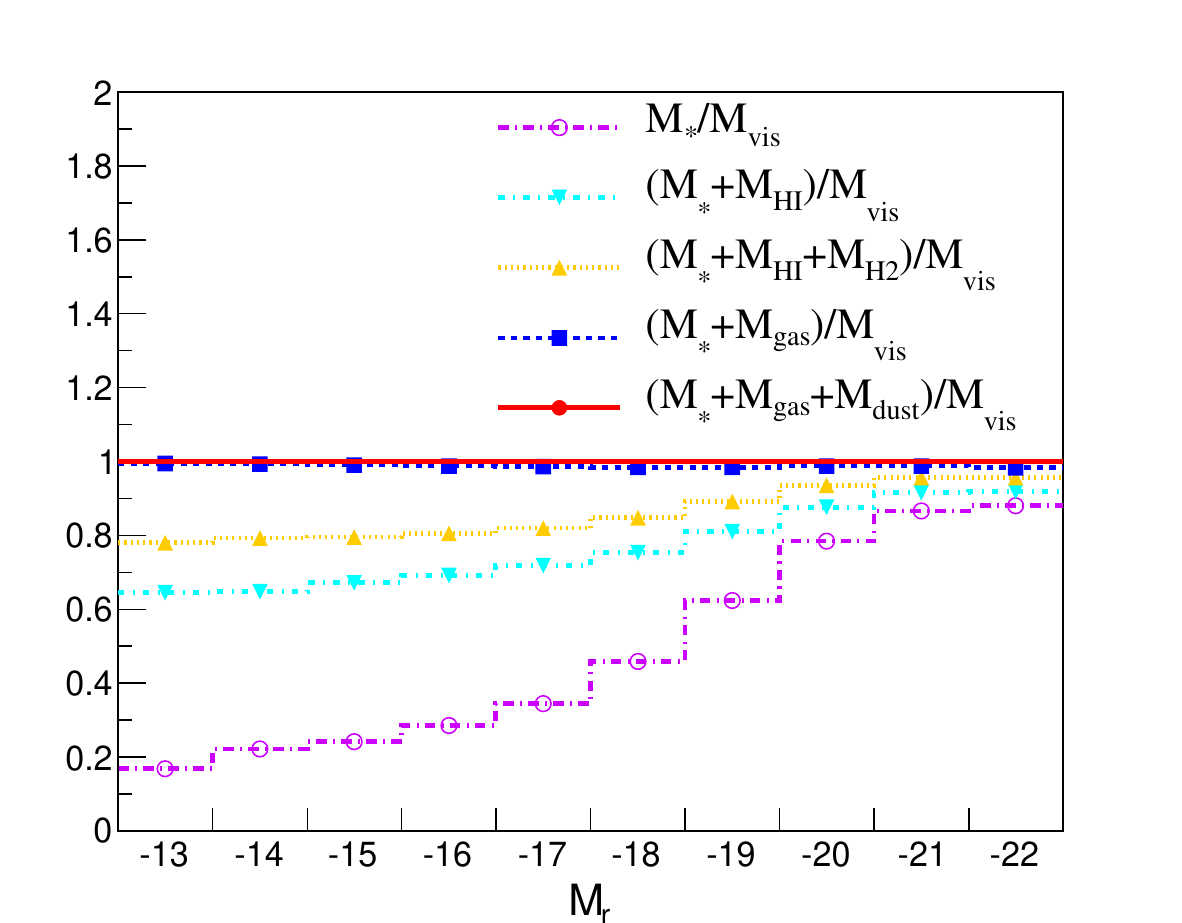}
    \caption{The relative contributions of each mass component to the total visible mass of SDSS~DR7 galaxies within $R_{90}$ as a function of $M_r$. For simplicity, we only show \MHtwo parameterized as a function of $M_r$ here.  \MHtwo is parameterized as a function of $M_r$, \MHe is added by fraction, and the other components are based on measurements. 
    }
    \label{fig:rel_comp}
\end{figure}

We define the total visible mass of a galaxy, $M_\text{vis}$, as the sum of the stellar mass, $M_*$, the gas mass, \Mgas (Equation~(\ref{eqn:Mgas})), and the heavy metals and dust mass, \Mdust (Equation~(\ref{eqn:Mdust})):
\begin{equation}
    \text{\Mvis} = M_* + \text{\MHI} + \text{\MHtwo} + \text{\MHe} + \text{\Mdust}.
\end{equation}
A summary of the relative contributions of each individual mass component to the total visible mass for SDSS~DR7 galaxies within $R_{90}$ as a function of the $r$-band luminosity, $M_r$, is shown in Figure~\ref{fig:rel_comp}.  For galaxies with $M_r > -17$, gas is the dominant component of the visible mass, whereas for galaxies with $M_r < -18$, the stellar mass dominates the visible mass. Heavy metals and dust contribute on the order of 1\% regardless of magnitude.

\section{Statistically modeling the rotational velocity}\label{sec:stat}

\begin{figure}
    \includegraphics[width=0.48\textwidth]{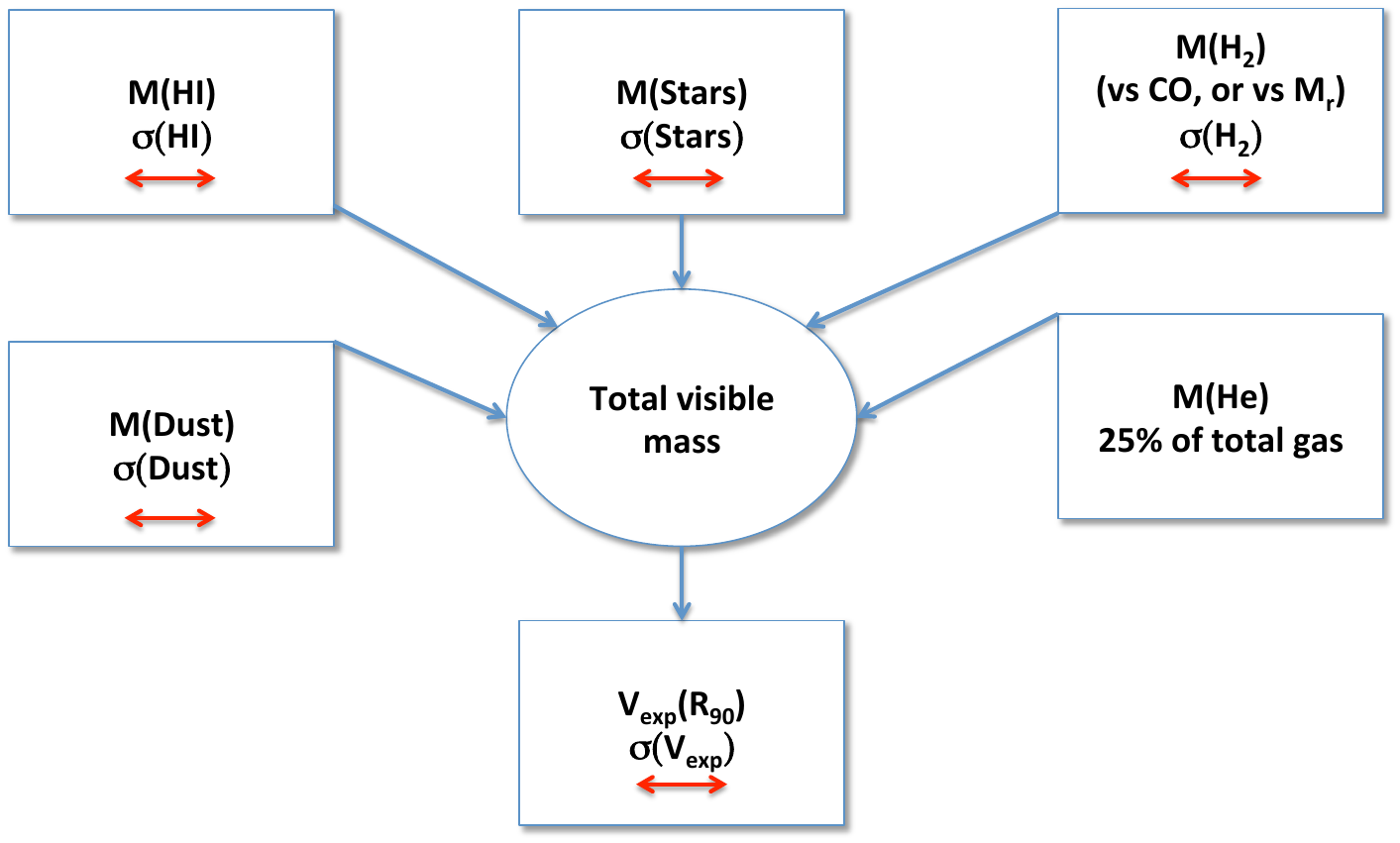}
    \caption{Illustration of the statistical model. The red horizontal arrows denote Gaussian smearing with the corresponding $\sigma$.}
    \label{fig:stat_model}
\end{figure}

To test our null hypothesis---that galaxies do not have a dark matter halo, so the observed rotational velocity at $R_{90}$ is due entirely to visible mass---we construct a statistical model to predict the expected rotational velocity of a galaxy given its total visible mass.  We choose the ratio of the expected to observed velocity evaluated at $R_{90}$, $V_\text{exp}/V_\text{obs}$, as our observable.  The expected velocity is evaluated based on the visible mass calculated using the methods described below.  A value of this observable close to unity signals consistency of the data with the null hypothesis. 

The resolution on this observable is determined by the measured uncertainty of each visible mass component and the uncertainty of the fitted rotation curve to the velocity map, from which we determine the velocity at $R_{90}$. We expect the velocity to be normally distributed around its true value with the uncertainty returned by the fit. 

To evaluate the effect of these uncertainties, we implement the following procedure. First, for each galaxy, we determine the mass of each component of the visible mass as described in Sections~\ref{sec:Star_mass}--\ref{sec:dust_mass}. Since \MHtwo is available from CO observations for only a small number of galaxies, we also use the parameterization as a function of $M_r$ to estimate \MHtwo as described in Section~\ref{sec:H2_mass}.  We estimate the total mass, \Mtot, from the best-fit rotation curve as described in Section~\ref{sec:Mtot}.  We then compute the ratio of visible to total mass, 
\begin{equation}
    F_\text{vis} = \frac{\text{\Mvis}}{\text{\Mtot}},
\end{equation}
for each galaxy. 

To statistically determine the rotational velocity, we smear each mass component according to its expected resolution.\footnote{Since we observe a Gaussian distribution in $\log M$ of the corresponding component, we randomly smear $\log M$ according to a Gaussian distribution and then invert to find the corresponding mass.} The expected velocity, $V_\text{exp}$, is then evaluated based on the sum of each of these smeared mass components and is smeared according to the velocity uncertainty from the fit to the rotation curve. This smearing procedure is repeated 1000 times for each galaxy. A schematic of this statistical model is illustrated in Figure~\ref{fig:stat_model}. From this procedure, we find the expected fraction of the instances where the observed rotational velocity is less than the rotational velocity expected from just the visible mass components, $F(V_\text{obs} < V_\text{exp})$, where $V_\text{obs}$ is the rotational velocity measured at $R_{90}$ from the best-fit rotation curve. This is the fraction of galaxies consistent with the null hypothesis.

\section{Studying the ratio of visible to total mass}\label{sec:results}

\begin{figure}
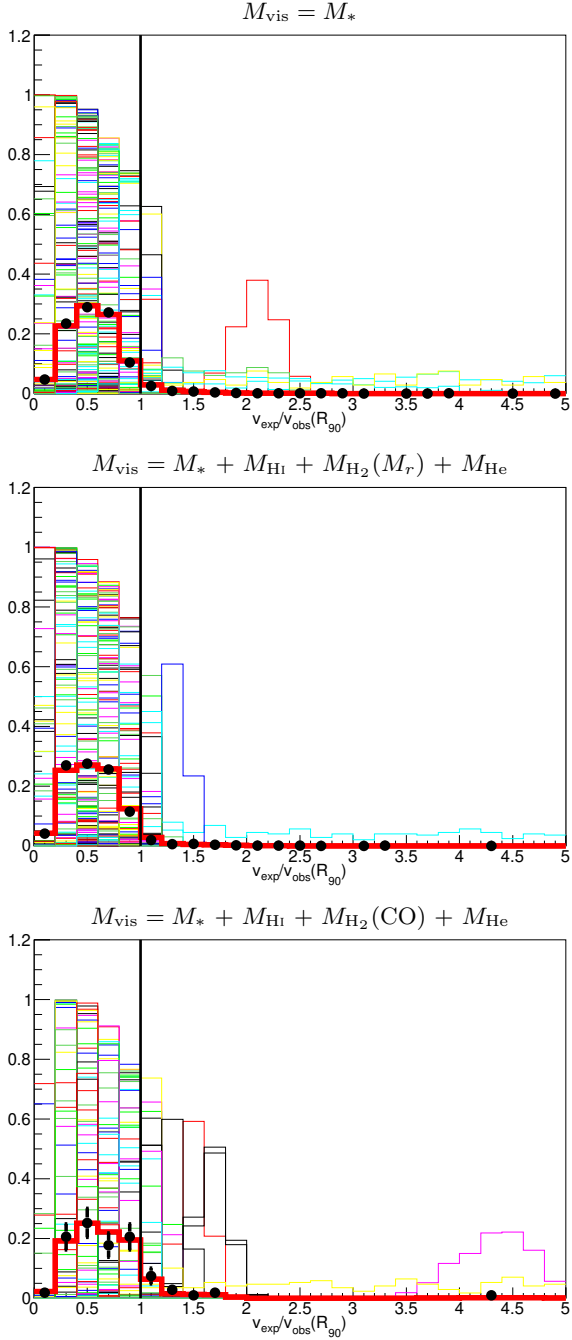

    \centering
    \begin{annotationimage}{width=0.49\textwidth}{Fig10a.eps}
        \draw[image label = {{$M_\text{vis} = M_*$} at north}];
    \end{annotationimage}
    \begin{annotationimage}{width=0.49\textwidth}{Fig10b.eps}
        \draw[image label = {{$M_\text{vis} = M_*$ + \MHI + $M_{\text{H}_2}(M_r)$ + \MHe} at north}];
    \end{annotationimage}
    \begin{annotationimage}{width=0.49\textwidth}{Fig10c.eps}
        \draw[image label = {{$M_\text{vis} = M_*$ + \MHI + $M_{\text{H}_2}(\text{CO})$ + \MHe} at north}];
   \end{annotationimage}
   \caption{The PDF of the ratio of expected to observed velocities at $R_{90}$.  The black points show the data with the expected velocity evaluated from the visible mass without smearing.  The colored histograms show the PDF evaluated based on the statistical model for a sample of randomly selected galaxies.  The red histogram is the normalized sum of the individual PDFs.  The vertical black line at 1 corresponds to the observed and expected velocities being equal. The integral of the PDFs to the right of this line corresponds to the observed (black points) and modeled (red histogram) $F(V_\text{obs} < V_\text{exp})$ listed in Table~\ref{tab:v90_fit}. Top row: only stellar mass contributes to the visible mass. Second row: gas mas is added to stellar mass, with \MHtwo determined from $M_r$. Third row: the same as the second row, but \MHtwo is determined from CO observations.}
   \label{fig:v90_fit}
\end{figure}

In Figure~\ref{fig:v90_fit}, we present the probability distribution functions (PDFs) of the ratio of expected to observed velocities at $R_{90}$ derived using the statistical model and the distribution observed in data. The integrals of these distributions above 1 correspond to the fractions of galaxies for which the expected velocity exceeds the observed one, $F(V_\text{obs} < V_\text{exp})$, listed in Table~\ref{tab:v90_fit}. In Table~\ref{tab:v90_fit}, we also present the mean and rms of $F_\text{vis}$ (the ratio of visible to total mass, as described in Section~\ref{sec:stat}). We break down the sample into a number of different subsets: by CMD class into blue cloud, green valley, and red sequence;  by MaNGA targeting sample (to check for possible systematic bias), and by $M_r$. Due to the limited statistics, we combine galaxies in the green valley and red sequence.  For each sample of galaxies, we consider three different mass ratios: $M_*/$\Mtot (labeled ``Only stars'' in Table~\ref{tab:v90_fit}); \Mvis/\Mtot, with \MHtwo inferred from $M_r$; and \Mvis/\Mtot, with \MHtwo measured with CO observations. 

\begin{deluxetable*}{lcccCC}
    \tablewidth{0pt}
    \tablecolumns{6}
    \tablecaption{Mass ratio statistics for MaNGA DR17 galaxies.\label{tab:v90_fit}}
    \tablehead{  & & \multicolumn{2}{c}{$F(V_\text{obs}<V_\text{exp})$} & \multicolumn{2}{c}{$F_\text{vis}$} \\
    \colhead{Sample} & \colhead{Count} & \colhead{Observed} & \colhead{Modeled} & \colhead{Mean} & \colhead{rms}  }
    \startdata
        \sidehead{\bf Only stars}
        All  & 5503 & 5.4\% & 5.7\% & 45\pm0.5\% & 35\pm0.4\% \\ 
        Blue cloud  & 3013 & 3.4\% & 3.7\% & 40\pm0.6\% & 30\pm0.4\% \\ 
        Green valley, red sequence  & 1943 & 8.2\% & 8.6\% & 52\pm0.9\% & 39\pm0.7\% \\ 
        MaNGA sample 1  & 2460 & 4.8\% & 5,3\% & 45\pm0.8\% & 35\pm0.6\% \\ 
        MaNGA sample 2  & 2073 & 5.5\% & 5.5\% & 44\pm0.8\% & 33\pm0.6\% \\ 
        MaNGA sample 3  & 942 & 6.5\% & 6.9\% & 46 \pm1.3\% & 37\pm0.9\% \\ 
        $M_r>-19$  & 1790 & 3.5\% & 3.7\% & 35\pm0.9\% & 34\pm0.7\% \\ 
        $M_r<-19$  & 3713 & 6.3\% & 6.6\% & 49\pm0.6\% & 34\pm0.4\% \\ 
        \sidehead{\bf Stars, dust, \HI, H$_2$($M_r$), He }
        All & 2575 & 4.3\% & 4.8\% & 44\pm0.7\% & 34\pm0.5\% \\ 
        Blue cloud  & 1734 & 3.2\% & 3.8\% & 41\pm0.8\% & 32\pm0.6\% \\ 
        Green valley, red sequence  & 559 & 6.4\% & 7.1\% & 50\pm1.8\% & 39\pm1.3\% \\ 
        MaNGA sample 1  & 1576 & 4.4\% & 5.1\% & 45\pm1.0 \% & 35\pm0.7\% \\ 
        MaNGA sample 2  & 560 & 2.1\% & 2.4\% & 38\pm1.2\% & 26\pm0.8\% \\ 
        MaNGA sample 3  & 430 & 6.3\% & 6.6\% &47\pm2.0\% & 40\pm1.4\% \\ 
        $M_r>-19$  & 1011 & 3.1\% & 3.3\% & 36\pm1.1\% & 32\pm0.8\% \\ 
        $M_r<-19$  & 1564 & 5.1\% & 5.7\% & 48\pm0.9\% & 34\pm0.7\% \\ 
    \hline
        \sidehead{\bf Stars, dust, \HI, H$_2$(CO), He  }
        All & 107 & 14.0\% & 11.8\% & 60\pm 5\% & 50\pm 4\% \\ 
        Blue cloud & 75 & 16\% & 12\% & 62\pm 7\% & 56\pm 5\% \\ 
        Green valley, red sequence & 28 & 11\% & 12\% & 60\pm 7\% & 36\pm5\% \\
        $M_r>-19$  & 6 &  &  &  &  \\ 
        $M_r<-19$  & 101 & 15\% & 12.5\% & 61\pm5\% & 51\pm4\% \\ 
    \enddata
    \tablecomments{ The observed velocity, $V_\text{obs}$, is evaluated at $R_{90}$ based on the fit to the rotation curve. The expected velocity, $V_\text{exp}$, is evaluated based on the visible mass. $F(V_\text{obs} < V_\text{exp})$ is the fraction of galaxies for which $V_\text{obs} < V_\text{exp}$. In the ``Modeled'' column, the visible mass and $V_\text{exp}$ are distributed according to the statistical model; in the ``Observed'' column, they are not smeared. $F_\text{vis}$ is the fraction of the visible mass, i.e., the ratio of the visible to total mass. Color classification and MaNGA sample information may not be available for all galaxies.}
\end{deluxetable*}

\begin{figure}
    \includegraphics[width=0.49\textwidth]{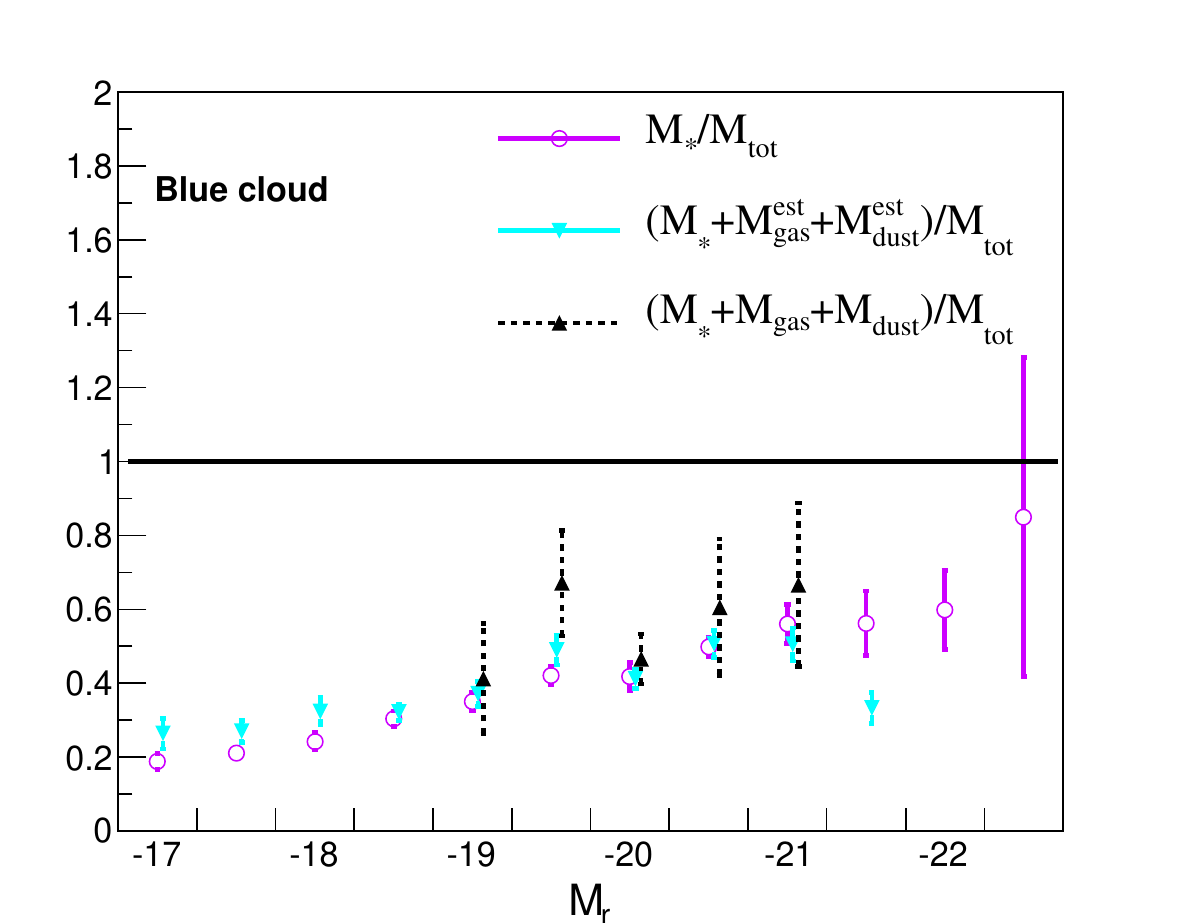} 
    \includegraphics[width=0.49\textwidth]{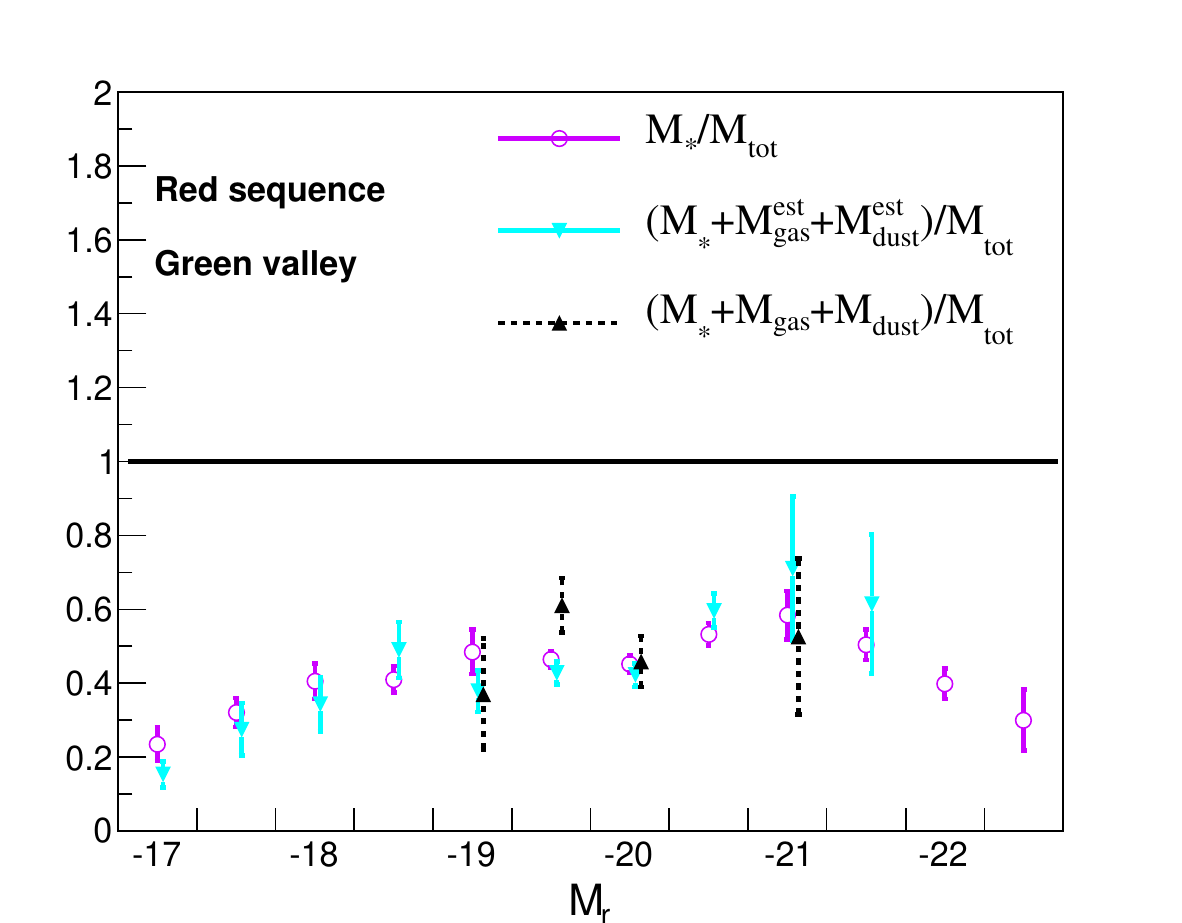}
    \caption{The dependence of various mass fractions on luminosity for blue-cloud galaxies (top) and green-valley and red-sequence galaxies (bottom). The purple circles compare just the stellar mass to total mass, the cyan triangles compare the visible mass (with \MHtwo estimated from $M_r$) to total mass, and the black triangles compare the visible mass (with \MHtwo inferred from CO observations) to total mass. The black line at 1 is where the visible mass is equal to the total mass. The points correspond to the mean of the distribution in $F_\text{vis}$ in each bin in $M_r$. The error bars represent uncertainties on the mean, which are significantly smaller than the rms values. Typical rms values are given in Table~\ref{tab:v90_fit}.}
    \label{fig:Frvis}
\end{figure}
 
The preferred value of $F_\text{vis}$ is 40--50\% for all of the galaxy samples, when only the stellar mass is included. When we include all visible mass, with \MHtwo parameterized by $M_r$, $F_\text{vis}$ does not change significantly. Finally, when we use \MHtwo estimated from CO observations, which is a more reliable method than our parameterization with $M_r$, we see $F_\text{vis}$ increase to $\sim$60\%. We must note that these galaxies tend to be on average brighter than the galaxies for which we estimate \MHtwo using the $M_r$ parameterization. As we go from just $M_*$ to \Mvis with all mass components and \MHtwo estimated from CO, we see an increase in the fraction of galaxies with $V_\text{obs} < V_\text{exp}$. We find that blue-cloud galaxies and the lower-brightness galaxies tend to have a lower fraction of visible mass compared to brighter galaxies or green-valley and red-sequence galaxies. As shown by the values in Table~\ref{tab:v90_fit}, we find no statistically significant difference between the three MaNGA targeting samples.

The remaining component of the baryonic mass that is missing from our analysis is ionized hydrogen, \HII. MaNGA has a spaxel resolution of only 0.5" \citep{Law15}, so regions of uniform density cannot be resolved in MaNGA observations. As a result, we cannot estimate the \HII mass without assuming an electron density distribution. We expect the \HII mass to be on the order of 1\% of the stellar mass \citep{Dettmar90, Sofue16}, with star-forming galaxies containing more \HII. We do not anticipate that the inclusion of \HII to significantly change our results, because its contribution to the visible mass is negligible. 

Finally, we show the dependence of $F_\text{vis}$ on luminosity for galaxies in the blue cloud, green valley and red sequence in Figure~\ref{fig:Frvis}. When only stellar mass is included in the visible mass estimation, the dependence of $F_\text{vis}$ on $M_r$ is rather flat for green-valley and red-sequence galaxies, while for the blue-cloud galaxies there is a notable upward trend, with brighter galaxies having a larger ratio of $M_*$/\Mtot.  This matches results from previous studies of the stellar-halo mass relation (SHMR), including \cite{Persic96, Strigari08, TorresFlores11, Karukes17, Behroozi19, DiPaolo19, Douglass19, Douglass22}, and from the simulations by \cite{Moster10}. We find that when gas and dust are added to the visible mass, these trends are preserved. The dependence of $F_\text{vis}$ on $M_r$ remains flat for green-valley and red-sequence galaxies and $F_\text{vis}$ increases with galaxy luminosity for blue-cloud galaxies.

\subsection{Comparison to Previous Results}

As shown in Figure~\ref{fig:Frvis}, we find that once we account for all of the visible mass components of a galaxy, the ratio of \Mvis/\Mtot shows an upward trend with galaxy luminosity. This is in agreement with the previous work by \cite{TorresFlores11}, who consider the relationship between \Mvis, defined as stellar mass and \HI mass, and total mass. \cite{TorresFlores11} find a correlation between the mass ratio and evolutionary stage, in that late-type low-mass spirals are dominated by dark matter in comparison to early-type high mass spirals. 

\Mvis/\Mtot is a version of the SHMR typically described as the ratio of stellar mass to halo mass. Models predict an SHMR that deviates from a flat distribution \citep[e.g.][]{Behroozi19}, with lower values for the faintest and brightest galaxies. These galaxies are thought to be dominated by dark matter. We find that the faint end of blue-cloud galaxies shows this expected decrease in \Mvis/\Mtot, suggesting an additional abundance of dark matter within the visible extent of these galaxies that is not present in brighter galaxies.

\section{Conclusions\label{sec:conclusion}}

We study the ratio of visible to total mass in spiral galaxies using rotation curves evaluated with the \Halpha velocity maps from SDSS MaNGA DR17. From the dependence of the rotational velocity on the distance from the center of a galaxy, we evaluate the velocity at the 90\% elliptical Petrosian radius, $R_{90}$, from the fitted rotation curves. We compute the visible mass of each galaxy, which includes stellar mass and the mass of atomic hydrogen (\HI) evaluated at the same radius, $R_{90}$, molecular hydrogen (\Htwo) evaluated based on the CO content, helium, and the heavy metals and dust mass. To increase the size of the sample under study, we also use a parameterization of \MHtwo as a function of the galaxy luminosity in the $r$~band, $M_r$, derived using the SDSS DR7 galaxy sample. The helium mass is added assuming that its mass fraction in the total gas amount is 25\%.

We construct a statistical model that predicts the velocity based on the visible mass and compares it to the observed velocity. If the expected velocity is evaluated based solely on the stellar mass, the expected velocity exceeds the observed velocity in only 3\%--9\% of the cases. After including all of the gas and dust mass, this fraction increases to 2\%--16\%, depending on the sample selection and method for estimating \MHtwo. Hence, the null hypothesis (no dark matter) cannot be excluded at a confidence level better than 95\% for the mass within the visible extent of disk galaxies. We find that when all of the visible mass is accounted for, the ratio of visible to total mass is independent of galaxy luminosity for green-valley and red-sequence galaxies and increases with galaxy luminosity for galaxies in the blue cloud. 

Future work will incorporate the mass of ionized hydrogen and extend the mass component analysis to elliptical galaxies. 

\section*{Acknowledgements}

The authors would like to thank Bob Cousins for insightful remarks on the statistical model, and Eric Blackman and Alice Quillen for careful reading and thoughtful comments, Am\'elie Saintonge for suggesting to scale the \HI gas mass to $R_{90}$, and the anonymous referee for detailed comments and suggestions.  

N.R. acknowledges support from the Feinberg Research Award through the Department of Physics \& Astronomy at the University of Rochester.  R.D. acknowledges support from the Department of Energy under the grant DE-SC0008475.0.

This project makes use of the MaNGA-Pipe3D data products.  We thank the IA-UNAM 
MaNGA team for creating this catalogue, and the Conacyt Project CB-285080 for 
supporting them.

Funding for the Sloan Digital Sky Survey IV has been provided by the Alfred P. 
Sloan Foundation, the U.S. Department of Energy Office of Science, and the 
Participating Institutions.  SDSS-IV acknowledges support and resources from the 
Center for High-Performance Computing at the University of Utah.  The SDSS web 
site is www.sdss.org.

SDSS-IV is managed by the Astrophysical Research Consortium for the 
Participating Institutions of the SDSS Collaboration including the Brazilian 
Participation Group, the Carnegie Institution for Science, Carnegie Mellon 
University, the Chilean Participation Group, the French Participation Group, 
Harvard-Smithsonian Center for Astrophysics, Instituto de Astrof\'isica de 
Canarias, The Johns Hopkins University, Kavli Institute for the Physics and 
Mathematics of the Universe (IPMU) / University of Tokyo, the Korean 
Participation Group, Lawrence Berkeley National Laboratory, Leibniz Institut 
f\"ur Astrophysik Potsdam (AIP),  Max-Planck-Institut f\"ur Astronomie (MPIA 
Heidelberg), Max-Planck-Institut f\"ur Astrophysik (MPA Garching), 
Max-Planck-Institut f\"ur Extraterrestrische Physik (MPE), National Astronomical 
Observatories of China, New Mexico State University, New York University, 
University of Notre Dame, Observat\'ario Nacional / MCTI, The Ohio State 
University, Pennsylvania State University, Shanghai Astronomical Observatory, 
United Kingdom Participation Group, Universidad Nacional Aut\'onoma de M\'exico, 
University of Arizona, University of Colorado Boulder, University of Oxford, 
University of Portsmouth, University of Utah, University of Virginia, University 
of Washington, University of Wisconsin, Vanderbilt University, and Yale 
University.

\bibliographystyle{aasjournal}
\bibliography{Ravi0624_sources}

\begin{thebibliography}{}
\expandafter\ifx\csname natexlab\endcsname\relax\def\natexlab#1{#1}\fi
\providecommand{\url}[1]{\href{#1}{#1}}
\providecommand{\dodoi}[1]{doi:~\href{http://doi.org/#1}{\nolinkurl{#1}}}
\providecommand{\doeprint}[1]{\href{http://ascl.net/#1}{\nolinkurl{http://ascl.net/#1}}}
\providecommand{\doarXiv}[1]{\href{https://arxiv.org/abs/#1}{\nolinkurl{https://arxiv.org/abs/#1}}}

\bibitem[{Aad {et~al.}(2023)}]{ATLAS:2022ygn}
Aad, G., {et~al.} 2023, Eur. Phys. J. C, 83, 503,
  \dodoi{10.1140/epjc/s10052-023-11477-z}

\bibitem[{{Abazajian} {et~al.}(2009){Abazajian}, {Adelman-McCarthy},
  {Ag{\"u}eros}, {Allam}, {Allende Prieto}, {An}, {Anderson}, {Anderson},
  {Annis}, {Bahcall}, {Bailer-Jones}, {Barentine}, {Bassett}, {Becker},
  {Beers}, {Bell}, {Belokurov}, {Berlind}, {Berman}, {Bernardi}, {Bickerton},
  {Bizyaev}, {Blakeslee}, {Blanton}, {Bochanski}, {Boroski}, {Brewington},
  {Brinchmann}, {Brinkmann}, {Brunner}, {Budav{\'a}ri}, {Carey}, {Carliles},
  {Carr}, {Castander}, {Cinabro}, {Connolly}, {Csabai}, {Cunha}, {Czarapata},
  {Davenport}, {de Haas}, {Dilday}, {Doi}, {Eisenstein}, {Evans}, {Evans},
  {Fan}, {Friedman}, {Frieman}, {Fukugita}, {G{\"a}nsicke}, {Gates},
  {Gillespie}, {Gilmore}, {Gonzalez}, {Gonzalez}, {Grebel}, {Gunn},
  {Gy{\"o}ry}, {Hall}, {Harding}, {Harris}, {Harvanek}, {Hawley}, {Hayes},
  {Heckman}, {Hendry}, {Hennessy}, {Hindsley}, {Hoblitt}, {Hogan}, {Hogg},
  {Holtzman}, {Hyde}, {Ichikawa}, {Ichikawa}, {Im}, {Ivezi{\'c}}, {Jester},
  {Jiang}, {Johnson}, {Jorgensen}, {Juri{\'c}}, {Kent}, {Kessler}, {Kleinman},
  {Knapp}, {Konishi}, {Kron}, {Krzesinski}, {Kuropatkin}, {Lampeitl},
  {Lebedeva}, {Lee}, {Lee}, {French Leger}, {L{\'e}pine}, {Li}, {Lima}, {Lin},
  {Long}, {Loomis}, {Loveday}, {Lupton}, {Magnier}, {Malanushenko},
  {Malanushenko}, {Mandelbaum}, {Margon}, {Marriner}, {Mart{\'\i}nez-Delgado},
  {Matsubara}, {McGehee}, {McKay}, {Meiksin}, {Morrison}, {Mullally}, {Munn},
  {Murphy}, {Nash}, {Nebot}, {Neilsen}, {Newberg}, {Newman}, {Nichol},
  {Nicinski}, {Nieto-Santisteban}, {Nitta}, {Okamura}, {Oravetz}, {Ostriker},
  {Owen}, {Padmanabhan}, {Pan}, {Park}, {Pauls}, {Peoples}, {Percival}, {Pier},
  {Pope}, {Pourbaix}, {Price}, {Purger}, {Quinn}, {Raddick}, {Re Fiorentin},
  {Richards}, {Richmond}, {Riess}, {Rix}, {Rockosi}, {Sako}, {Schlegel},
  {Schneider}, {Scholz}, {Schreiber}, {Schwope}, {Seljak}, {Sesar}, {Sheldon},
  {Shimasaku}, {Sibley}, {Simmons}, {Sivarani}, {Allyn Smith}, {Smith},
  {Smol{\v{c}}i{\'c}}, {Snedden}, {Stebbins}, {Steinmetz}, {Stoughton},
  {Strauss}, {SubbaRao}, {Suto}, {Szalay}, {Szapudi}, {Szkody}, {Tanaka},
  {Tegmark}, {Teodoro}, {Thakar}, {Tremonti}, {Tucker}, {Uomoto}, {Vanden
  Berk}, {Vandenberg}, {Vidrih}, {Vogeley}, {Voges}, {Vogt}, {Wadadekar},
  {Watters}, {Weinberg}, {West}, {White}, {Wilhite}, {Wonders}, {Yanny},
  {Yocum}, {York}, {Zehavi}, {Zibetti}, \& {Zucker}}]{SDSS7}
{Abazajian}, K.~N., {Adelman-McCarthy}, J.~K., {Ag{\"u}eros}, M.~A., {et~al.}
  2009, \apjs, 182, 543, \dodoi{10.1088/0067-0049/182/2/543}

\bibitem[{{Abdurro'uf} {et~al.}(2022){Abdurro'uf}, {Accetta}, {Aerts}, {Silva
  Aguirre}, {Ahumada}, {Ajgaonkar}, {Filiz Ak}, {Alam}, {Allende Prieto},
  {Almeida}, {Anders}, {Anderson}, {Andrews}, {Anguiano}, {Aquino-Ort{\'\i}z},
  {Arag{\'o}n-Salamanca}, {Argudo-Fern{\'a}ndez}, {Ata}, {Aubert},
  {Avila-Reese}, {Badenes}, {Barb{\'a}}, {Barger}, {Barrera-Ballesteros},
  {Beaton}, {Beers}, {Belfiore}, {Bender}, {Bernardi}, {Bershady}, {Beutler},
  {Bidin}, {Bird}, {Bizyaev}, {Blanc}, {Blanton}, {Boardman}, {Bolton},
  {Boquien}, {Borissova}, {Bovy}, {Brandt}, {Brown}, {Brownstein}, {Brusa},
  {Buchner}, {Bundy}, {Burchett}, {Bureau}, {Burgasser}, {Cabang}, {Campbell},
  {Cappellari}, {Carlberg}, {Wanderley}, {Carrera}, {Cash}, {Chen}, {Chen},
  {Cherinka}, {Chiappini}, {Choi}, {Chojnowski}, {Chung}, {Clerc}, {Cohen},
  {Comerford}, {Comparat}, {da Costa}, {Covey}, {Crane}, {Cruz-Gonzalez},
  {Culhane}, {Cunha}, {Dai}, {Damke}, {Darling}, {Davidson}, {Davies},
  {Dawson}, {De Lee}, {Diamond-Stanic}, {Cano-D{\'\i}az}, {S{\'a}nchez},
  {Donor}, {Duckworth}, {Dwelly}, {Eisenstein}, {Elsworth}, {Emsellem},
  {Eracleous}, {Escoffier}, {Fan}, {Farr}, {Feng}, {Fern{\'a}ndez-Trincado},
  {Feuillet}, {Filipp}, {Fillingham}, {Frinchaboy}, {Fromenteau}, {Galbany},
  {Garc{\'\i}a}, {Garc{\'\i}a-Hern{\'a}ndez}, {Ge}, {Geisler}, {Gelfand},
  {G{\'e}ron}, {Gibson}, {Goddy}, {Godoy-Rivera}, {Grabowski}, {Green},
  {Greener}, {Grier}, {Griffith}, {Guo}, {Guy}, {Hadjara}, {Harding},
  {Hasselquist}, {Hayes}, {Hearty}, {Hern{\'a}ndez}, {Hill}, {Hogg},
  {Holtzman}, {Horta}, {Hsieh}, {Hsu}, {Hsu}, {Huber}, {Huertas-Company},
  {Hutchinson}, {Hwang}, {Ibarra-Medel}, {Chitham}, {Ilha}, {Imig}, {Jaekle},
  {Jayasinghe}, {Ji}, {Johnson}, {Jones}, {J{\"o}nsson}, {Katkov}, {Khalatyan},
  {Kinemuchi}, {Kisku}, {Knapen}, {Kneib}, {Kollmeier}, {Kong}, {Kounkel},
  {Kreckel}, {Krishnarao}, {Lacerna}, {Lane}, {Langgin}, {Lavender}, {Law},
  {Lazarz}, {Leung}, {Leung}, {Lewis}, {Li}, {Li}, {Lian}, {Liang}, {Lin},
  {Lin}, {Lin}, {Lintott}, {Long}, {Longa-Pe{\~n}a}, {L{\'o}pez-Cob{\'a}},
  {Lu}, {Lundgren}, {Luo}, {Mackereth}, {de la Macorra}, {Mahadevan},
  {Majewski}, {Manchado}, {Mandeville}, {Maraston}, {Margalef-Bentabol},
  {Masseron}, {Masters}, {Mathur}, {McDermid}, {Mckay}, {Merloni},
  {Merrifield}, {Meszaros}, {Miglio}, {Di Mille}, {Minniti}, {Minsley},
  {Monachesi}, {Moon}, {Mosser}, {Mulchaey}, {Muna}, {Mu{\~n}oz}, {Myers},
  {Myers}, {Nadathur}, {Nair}, {Nandra}, {Neumann}, {Newman}, {Nidever},
  {Nikakhtar}, {Nitschelm}, {O'Connell}, {Garma-Oehmichen}, {Luan Souza de
  Oliveira}, {Olney}, {Oravetz}, {Ortigoza-Urdaneta}, {Osorio}, {Otter},
  {Pace}, {Padilla}, {Pan}, {Pan}, {Parikh}, {Parker}, {Peirani}, {Pe{\~n}a
  Ram{\'\i}rez}, {Penny}, {Percival}, {Perez-Fournon}, {Pinsonneault},
  {Poidevin}, {Poovelil}, {Price-Whelan}, {B{\'a}rbara de Andrade Queiroz},
  {Raddick}, {Ray}, {Rembold}, {Riddle}, {Riffel}, {Riffel}, {Rix}, {Robin},
  {Rodr{\'\i}guez-Puebla}, {Roman-Lopes}, {Rom{\'a}n-Z{\'u}{\~n}iga}, {Rose},
  {Ross}, {Rossi}, {Rubin}, {Salvato}, {S{\'a}nchez}, {S{\'a}nchez-Gallego},
  {Sanderson}, {Santana Rojas}, {Sarceno}, {Sarmiento}, {Sayres}, {Sazonova},
  {Schaefer}, {Schiavon}, {Schlegel}, {Schneider}, {Schultheis}, {Schwope},
  {Serenelli}, {Serna}, {Shao}, {Shapiro}, {Sharma}, {Shen}, {Shetrone}, {Shu},
  {Simon}, {Skrutskie}, {Smethurst}, {Smith}, {Sobeck}, {Spoo}, {Sprague},
  {Stark}, {Stassun}, {Steinmetz}, {Stello}, {Stone-Martinez},
  {Storchi-Bergmann}, {Stringfellow}, {Stutz}, {Su}, {Taghizadeh-Popp},
  {Talbot}, {Tayar}, {Telles}, {Teske}, {Thakar}, {Theissen}, {Tkachenko},
  {Thomas}, {Tojeiro}, {Hernandez Toledo}, {Troup}, {Trump}, {Trussler},
  {Turner}, {Tuttle}, {Unda-Sanzana}, {V{\'a}zquez-Mata}, {Valentini},
  {Valenzuela}, {Vargas-Gonz{\'a}lez}, {Vargas-Maga{\~n}a}, {Alfaro},
  {Villanova}, {Vincenzo}, {Wake}, {Warfield}, {Washington}, {Weaver},
  {Weijmans}, {Weinberg}, {Weiss}, {Westfall}, {Wild}, {Wilde}, {Wilson},
  {Wilson}, {Wilson}, {Wolf}, {Wood-Vasey}, {Yan}, {Zamora}, {Zasowski},
  {Zhang}, {Zhao}, {Zheng}, {Zheng}, \& {Zhu}}]{SDSS17}
{Abdurro'uf}, {Accetta}, K., {Aerts}, C., {et~al.} 2022, \apjs, 259, 35,
  \dodoi{10.3847/1538-4365/ac4414}

\bibitem[{{Aguado} {et~al.}(2019){Aguado}, {Ahumada}, {Almeida}, {Anderson},
  {Andrews}, {Anguiano}, {Aquino Ort{\'\i}z}, {Arag{\'o}n-Salamanca},
  {Argudo-Fern{\'a}ndez}, {Aubert}, {Avila-Reese}, {Badenes}, {Barboza
  Rembold}, {Barger}, {Barrera-Ballesteros}, {Bates}, {Bautista}, {Beaton},
  {Beers}, {Belfiore}, {Bernardi}, {Bershady}, {Beutler}, {Bird}, {Bizyaev},
  {Blanc}, {Blanton}, {Blomqvist}, {Bolton}, {Boquien}, {Borissova}, {Bovy},
  {Nielsen Brandt}, {Brinkmann}, {Brownstein}, {Bundy}, {Burgasser}, {Byler},
  {Cano Diaz}, {Cappellari}, {Carrera}, {Cervantes Sodi}, {Chen}, {Cherinka},
  {Doohyun Choi}, {Chung}, {Coffey}, {Comerford}, {Comparat}, {Covey}, {da
  Silva Ilha}, {da Costa}, {Dai}, {Damke}, {Darling}, {Davies}, {Dawson}, {de
  Sainte Agathe}, {Deconto Machado}, {Del Moro}, {De Lee}, {Diamond-Stanic},
  {Dom{\'\i}nguez S{\'a}nchez}, {Donor}, {Drory}, {du Mas des Bourboux},
  {Duckworth}, {Dwelly}, {Ebelke}, {Emsellem}, {Escoffier},
  {Fern{\'a}ndez-Trincado}, {Feuillet}, {Fischer}, {Fleming},
  {Fraser-McKelvie}, {Freischlad}, {Frinchaboy}, {Fu}, {Galbany},
  {Garcia-Dias}, {Garc{\'\i}a-Hern{\'a}ndez}, {Garma Oehmichen}, {Geimba Maia},
  {Gil-Mar{\'\i}n}, {Grabowski}, {Gu}, {Guo}, {Ha}, {Harrington},
  {Hasselquist}, {Hayes}, {Hearty}, {Hernandez Toledo}, {Hicks}, {Hogg},
  {Holley-Bockelmann}, {Holtzman}, {Hsieh}, {Hunt}, {Hwang}, {Ibarra-Medel},
  {Jimenez Angel}, {Johnson}, {Jones}, {J{\"o}nsson}, {Kinemuchi}, {Kollmeier},
  {Krawczyk}, {Kreckel}, {Kruk}, {Lacerna}, {Lan}, {Lane}, {Law}, {Lee}, {Li},
  {Lian}, {Lin}, {Lin}, {Lintott}, {Long}, {Longa-Pe{\~n}a}, {Mackereth}, {de
  la Macorra}, {Majewski}, {Malanushenko}, {Manchado}, {Maraston}, {Mariappan},
  {Marinelli}, {Marques-Chaves}, {Masseron}, {Masters}, {McDermid}, {Medina
  Pe{\~n}a}, {Meneses-Goytia}, {Merloni}, {Merrifield}, {Meszaros}, {Minniti},
  {Minsley}, {Muna}, {Myers}, {Nair}, {Correa do Nascimento}, {Newman},
  {Nitschelm}, {Olmstead}, {Oravetz}, {Oravetz}, {Ortega Minakata}, {Pace},
  {Padilla}, {Palicio}, {Pan}, {Pan}, {Parikh}, {Parker}, {Peirani}, {Penny},
  {Percival}, {Perez-Fournon}, {Peterken}, {Pinsonneault}, {Prakash},
  {Raddick}, {Raichoor}, {Riffel}, {Riffel}, {Rix}, {Robin}, {Roman-Lopes},
  {Rose}, {Ross}, {Rossi}, {Rowlands}, {Rubin}, {S{\'a}nchez},
  {S{\'a}nchez-Gallego}, {Sayres}, {Schaefer}, {Schiavon}, {Schimoia},
  {Schlafly}, {Schlegel}, {Schneider}, {Schultheis}, {Seo}, {Shamsi}, {Shao},
  {Shen}, {Shetty}, {Simonian}, {Smethurst}, {Sobeck}, {Souter}, {Spindler},
  {Stark}, {Stassun}, {Steinmetz}, {Storchi-Bergmann}, {Stringfellow},
  {Su{\'a}rez}, {Sun}, {Taghizadeh-Popp}, {Talbot}, {Tayar}, {Thakar},
  {Thomas}, {Tissera}, {Tojeiro}, {Troup}, {Unda-Sanzana}, {Valenzuela},
  {Vargas-Maga{\~n}a}, {V{\'a}zquez-Mata}, {Wake}, {Weaver}, {Weijmans},
  {Westfall}, {Wild}, {Wilson}, {Woods}, {Yan}, {Yang}, {Zamora}, {Zasowski},
  {Zhang}, {Zheng}, {Zheng}, {Zhu}, {Zinn}, \& {Zou}}]{SDSS15}
{Aguado}, D.~S., {Ahumada}, R., {Almeida}, A., {et~al.} 2019, The Astrophysical
  Journal Supplement Series, 240, 23, \dodoi{10.3847/1538-4365/aaf651}

\bibitem[{{ATLAS collaboration}(2023)}]{ATLAS23}
{ATLAS collaboration}. 2023, in ATLAS-CONF-2023 No. 055

\bibitem[{{Barrera-Ballesteros} {et~al.}(2018){Barrera-Ballesteros}, {Heckman},
  {S{\'a}nchez}, {Zakamska}, {Cleary}, {Zhu}, {Brinkmann}, {Drory}, \& {THE
  MaNGA TEAM}}]{BarreraBallesteros18}
{Barrera-Ballesteros}, J.~K., {Heckman}, T., {S{\'a}nchez}, S.~F., {et~al.}
  2018, \apj, 852, 74, \dodoi{10.3847/1538-4357/aa9b31}

\bibitem[{{Bartelmann}(2010)}]{Bartelmann10}
{Bartelmann}, M. 2010, Classical and Quantum Gravity, 27, 233001,
  \dodoi{10.1088/0264-9381/27/23/233001}

\bibitem[{{Begeman}(1989)}]{Begeman89}
{Begeman}, K.~G. 1989, \aap, 223, 47

\bibitem[{{Behroozi} {et~al.}(2019){Behroozi}, {Wechsler}, {Hearin}, \&
  {Conroy}}]{Behroozi19}
{Behroozi}, P., {Wechsler}, R.~H., {Hearin}, A.~P., \& {Conroy}, C. 2019,
  \mnras, 1134, \dodoi{10.1093/mnras/stz1182}

\bibitem[{{Blanton} {et~al.}(2011){Blanton}, {Kazin}, {Muna}, {Weaver}, \&
  {Price-Whelan}}]{Blanton11}
{Blanton}, M.~R., {Kazin}, E., {Muna}, D., {Weaver}, B.~A., \& {Price-Whelan},
  A. 2011, \aj, 142, 31, \dodoi{10.1088/0004-6256/142/1/31}

\bibitem[{{Blanton} {et~al.}(2005){Blanton}, {Schlegel}, {Strauss},
  {Brinkmann}, {Finkbeiner}, {Fukugita}, {Gunn}, {Hogg}, {Ivezi{\'c}}, {Knapp},
  {Lupton}, {Munn}, {Schneider}, {Tegmark}, \& {Zehavi}}]{Blanton05}
{Blanton}, M.~R., {Schlegel}, D.~J., {Strauss}, M.~A., {et~al.} 2005, \aj, 129,
  2562, \dodoi{10.1086/429803}

\bibitem[{{Bosma}(1978)}]{Bosma78}
{Bosma}, A. 1978, PhD thesis, Groningen University

\bibitem[{{Boveia} \& {Doglioni}(2018)}]{Boveia18}
{Boveia}, A., \& {Doglioni}, C. 2018, Annual Review of Nuclear and Particle
  Science, 68, 429, \dodoi{10.1146/annurev-nucl-101917-021008}

\bibitem[{{Bundy} {et~al.}(2015){Bundy}, {Bershady}, {Law}, {Yan}, {Drory},
  {MacDonald}, {Wake}, {Cherinka}, {S{\'a}nchez-Gallego}, {Weijmans}, {Thomas},
  {Tremonti}, {Masters}, {Coccato}, {Diamond-Stanic}, {Arag{\'o}n-Salamanca},
  {Avila-Reese}, {Badenes}, {Falc{\'o}n-Barroso}, {Belfiore}, {Bizyaev},
  {Blanc}, {Bland-Hawthorn}, {Blanton}, {Brownstein}, {Byler}, {Cappellari},
  {Conroy}, {Dutton}, {Emsellem}, {Etherington}, {Frinchaboy}, {Fu}, {Gunn},
  {Harding}, {Johnston}, {Kauffmann}, {Kinemuchi}, {Klaene}, {Knapen},
  {Leauthaud}, {Li}, {Lin}, {Maiolino}, {Malanushenko}, {Malanushenko}, {Mao},
  {Maraston}, {McDermid}, {Merrifield}, {Nichol}, {Oravetz}, {Pan}, {Parejko},
  {Sanchez}, {Schlegel}, {Simmons}, {Steele}, {Steinmetz}, {Thanjavur},
  {Thompson}, {Tinker}, {van den Bosch}, {Westfall}, {Wilkinson}, {Wright},
  {Xiao}, \& {Zhang}}]{MaNGA}
{Bundy}, K., {Bershady}, M.~A., {Law}, D.~R., {et~al.} 2015, \apj, 798, 7,
  \dodoi{10.1088/0004-637X/798/1/7}

\bibitem[{{Carignan} \& {Freeman}(1985)}]{Carignan85}
{Carignan}, C., \& {Freeman}, K.~C. 1985, \apj, 294, 494,
  \dodoi{10.1086/163316}

\bibitem[{{Catinella} {et~al.}(2006){Catinella}, {Giovanelli}, \&
  {Haynes}}]{Catinella06}
{Catinella}, B., {Giovanelli}, R., \& {Haynes}, M.~P. 2006, \apj, 640, 751,
  \dodoi{10.1086/500171}

\bibitem[{{Cherinka} {et~al.}(2019){Cherinka}, {Andrews},
  {S{\'a}nchez-Gallego}, {Brownstein}, {Argudo-Fern{\'a}ndez}, {Blanton},
  {Bundy}, {Jones}, {Masters}, {Law}, {Rowlands}, {Weijmans}, {Westfall}, \&
  {Yan}}]{SDSS-Marvin}
{Cherinka}, B., {Andrews}, B.~H., {S{\'a}nchez-Gallego}, J., {et~al.} 2019,
  \aj, 158, 74, \dodoi{10.3847/1538-3881/ab2634}

\bibitem[{{Choi} {et~al.}(2010){Choi}, {Han}, \& {Kim}}]{Choi10}
{Choi}, Y.-Y., {Han}, D.-H., \& {Kim}, S.~S. 2010, Journal of Korean
  Astronomical Society, 43, 191, \dodoi{10.5303/JKAS.2010.43.6.191}

\bibitem[{{Cooke} \& {Fumagalli}(2018)}]{Cooke18}
{Cooke}, R.~J., \& {Fumagalli}, M. 2018, Nature Astronomy, 2, 957,
  \dodoi{10.1038/s41550-018-0584-z}

\bibitem[{{de Blok} {et~al.}(2008){de Blok}, {Walter}, {Brinks},
  {Trachternach}, {Oh}, \& {Kennicutt}}]{deBlok08}
{de Blok}, W.~J.~G., {Walter}, F., {Brinks}, E., {et~al.} 2008, \aj, 136, 2648,
  \dodoi{10.1088/0004-6256/136/6/2648}

\bibitem[{{De Vis} {et~al.}(2019){De Vis}, {Jones}, {Viaene}, {Casasola},
  {Clark}, {Baes}, {Bianchi}, {Cassara}, {Davies}, {De Looze}, {Galametz},
  {Galliano}, {Lianou}, {Madden}, {Manilla-Robles}, {Mosenkov}, {Nersesian},
  {Roychowdhury}, {Xilouris}, \& {Ysard}}]{deVis19}
{De Vis}, P., {Jones}, A., {Viaene}, S., {et~al.} 2019, \aap, 623, A5,
  \dodoi{10.1051/0004-6361/201834444}

\bibitem[{{Dettmar}(1990)}]{Dettmar90}
{Dettmar}, R.~J. 1990, \aap, 232, L15

\bibitem[{{Di Paolo} {et~al.}(2019){Di Paolo}, {Salucci}, \&
  {Erkurt}}]{DiPaolo19}
{Di Paolo}, C., {Salucci}, P., \& {Erkurt}, A. 2019, \mnras, 490, 5451,
  \dodoi{10.1093/mnras/stz2700}

\bibitem[{{Di Teodoro} {et~al.}(2021){Di Teodoro}, {Posti}, {Ogle}, {Fall}, \&
  {Jarrett}}]{DiTeodoro21}
{Di Teodoro}, E.~M., {Posti}, L., {Ogle}, P.~M., {Fall}, S.~M., \& {Jarrett},
  T. 2021, \mnras, 507, 5820, \dodoi{10.1093/mnras/stab2549}

\bibitem[{{Dom{\'\i}nguez S{\'a}nchez} {et~al.}(2022){Dom{\'\i}nguez
  S{\'a}nchez}, {Margalef}, {Bernardi}, \&
  {Huertas-Company}}]{DominguezSanchez22}
{Dom{\'\i}nguez S{\'a}nchez}, H., {Margalef}, B., {Bernardi}, M., \&
  {Huertas-Company}, M. 2022, \mnras, 509, 4024, \dodoi{10.1093/mnras/stab3089}

\bibitem[{{Douglass} \& {Demina}(2022)}]{Douglass22}
{Douglass}, K.~A., \& {Demina}, R. 2022, \apj, 925, 127,
  \dodoi{10.3847/1538-4357/ac3b56}

\bibitem[{{Douglass} {et~al.}(2019){Douglass}, {Smith}, \&
  {Demina}}]{Douglass19}
{Douglass}, K.~A., {Smith}, J.~A., \& {Demina}, R. 2019, \apj, 886, 153,
  \dodoi{10.3847/1538-4357/ab4bce}

\bibitem[{{Drory} {et~al.}(2015){Drory}, {MacDonald}, {Bershady}, {Bundy},
  {Gunn}, {Law}, {Smith}, {Stoll}, {Tremonti}, {Wake}, {Yan}, {Weijmans},
  {Byler}, {Cherinka}, {Cope}, {Eigenbrot}, {Harding}, {Holder}, {Huehnerhoff},
  {Jaehnig}, {Jansen}, {Klaene}, {Paat}, {Percival}, \& {Sayres}}]{Drory15}
{Drory}, N., {MacDonald}, N., {Bershady}, M.~A., {et~al.} 2015, \aj, 149, 77,
  \dodoi{10.1088/0004-6256/149/2/77}

\bibitem[{{Freeman}(1970)}]{Freeman70}
{Freeman}, K.~C. 1970, \apj, 160, 811, \dodoi{10.1086/150474}

\bibitem[{{Fukugita} {et~al.}(1996){Fukugita}, {Ichikawa}, {Gunn}, {Doi},
  {Shimasaku}, \& {Schneider}}]{Fukugita96}
{Fukugita}, M., {Ichikawa}, T., {Gunn}, J.~E., {et~al.} 1996, \aj, 111, 1748,
  \dodoi{10.1086/117915}

\bibitem[{{Gunn} {et~al.}(1998){Gunn}, {Carr}, {Rockosi}, {Sekiguchi}, {Berry},
  {Elms}, {de Haas}, {Ivezi{\'c}}, {Knapp}, {Lupton}, {Pauls}, {Simcoe},
  {Hirsch}, {Sanford}, {Wang}, {York}, {Harris}, {Annis}, {Bartozek},
  {Boroski}, {Bakken}, {Haldeman}, {Kent}, {Holm}, {Holmgren}, {Petravick},
  {Prosapio}, {Rechenmacher}, {Doi}, {Fukugita}, {Shimasaku}, {Okada}, {Hull},
  {Siegmund}, {Mannery}, {Blouke}, {Heidtman}, {Schneider}, {Lucinio}, \&
  {Brinkman}}]{Gunn98}
{Gunn}, J.~E., {Carr}, M., {Rockosi}, C., {et~al.} 1998, \aj, 116, 3040,
  \dodoi{10.1086/300645}

\bibitem[{{Haynes} {et~al.}(2018){Haynes}, {Giovanelli}, {Kent}, {Adams},
  {Balonek}, {Craig}, {Fertig}, {Finn}, {Giovanardi}, {Hallenbeck}, {Hess},
  {Hoffman}, {Huang}, {Jones}, {Koopmann}, {Kornreich}, {Leisman}, {Miller},
  {Moorman}, {O'Connor}, {O'Donoghue}, {Papastergis}, {Troischt}, {Stark}, \&
  {Xiao}}]{Haynes18}
{Haynes}, M.~P., {Giovanelli}, R., {Kent}, B.~R., {et~al.} 2018, \apj, 861, 49,
  \dodoi{10.3847/1538-4357/aac956}

\bibitem[{{Hirashita} {et~al.}(2003){Hirashita}, {Buat}, \&
  {Inoue}}]{Hirashita03}
{Hirashita}, H., {Buat}, V., \& {Inoue}, A.~K. 2003, \aap, 410, 83,
  \dodoi{10.1051/0004-6361:20031144}

\bibitem[{{Kalinova} {et~al.}(2017){Kalinova}, {van de Ven}, {Lyubenova},
  {Falc{\'o}n-Barroso}, {Colombo}, \& {Rosolowsky}}]{Kalinova17}
{Kalinova}, V., {van de Ven}, G., {Lyubenova}, M., {et~al.} 2017, \mnras, 464,
  1903, \dodoi{10.1093/mnras/stw2448}

\bibitem[{{Kamphuis} \& {Briggs}(1992)}]{Kamphuis92}
{Kamphuis}, J., \& {Briggs}, F. 1992, \aap, 253, 335

\bibitem[{{Karukes} \& {Salucci}(2017)}]{Karukes17}
{Karukes}, E.~V., \& {Salucci}, P. 2017, \mnras, 465, 4703,
  \dodoi{10.1093/mnras/stw3055}

\bibitem[{{Komatsu} {et~al.}(2011){Komatsu}, {Smith}, {Dunkley}, {Bennett},
  {Gold}, {Hinshaw}, {Jarosik}, {Larson}, {Nolta}, {Page}, {Spergel},
  {Halpern}, {Hill}, {Kogut}, {Limon}, {Meyer}, {Odegard}, {Tucker}, {Weiland},
  {Wollack}, \& {Wright}}]{Komatsu11}
{Komatsu}, E., {Smith}, K.~M., {Dunkley}, J., {et~al.} 2011, \apjs, 192, 18,
  \dodoi{10.1088/0067-0049/192/2/18}

\bibitem[{{Law} {et~al.}(2015){Law}, {Yan}, {Bershady}, {Bundy}, {Cherinka},
  {Drory}, {MacDonald}, {S{\'a}nchez-Gallego}, {Wake}, {Weijmans}, {Blanton},
  {Klaene}, {Moran}, {Sanchez}, \& {Zhang}}]{Law15}
{Law}, D.~R., {Yan}, R., {Bershady}, M.~A., {et~al.} 2015, \aj, 150, 19,
  \dodoi{10.1088/0004-6256/150/1/19}

\bibitem[{{Lupton} {et~al.}(2001){Lupton}, {Gunn}, {Ivezi{\'c}}, {Knapp}, \&
  {Kent}}]{Lupton01}
{Lupton}, R., {Gunn}, J.~E., {Ivezi{\'c}}, Z., {Knapp}, G.~R., \& {Kent}, S.
  2001, in Astronomical Society of the Pacific Conference Series, Vol. 238,
  Astronomical Data Analysis Software and Systems X, ed. J.~{Harnden}, F.~R.,
  F.~A. {Primini}, \& H.~E. {Payne}, 269

\bibitem[{{Martin} {et~al.}(2007){Martin}, {Wyder}, {Schiminovich}, {Barlow},
  {Forster}, {Friedman}, {Morrissey}, {Neff}, {Seibert}, {Small}, {Welsh},
  {Bianchi}, {Donas}, {Heckman}, {Lee}, {Madore}, {Milliard}, {Rich}, {Szalay},
  \& {Yi}}]{Martin07}
{Martin}, D.~C., {Wyder}, T.~K., {Schiminovich}, D., {et~al.} 2007, \apjs, 173,
  342, \dodoi{10.1086/516639}

\bibitem[{{Mathewson} {et~al.}(1992){Mathewson}, {Ford}, \&
  {Buchhorn}}]{Mathewson92}
{Mathewson}, D.~S., {Ford}, V.~L., \& {Buchhorn}, M. 1992, \apjs, 81, 413,
  \dodoi{10.1086/191700}

\bibitem[{{Moster} {et~al.}(2010){Moster}, {Somerville}, {Maulbetsch}, {van den
  Bosch}, {Macci{\`o}}, {Naab}, \& {Oser}}]{Moster10}
{Moster}, B.~P., {Somerville}, R.~S., {Maulbetsch}, C., {et~al.} 2010, \apj,
  710, 903, \dodoi{10.1088/0004-637X/710/2/903}

\bibitem[{{Osterbrock} \& {Ferland}(2006)}]{Osterbrock06}
{Osterbrock}, D.~E., \& {Ferland}, G.~J. 2006, {Astrophysics of gaseous nebulae
  and active galactic nuclei} (University Science Books)

\bibitem[{{Ostriker} {et~al.}(1974){Ostriker}, {Peebles}, \&
  {Yahil}}]{Ostriker74}
{Ostriker}, J.~P., {Peebles}, P.~J.~E., \& {Yahil}, A. 1974, \apjl, 193, L1,
  \dodoi{10.1086/181617}

\bibitem[{{Park} \& {Choi}(2005)}]{Park05}
{Park}, C., \& {Choi}, Y.-Y. 2005, \apjl, 635, L29, \dodoi{10.1086/499243}

\bibitem[{{Persic} {et~al.}(1996){Persic}, {Salucci}, \& {Stel}}]{Persic96}
{Persic}, M., {Salucci}, P., \& {Stel}, F. 1996, \mnras, 281, 27,
  \dodoi{10.1093/mnras/281.1.27}

\bibitem[{{Pilyugin} \& {Grebel}(2016)}]{Pilyugin16}
{Pilyugin}, L.~S., \& {Grebel}, E.~K. 2016, \mnras, 457, 3678,
  \dodoi{10.1093/mnras/stw238}

\bibitem[{{Planck Collaboration} {et~al.}(2020){Planck Collaboration},
  {Aghanim}, {Akrami}, {Ashdown}, {Aumont}, {Baccigalupi}, {Ballardini},
  {Banday}, {Barreiro}, {Bartolo}, {Basak}, {Battye}, {Benabed}, {Bernard},
  {Bersanelli}, {Bielewicz}, {Bock}, {Bond}, {Borrill}, {Bouchet}, {Boulanger},
  {Bucher}, {Burigana}, {Butler}, {Calabrese}, {Cardoso}, {Carron},
  {Challinor}, {Chiang}, {Chluba}, {Colombo}, {Combet}, {Contreras}, {Crill},
  {Cuttaia}, {de Bernardis}, {de Zotti}, {Delabrouille}, {Delouis}, {Di
  Valentino}, {Diego}, {Dor{\'e}}, {Douspis}, {Ducout}, {Dupac}, {Dusini},
  {Efstathiou}, {Elsner}, {En{\ss}lin}, {Eriksen}, {Fantaye}, {Farhang},
  {Fergusson}, {Fernandez-Cobos}, {Finelli}, {Forastieri}, {Frailis},
  {Fraisse}, {Franceschi}, {Frolov}, {Galeotta}, {Galli}, {Ganga},
  {G{\'e}nova-Santos}, {Gerbino}, {Ghosh}, {Gonz{\'a}lez-Nuevo}, {G{\'o}rski},
  {Gratton}, {Gruppuso}, {Gudmundsson}, {Hamann}, {Handley}, {Hansen},
  {Herranz}, {Hildebrandt}, {Hivon}, {Huang}, {Jaffe}, {Jones}, {Karakci},
  {Keih{\"a}nen}, {Keskitalo}, {Kiiveri}, {Kim}, {Kisner}, {Knox},
  {Krachmalnicoff}, {Kunz}, {Kurki-Suonio}, {Lagache}, {Lamarre}, {Lasenby},
  {Lattanzi}, {Lawrence}, {Le Jeune}, {Lemos}, {Lesgourgues}, {Levrier},
  {Lewis}, {Liguori}, {Lilje}, {Lilley}, {Lindholm}, {L{\'o}pez-Caniego},
  {Lubin}, {Ma}, {Mac{\'\i}as-P{\'e}rez}, {Maggio}, {Maino}, {Mandolesi},
  {Mangilli}, {Marcos-Caballero}, {Maris}, {Martin}, {Martinelli},
  {Mart{\'\i}nez-Gonz{\'a}lez}, {Matarrese}, {Mauri}, {McEwen}, {Meinhold},
  {Melchiorri}, {Mennella}, {Migliaccio}, {Millea}, {Mitra},
  {Miville-Desch{\^e}nes}, {Molinari}, {Montier}, {Morgante}, {Moss}, {Natoli},
  {N{\o}rgaard-Nielsen}, {Pagano}, {Paoletti}, {Partridge}, {Patanchon},
  {Peiris}, {Perrotta}, {Pettorino}, {Piacentini}, {Polastri}, {Polenta},
  {Puget}, {Rachen}, {Reinecke}, {Remazeilles}, {Renzi}, {Rocha}, {Rosset},
  {Roudier}, {Rubi{\~n}o-Mart{\'\i}n}, {Ruiz-Granados}, {Salvati}, {Sandri},
  {Savelainen}, {Scott}, {Shellard}, {Sirignano}, {Sirri}, {Spencer},
  {Sunyaev}, {Suur-Uski}, {Tauber}, {Tavagnacco}, {Tenti}, {Toffolatti},
  {Tomasi}, {Trombetti}, {Valenziano}, {Valiviita}, {Van Tent}, {Vibert},
  {Vielva}, {Villa}, {Vittorio}, {Wandelt}, {Wehus}, {White}, {White},
  {Zacchei}, \& {Zonca}}]{Planck20}
{Planck Collaboration}, {Aghanim}, N., {Akrami}, Y., {et~al.} 2020, \aap, 641,
  A6, \dodoi{10.1051/0004-6361/201833910}

\bibitem[{{Pohlen} {et~al.}(2010){Pohlen}, {Cortese}, {Smith}, {Eales},
  {Boselli}, {Bendo}, {Gomez}, {Papageorgiou}, {Auld}, {Baes}, {Bock},
  {Bradford}, {Buat}, {Castro-Rodriguez}, {Chanial}, {Charlot}, {Ciesla},
  {Clements}, {Cooray}, {Cormier}, {Dwek}, {Eales}, {Elbaz}, {Galametz},
  {Galliano}, {Gear}, {Glenn}, {Griffin}, {Hony}, {Isaak}, {Levenson}, {Lu},
  {Madden}, {O'Halloran}, {Okumura}, {Oliver}, {Page}, {Panuzzo}, {Parkin},
  {Perez-Fournon}, {Rangwala}, {Rigby}, {Roussel}, {Rykala}, {Sacchi},
  {Sauvage}, {Schulz}, {Schirm}, {Smith}, {Spinoglio}, {Stevens}, {Srinivasan},
  {Symeonidis}, {Trichas}, {Vaccari}, {Vigroux}, {Wilson}, {Wozniak}, {Wright},
  \& {Zeilinger}}]{Pohlen10}
{Pohlen}, M., {Cortese}, L., {Smith}, M.~W.~L., {et~al.} 2010, \aap, 518, L72,
  \dodoi{10.1051/0004-6361/201014554}

\bibitem[{{Porter} {et~al.}(2011){Porter}, {Johnson}, \& {Graham}}]{Porter11}
{Porter}, T.~A., {Johnson}, R.~P., \& {Graham}, P.~W. 2011, \araa, 49, 155,
  \dodoi{10.1146/annurev-astro-081710-102528}

\bibitem[{{Robertson} \& {Kravtsov}(2008)}]{Robertson08}
{Robertson}, B.~E., \& {Kravtsov}, A.~V. 2008, \apj, 680, 1083,
  \dodoi{10.1086/587796}

\bibitem[{Rubin {et~al.}(1985)Rubin, Burstein, Ford, \&
  Thonnard}]{Rubin:1985ze}
Rubin, V.~C., Burstein, D., Ford, Jr., W.~K., \& Thonnard, N. 1985, Astrophys.
  J., 289, 81, \dodoi{10.1086/162866}

\bibitem[{Rubin \& Ford(1970)}]{Rubin:1970zza}
Rubin, V.~C., \& Ford, Jr., W.~K. 1970, Astrophys. J., 159, 379,
  \dodoi{10.1086/150317}

\bibitem[{Rubin {et~al.}(1982)Rubin, Ford, Thonnard, \&
  Burstein}]{Rubin:1982kyu}
Rubin, V.~C., Ford, Jr., W.~K., Thonnard, N., \& Burstein, D. 1982, Astrophys.
  J., 261, 439, \dodoi{10.1086/160355}

\bibitem[{Rubin {et~al.}(1980)Rubin, Thonnard, \& Ford}]{Rubin:1980zd}
Rubin, V.~C., Thonnard, N., \& Ford, Jr., W.~K. 1980, Astrophys. J., 238, 471,
  \dodoi{10.1086/158003}

\bibitem[{{Saintonge} {et~al.}(2017){Saintonge}, {Catinella}, {Tacconi},
  {Kauffmann}, {Genzel}, {Cortese}, {Dav{\'e}}, {Fletcher},
  {Graci{\'a}-Carpio}, {Kramer}, {Heckman}, {Janowiecki}, {Lutz}, {Rosario},
  {Schiminovich}, {Schuster}, {Wang}, {Wuyts}, {Borthakur}, {Lamperti}, \&
  {Roberts-Borsani}}]{Saintonge17}
{Saintonge}, A., {Catinella}, B., {Tacconi}, L.~J., {et~al.} 2017, \apjs, 233,
  22, \dodoi{10.3847/1538-4365/aa97e0}

\bibitem[{{Salucci}(2019)}]{Salucci19}
{Salucci}, P. 2019, \aapr, 27, 2, \dodoi{10.1007/s00159-018-0113-1}

\bibitem[{{S{\'a}nchez} {et~al.}(2016){S{\'a}nchez}, {P{\'e}rez},
  {S{\'a}nchez-Bl{\'a}zquez}, {Garc{\'{\i}}a-Benito}, {Ibarra-Mede},
  {Gonz{\'a}lez}, {Rosales-Ortega}, {S{\'a}nchez-Menguiano}, {Ascasibar},
  {Bitsakis}, {Law}, {Cano-D{\'{\i}}az}, {L{\'o}pez-Cob{\'a}}, {Marino}, {Gil
  de Paz}, {L{\'o}pez-S{\'a}nchez}, {Barrera-Ballesteros}, {Galbany}, {Mast},
  {Abril-Melgarejo}, \& {Roman-Lopes}}]{Sanchez16}
{S{\'a}nchez}, S.~F., {P{\'e}rez}, E., {S{\'a}nchez-Bl{\'a}zquez}, P., {et~al.}
  2016, \rmxaa, 52, 171

\bibitem[{{S{\'a}nchez} {et~al.}(2018){S{\'a}nchez}, {Avila-Reese},
  {Hernandez-Toledo}, {Cortes-Su{\'a}rez}, {Rodr{\'{\i}}guez-Puebla},
  {Ibarra-Medel}, {Cano-D{\'{\i}}az}, {Barrera-Ballesteros}, {Negrete},
  {Calette}, {de Lorenzo-C{\'a}ceres}, {Ortega-Minakata}, {Aquino},
  {Valenzuela}, {Clemente}, {Storchi-Bergmann}, {Riffel}, {Schimoia}, {Riffel},
  {Rembold}, {Brownstein}, {Pan}, {Yates}, {Mallmann}, \&
  {Bitsakis}}]{Sanchez18}
{S{\'a}nchez}, S.~F., {Avila-Reese}, V., {Hernandez-Toledo}, H., {et~al.} 2018,
  \rmxaa, 54, 217

\bibitem[{{Schmidt} {et~al.}(2023){Schmidt}, {Mast}, {Gaspar}, \&
  {Weidmann}}]{Schmidt23}
{Schmidt}, E.~O., {Mast}, D., {Gaspar}, G., \& {Weidmann}, W. 2023, \mnras,
  523, 1885, \dodoi{10.1093/mnras/stad1531}

\bibitem[{{Sellwood} \& {Evans}(2001)}]{Sellwood01}
{Sellwood}, J.~A., \& {Evans}, N.~W. 2001, \apj, 546, 176,
  \dodoi{10.1086/318228}

\bibitem[{{Smee} {et~al.}(2013){Smee}, {Gunn}, {Uomoto}, {Roe}, {Schlegel},
  {Rockosi}, {Carr}, {Leger}, {Dawson}, {Olmstead}, {Brinkmann}, {Owen},
  {Barkhouser}, {Honscheid}, {Harding}, {Long}, {Lupton}, {Loomis}, {Anderson},
  {Annis}, {Bernardi}, {Bhardwaj}, {Bizyaev}, {Bolton}, {Brewington}, {Briggs},
  {Burles}, {Burns}, {Castander}, {Connolly}, {Davenport}, {Ebelke}, {Epps},
  {Feldman}, {Friedman}, {Frieman}, {Heckman}, {Hull}, {Knapp}, {Lawrence},
  {Loveday}, {Mannery}, {Malanushenko}, {Malanushenko}, {Merrelli}, {Muna},
  {Newman}, {Nichol}, {Oravetz}, {Pan}, {Pope}, {Ricketts}, {Shelden},
  {Sandford}, {Siegmund}, {Simmons}, {Smith}, {Snedden}, {Schneider},
  {SubbaRao}, {Tremonti}, {Waddell}, \& {York}}]{Smee13}
{Smee}, S.~A., {Gunn}, J.~E., {Uomoto}, A., {et~al.} 2013, \aj, 146, 32,
  \dodoi{10.1088/0004-6256/146/2/32}

\bibitem[{{Sofue}(2013)}]{Sofue13}
{Sofue}, Y. 2013, {Mass Distribution and Rotation Curve in the Galaxy}, ed.
  T.~D. {Oswalt} \& G.~{Gilmore}, Vol.~5, 985

\bibitem[{{Sofue}(2016)}]{Sofue16}
---. 2016, Publications of the Astronomical Society of Japan, 68, 2,
  \dodoi{10.1093/pasj/psv103}

\bibitem[{{Sofue}(2017)}]{Sofue17}
---. 2017, Publications of the Astronomical Society of Japan, 69, R1,
  \dodoi{10.1093/pasj/psw103}

\bibitem[{{Springel} {et~al.}(2005){Springel}, {White}, {Jenkins}, {Frenk},
  {Yoshida}, {Gao}, {Navarro}, {Thacker}, {Croton}, {Helly}, {Peacock}, {Cole},
  {Thomas}, {Couchman}, {Evrard}, {Colberg}, \& {Pearce}}]{Springel05}
{Springel}, V., {White}, S. D.~M., {Jenkins}, A., {et~al.} 2005, \nat, 435,
  629, \dodoi{10.1038/nature03597}

\bibitem[{{Stark} {et~al.}(2021){Stark}, {Masters}, {Avila-Reese}, {Riffel},
  {Riffel}, {Boardman}, {Zheng}, {Weijmans}, {Dillon}, {Fielder}, {Finnegan},
  {Fofie}, {Goddy}, {Harrington}, {Pace}, {Rujopakarn}, {Samanso}, {Shamsi},
  {Sharma}, {Warrick}, {Witherspoon}, \& {Wolthuis}}]{Stark21}
{Stark}, D.~V., {Masters}, K.~L., {Avila-Reese}, V., {et~al.} 2021, \mnras,
  503, 1345, \dodoi{10.1093/mnras/stab566}

\bibitem[{{Strauss} {et~al.}(2002){Strauss}, {Weinberg}, {Lupton}, {Narayanan},
  {Annis}, {Bernardi}, {Blanton}, {Burles}, {Connolly}, {Dalcanton}, {Doi},
  {Eisenstein}, {Frieman}, {Fukugita}, {Gunn}, {Ivezi{\'c}}, {Kent}, {Kim},
  {Knapp}, {Kron}, {Munn}, {Newberg}, {Nichol}, {Okamura}, {Quinn}, {Richmond},
  {Schlegel}, {Shimasaku}, {SubbaRao}, {Szalay}, {Vanden Berk}, {Vogeley},
  {Yanny}, {Yasuda}, {York}, \& {Zehavi}}]{Strauss02}
{Strauss}, M.~A., {Weinberg}, D.~H., {Lupton}, R.~H., {et~al.} 2002, \aj, 124,
  1810, \dodoi{10.1086/342343}

\bibitem[{{Strigari} {et~al.}(2008){Strigari}, {Bullock}, {Kaplinghat},
  {Simon}, {Geha}, {Willman}, \& {Walker}}]{Strigari08}
{Strigari}, L.~E., {Bullock}, J.~S., {Kaplinghat}, M., {et~al.} 2008, \nat,
  454, 1096, \dodoi{10.1038/nature07222}

\bibitem[{{Thomas} {et~al.}(2013){Thomas}, {Steele}, {Maraston}, {Johansson},
  {Beifiori}, {Pforr}, {Str{\"o}mb{\"a}ck}, {Tremonti}, {Wake}, {Bizyaev},
  {Bolton}, {Brewington}, {Brownstein}, {Comparat}, {Kneib}, {Malanushenko},
  {Malanushenko}, {Oravetz}, {Pan}, {Parejko}, {Schneider}, {Shelden},
  {Simmons}, {Snedden}, {Tanaka}, {Weaver}, \& {Yan}}]{Thomas13}
{Thomas}, D., {Steele}, O., {Maraston}, C., {et~al.} 2013, \mnras, 431, 1383,
  \dodoi{10.1093/mnras/stt261}

\bibitem[{{Torres-Flores} {et~al.}(2011){Torres-Flores}, {Epinat}, {Amram},
  {Plana}, \& {Mendes de Oliveira}}]{TorresFlores11}
{Torres-Flores}, S., {Epinat}, B., {Amram}, P., {Plana}, H., \& {Mendes de
  Oliveira}, C. 2011, \mnras, 416, 1936,
  \dodoi{10.1111/j.1365-2966.2011.19169.x}

\bibitem[{Tumasyan {et~al.}(2022)}]{CMS:2022qva}
Tumasyan, A., {et~al.} 2022, Phys. Rev. D, 105, 092007,
  \dodoi{10.1103/PhysRevD.105.092007}

\bibitem[{{Tumasyan} {et~al.}(2023){Tumasyan}, {Adam}, {Andrejkovic},
  {Bergauer}, {Chatterjee}, {Damanakis}, {Dragicevic}, {Escalante Del Valle},
  {Hussain}, {Jeitler}, {Krammer}, {Lechner}, {Liko}, {Mikulec}, {Paulitsch},
  {Schieck}, {Sch{\"o}fbeck}, {Schwarz}, {Sonawane}, {Templ}, {Waltenberger},
  {Wulz}, {Darwish}, {Janssen}, {Kello}, {Rejeb Sfar}, {Van Mechelen}, {Bols},
  {D'Hondt}, {De Moor}, {Delcourt}, {Faham}, {Lowette}, {Morton}, {M{\"u}ller},
  {Sahasransu}, {Tavernier}, {Van Doninck}, {Van Putte}, {Vannerom},
  {Clerbaux}, {Dansana}, {De Lentdecker}, {Favart}, {Hohov}, {Jaramillo},
  {Lee}, {Mahdavikhorrami}, {Makarenko}, {Malara}, {Paredes}, {P{\'e}tr{\'e}},
  {Postiau}, {Thomas}, {Vanden Bemden}, {Vander Velde}, {Vanlaer}, {Dobur},
  {Knolle}, {Lambrecht}, {Mestdach}, {Rend{\'o}n}, {Samalan}, {Skovpen},
  {Tytgat}, {Van Den Bossche}, {Vermassen}, {Wezenbeek}, {Benecke}, {Bruno},
  {Bury}, {Caputo}, {David}, {Delaere}, {Donertas}, {Giammanco}, {Jaffel},
  {Jain}, {Lemaitre}, {Mondal}, {Taliercio}, {Tran}, {Vischia}, {Wertz},
  {Alves}, {Coelho}, {Hensel}, {Moraes}, {Rebello Teles}, {Ald{\'a}
  J{\'u}nior}, {Alves Gallo Pereira}, {Barroso Ferreira Filho}, {Brandao
  Malbouisson}, {Carvalho}, {Chinellato}, {Da Costa}, {Da Silveira}, {De Jesus
  Damiao}, {Dos Santos Sousa}, {Fonseca De Souza}, {Martins}, {Mora Herrera},
  {Mota Amarilo}, {Mundim}, {Nogima}, {Santoro}, {Silva Do Amaral}, {Sznajder},
  {Thiel}, {Vilela Pereira}, {Bernardes}, {Calligaris}, {Fernandez Perez
  Tomei}, {Gregores}, {Mercadante}, {Novaes}, {Padula}, {Aleksandrov},
  {Antchev}, {Hadjiiska}, {Iaydjiev}, {Misheva}, {Rodozov}, {Shopova},
  {Sultanov}, {Dimitrov}, {Ivanov}, {Litov}, {Pavlov}, {Petkov}, {Petrov},
  {Shumka}, {Thakur}, {Cheng}, {Javaid}, {Mittal}, {Yuan}, {Ahmad}, {Bauer},
  {Hu}, {Lezki}, {Yi}, {Chen}, {Chen}, {Chen}, {Iemmi}, {Jiang}, {Kapoor},
  {Liao}, {Liu}, {Milosevic}, {Monti}, {Sharma}, {Tao}, {Thomas-Wilsker},
  {Wang}, {Zhang}, {Zhao}, {Agapitos}, {An}, {Ban}, {Levin}, {Li}, {Li}, {Lyu},
  {Mao}, {Qian}, {Sun}, {Wang}, {Xiao}, {Yang}, {Lu}, {You}, {Lu}, {Gao},
  {Leggat}, {Okawa}, {Zhang}, {Lin}, {Lu}, {Xiao}, {Avila}, {Barbosa Trujillo},
  {Cabrera}, {Florez}, {Fraga}, {Mejia Guisao}, {Ramirez}, {Rodriguez}, {Ruiz
  Alvarez}, {Giljanovic}, {Godinovic}, {Lelas}, {Puljak}, {Antunovic}, {Kovac},
  {Sculac}, {Brigljevic}, {Chitroda}, {Ferencek}, {Mishra}, {Roguljic},
  {Starodumov}, {Susa}, {Attikis}, {Christoforou}, {Konstantinou}, {Mousa},
  {Nicolaou}, {Ptochos}, {Razis}, {Rykaczewski}, {Saka}, {Stepennov}, {Finger},
  {Finger}, {Kveton}, {Ayala}, {Carrera Jarrin}, {Abdelalim}, {Salama},
  {Abdullah Al-Mashad}, {Mahmoud}, {Bhowmik}, {Dewanjee}, {Ehataht},
  {Kadastik}, {Lange}, {Nandan}, {Nielsen}, {Pata}, {Raidal}, {Tani},
  {Veelken}, {Eerola}, {Kirschenmann}, {Osterberg}, {Voutilainen}, {Bharthuar},
  {Br{\"u}cken}, {Garcia}, {Havukainen}, {Kim}, {Kinnunen}, {Lamp{\'e}n},
  {Lassila-Perini}, {Lehti}, {Lind{\'e}n}, {Lotti}, {Martikainen},
  {Myllym{\"a}ki}, {Rantanen}, {Siikonen}, {Tuominen}, {Tuominiemi}, {Luukka},
  {Petrow}, {Tuuva}, {Amendola}, {Besancon}, {Couderc}, {Dejardin}, {Denegri},
  {Faure}, {Ferri}, {Ganjour}, {Gras}, {Hamel de Monchenault}, {Lohezic},
  {Malcles}, {Rander}, {Rosowsky}, {Sahin}, {Savoy-Navarro}, {Simkina},
  {Titov}, {Baldenegro Barrera}, {Beaudette}, {Buchot Perraguin}, {Busson},
  {Cappati}, {Charlot}, {Damas}, {Davignon}, {Diab}, {Falmagne}, {Fontana
  Santos Alves}, {Ghosh}, {Granier de Cassagnac}, {Hakimi}, {Harikrishnan},
  {Liu}, {Motta}, {Nguyen}, {Ochando}, {Portales}, {Salerno}, {Sarkar},
  {Sauvan}, {Sirois}, {Tarabini}, {Vernazza}, {Zabi}, {Zghiche}, {Agram},
  {Andrea}, {Apparu}, {Bloch}, {Bourgatte}, {Brom}, {Chabert}, {Collard},
  {Darej}, {Goerlach}, {Grimault}, {Le Bihan}, {Van Hove}, {Beauceron},
  {Blancon}, {Boudoul}, {Carle}, {Chanon}, {Choi}, {Contardo}, {Depasse},
  {Dozen}, {El Mamouni}, {Fay}, {Gascon}, {Gouzevitch}, {Grenier}, {Ille},
  {Laktineh}, {Lethuillier}, {Mirabito}, {Perries}, {Torterotot}, {Vander
  Donckt}, {Verdier}, {Viret}, {Lomidze}, {Lomidze}, {Tsamalaidze}, {Botta},
  {Feld}, {Klein}, {Lipinski}, {Meuser}, {Pauls}, {R{\"o}wert}, {Teroerde},
  {Diekmann}, {Dodonova}, {Eich}, {Eliseev}, {Erdmann}, {Fackeldey},
  {Fasanella}, {Fischer}, {Hebbeker}, {Hoepfner}, {Ivone}, {Lee},
  {Mastrolorenzo}, {Merschmeyer}, {Meyer}, {Mondal}, {Mukherjee}, {Noll},
  {Novak}, {Nowotny}, {Pozdnyakov}, {Rath}, {Redjeb}, {Rehm}, {Reithler},
  {Schmidt}, {Schuler}, {Sharma}, {Stein}, {Torres Da Silva De Araujo},
  {Vigilante}, {Wiedenbeck}, {Zaleski}, {Dziwok}, {Fl{\"u}gge}, {Haj Ahmad},
  {Hlushchenko}, {Kress}, {Nowack}, {Pooth}, {Stahl}, {Ziemons}, {Zotz},
  {Petersen}, {Martin}, {Alimena}, {Asmuss}, {Baxter}, {Bayatmakou}, {Becerril
  Gonzalez}, {Behnke}, {Bhattacharya}, {Blekman}, {Borras}, {Brunner},
  {Campbell}, {Cardini}, {Cheng}, {Colombina}, {Consuegra Rodr{\'\i}guez},
  {Silva}, {De Silva}, {Eckerlin}, {Eckstein}, {Estevez Banos}, {Filatov},
  {Gallo}, {Geiser}, {Giraldi}, {Greau}, {Grohsjean}, {Guglielmi}, {Guthoff},
  {Jafari}, {Jomhari}, {Kaech}, {Kasemann}, {Kaveh}, {Kleinwort}, {Kogler},
  {Komm}, {Kr{\"u}cker}, {Lange}, {Leyva Pernia}, {Lipka}, {Lohmann}, {Mankel},
  {Melzer-Pellmann}, {Mendizabal Morentin}, {Metwally}, {Meyer}, {Milella},
  {Mormile}, {Mussgiller}, {N{\"u}rnberg}, {Otarid}, {P{\'e}rez Ad{\'a}n},
  {Ranken}, {Raspereza}, {Ribeiro Lopes}, {R{\"u}benach}, {Saggio},
  {Savitskyi}, {Scham}, {Scheurer}, {Schnake}, {Sch{\"u}tze}, {Schwanenberger},
  {Shchedrolosiev}, {Sosa Ricardo}, {Stafford}, {Tonon}, {Van De Klundert},
  {Vazzoler}, {Ventura Barroso}, {Walsh}, {Walter}, {Wang}, {Wen}, {Wichmann},
  {Wiens}, {Wissing}, {Wuchterl}, {Yang}, {Zimermmane Castro Santos},
  {Albrecht}, {Albrecht}, {Antonello}, {Bein}, {Benato}, {Bonanomi}, {Connor},
  {De Leo}, {Eich}, {El Morabit}, {Feindt}, {Fr{\"o}hlich}, {Garbers},
  {Garutti}, {Hajheidari}, {Haller}, {Hinzmann}, {Jabusch}, {Kasieczka},
  {Keicher}, {Klanner}, {Korcari}, {Kramer}, {Kutzner}, {Labe}, {Lange},
  {Lobanov}, {Matthies}, {Mehta}, {Moureaux}, {Mrowietz}, {Nigamova}, {Nissan},
  {Paasch}, {Pena Rodriguez}, {Quadfasel}, {Rieger}, {Savoiu}, {Schindler},
  {Schleper}, {Schr{\"o}der}, {Schwandt}, {Sommerhalder}, {Stadie},
  {Steinbr{\"u}ck}, {Tews}, {Wolf}, {Brommer}, {Burkart}, {Butz}, {Chwalek},
  {Dierlamm}, {Droll}, {Faltermann}, {Giffels}, {Gosewisch}, {Gottmann},
  {Hartmann}, {Horzela}, {Husemann}, {Klute}, {Koppenh{\"o}fer}, {Link},
  {Lintuluoto}, {Maier}, {Mitra}, {M{\"u}ller}, {Neukum}, {Oh}, {Quast},
  {Rabbertz}, {Shvetsov}, {Simonis}, {Trevisani}, {Ulrich}, {van der Linden},
  {Von Cube}, {Wassmer}, {Wieland}, {Wolf}, {Wozniewski}, {Wunsch}, {Zuo},
  {Anagnostou}, {Assiouras}, {Daskalakis}, {Kyriakis}, {Loukas}, {Stakia},
  {Diamantopoulou}, {Karasavvas}, {Kontaxakis}, {Manousakis-Katsikakis},
  {Panagiotou}, {Papavergou}, {Saoulidou}, {Theofilatos}, {Tziaferi},
  {Vellidis}, {Zisopoulos}, {Bakas}, {Chatzistavrou}, {Karapostoli},
  {Kousouris}, {Papakrivopoulos}, {Tsipolitis}, {Zacharopoulou}, {Adamidis},
  {Bestintzanos}, {Evangelou}, {Foudas}, {Gianneios}, {Kamtsikis}, {Katsoulis},
  {Kokkas}, {Kosmoglou Kioseoglou}, {Manthos}, {Papadopoulos}, {Strologas},
  {Csan{\'a}d}, {Farkas}, {Gadallah}, {Major}, {Mandal}, {P{\'a}sztor},
  {R{\'a}dl}, {Sur{\'a}nyi}, {Veres}, {Bart{\'o}k}, {Bencze}, {Hajdu},
  {Horvath}, {Sikler}, {Veszpremi}, {Beni}, {Czellar}, {Karancsi}, {Molnar},
  {Szillasi}, {Teyssier}, {Raics}, {Ujvari}, {Zilizi}, {Csorgo}, {Nemes},
  {Novak}, {Babbar}, {Bansal}, {Beri}, {Bhatnagar}, {Chaudhary}, {Chauhan},
  {Dhingra}, {Gupta}, {Kaur}, {Kaur}, {Kaur}, {Kaur}, {Kumar}, {Kumari},
  {Meena}, {Sandeep}, {Sheokand}, {Singh}, {Singla}, {Ahmed}, {Bhardwaj},
  {Chhetri}, {Choudhary}, {Kumar}, {Naimuddin}, {Ranjan}, {Saumya}, {Baradia},
  {Barman}, {Bhattacharya}, {Bhowmik}, {Dutta}, {Dutta}, {Gomber}, {Maity},
  {Palit}, {Saha}, {Sahu}, {Sarkar}, {Behera}, {Behera}, {Chatterjee},
  {Kalbhor}, {Komaragiri}, {Kumar}, {Muhammad}, {Panwar}, {Pradhan},
  {Pujahari}, {Saha}, {Sharma}, {Sikdar}, {Verma}, {Naskar}, {Aziz}, {Das},
  {Dugad}, {Kumar}, {Mohanty}, {Suryadevara}, {Banerjee}, {Guchait},
  {Karmakar}, {Kumar}, {Majumder}, {Mazumdar}, {Mukherjee}, {Thachayath},
  {Bahinipati}, {Das}, {Kar}, {Mal}, {Mishra}, {Muraleedharan Nair Bindhu},
  {Nayak}, {Saha}, {Swain}, {Vats}, {Alpana}, {Dube}, {Kansal}, {Laha},
  {Pandey}, {Rastogi}, {Sharma}, {Bakhshiansohi}, {Khazaie}, {Zeinali},
  {Chenarani}, {Etesami}, {Khakzad}, {Mohammadi Najafabadi}, {Grunewald},
  {Abbrescia}, {Aly}, {Aruta}, {Colaleo}, {Creanza}, {Cristella}, {De
  Filippis}, {De Palma}, {Di Florio}, {Elmetenawee}, {Errico}, {Fiore},
  {Iaselli}, {Maggi}, {Maggi}, {Margjeka}, {Mastrapasqua}, {My}, {Nuzzo},
  {Pellecchia}, {Pompili}, {Pugliese}, {Radogna}, {Ramos}, {Ranieri},
  {Selvaggi}, {Silvestris}, {Simone}, {S{\"o}zbilir}, {Stamerra}, {Venditti},
  {Verwilligen}, {Abbiendi}, {Battilana}, {Bonacorsi}, {Borgonovi},
  {Brigliadori}, {Campanini}, {Capiluppi}, {Castro}, {Cavallo}, {Cuffiani},
  {Dallavalle}, {Diotalevi}, {Fabbri}, {Fanfani}, {Giacomelli}, {Giommi},
  {Grandi}, {Guiducci}, {Lo Meo}, {Lunerti}, {Marcellini}, {Masetti},
  {Navarria}, {Perrotta}, {Primavera}, {Rossi}, {Rovelli}, {Siroli}, {Costa},
  {Di Mattia}, {Potenza}, {Tricomi}, {Tuve}, {Barbagli}, {Bardelli},
  {Camaiani}, {Cassese}, {Ceccarelli}, {Ciulli}, {Civinini}, {D'Alessandro},
  {Focardi}, {Latino}, {Lenzi}, {Lizzo}, {Meschini}, {Paoletti}, {Sguazzoni},
  {Viliani}, {Benussi}, {Bianco}, {Meola}, {Piccolo}, {Bozzo}, {Chatagnon},
  {Ferro}, {Robutti}, {Tosi}, {Benaglia}, {Boldrini}, {Brivio}, {Cetorelli},
  {De Guio}, {Dinardo}, {Dini}, {Gennai}, {Ghezzi}, {Govoni}, {Guzzi},
  {Lucchini}, {Malberti}, {Malvezzi}, {Massironi}, {Menasce}, {Moroni},
  {Paganoni}, {Pedrini}, {Pinolini}, {Ragazzi}, {Redaelli}, {Tabarelli de
  Fatis}, {Zuolo}, {Buontempo}, {Carnevali}, {Cavallo}, {De Iorio}, {Fabozzi},
  {Iorio}, {Lista}, {Paolucci}, {Rossi}, {Sciacca}, {Azzi}, {Bacchetta},
  {Bortignon}, {Bragagnolo}, {Carlin}, {Checchia}, {Dorigo}, {Gasparini},
  {Gasparini}, {Grosso}, {Layer}, {Lusiani}, {Margoni}, {Maron}, {Meneguzzo},
  {Pazzini}, {Ronchese}, {Rossin}, {Simonetto}, {Strong}, {Tosi}, {Yarar},
  {Zanetti}, {Zotto}, {Zucchetta}, {Zumerle}, {Zeid}, {Aim{\`e}}, {Braghieri},
  {Calzaferri}, {Fiorina}, {Montagna}, {Re}, {Riccardi}, {Salvini}, {Vai},
  {Vitulo}, {Asenov}, {Bilei}, {Ciangottini}, {Fan{\`o}}, {Magherini},
  {Mantovani}, {Mariani}, {Menichelli}, {Moscatelli}, {Piccinelli}, {Presilla},
  {Rossi}, {Santocchia}, {Spiga}, {Tedeschi}, {Azzurri}, {Bagliesi},
  {Bertacchi}, {Bhattacharya}, {Bianchini}, {Boccali}, {Bossini}, {Bruschini},
  {Castaldi}, {Ciocci}, {D'Amante}, {Dell'Orso}, {Donato}, {Giassi}, {Ligabue},
  {Matos Figueiredo}, {Messineo}, {Musich}, {Palla}, {Parolia},
  {Ramirez-Sanchez}, {Rizzi}, {Rolandi}, {Roy Chowdhury}, {Sarkar}, {Scribano},
  {Spagnolo}, {Tenchini}, {Tonelli}, {Turini}, {Venturi}, {Verdini}, {Barria},
  {Campana}, {Cavallari}, {Del Re}, {Di Marco}, {Diemoz}, {Longo}, {Meridiani},
  {Organtini}, {Pandolfi}, {Paramatti}, {Quaranta}, {Rahatlou}, {Rovelli},
  {Santanastasio}, {Soffi}, {Tramontano}, {Amapane}, {Arcidiacono}, {Argiro},
  {Arneodo}, {Bartosik}, {Bellan}, {Bellora}, {Biino}, {Cartiglia}, {Costa},
  {Covarelli}, {Demaria}, {Grippo}, {Kiani}, {Legger}, {Luongo}, {Mariotti},
  {Maselli}, {Mecca}, {Migliore}, {Monteno}, {Mulargia}, {Obertino}, {Ortona},
  {Pacher}, {Pastrone}, {Pelliccioni}, {Ruspa}, {Shchelina}, {Siviero}, {Sola},
  {Solano}, {Soldi}, {Staiano}, {Tarricone}, {Tornago}, {Trocino}, {Umoret},
  {Vagnerini}, {Vlasov}, {Belforte}, {Candelise}, {Casarsa}, {Cossutti}, {Della
  Ricca}, {Sorrentino}, {Dogra}, {Huh}, {Kim}, {Kim}, {Kim}, {Kim}, {Lee},
  {Lee}, {Moon}, {Oh}, {Pak}, {Ryu}, {Sekmen}, {Yang}, {Kim}, {Moon}, {Asilar},
  {Kim}, {Park}, {Choi}, {Han}, {Hong}, {Lee}, {Lee}, {Lim}, {Park}, {Park},
  {Yoo}, {Goh}, {Kim}, {Kim}, {Lee}, {Almond}, {Bhyun}, {Choi}, {Jeon}, {Kim},
  {Kim}, {Ko}, {Kwon}, {Lee}, {Lee}, {Oh}, {Oh}, {Seo}, {Yang}, {Yoon}, {Jang},
  {Kang}, {Kang}, {Kim}, {Kim}, {Ko}, {Lee}, {Lee}, {Merlin}, {Park}, {Roh},
  {Song}, {Watson}, {Yang}, {Ha}, {Yoo}, {Choi}, {Kim}, {Lee}, {Lee}, {Yu},
  {Beyrouthy}, {Maghrbi}, {Dreimanis}, {Pikurs}, {Potrebko}, {Seidel},
  {Veckalns}, {Ambrozas}, {Carvalho Antunes De Oliveira}, {Juodagalvis},
  {Rinkevicius}, {Tamulaitis}, {Norjoharuddeen}, {Hoh}, {Yusuff}, {Zolkapli},
  {Benitez}, {Castaneda Hernandez}, {Encinas Acosta}, {Gallegos
  Mar{\'\i}{\~n}ez}, {Le{\'o}n Coello}, {Murillo Quijada}, {Sehrawat},
  {Valencia Palomo}, {Ayala}, {Castilla-Valdez}, {Heredia-De La Cruz},
  {Lopez-Fernandez}, {Mondragon Herrera}, {Perez Navarro}, {S{\'a}nchez
  Hern{\'a}ndez}, {Oropeza Barrera}, {Vazquez Valencia}, {Pedraza}, {Salazar
  Ibarguen}, {Uribe Estrada}, {Bubanja}, {Mijuskovic}, {Raicevic}, {Ahmad},
  {Asghar}, {Awais}, {Awan}, {Gul}, {Hoorani}, {Khan}, {Avati}, {Grzanka},
  {Malawski}, {Bialkowska}, {Bluj}, {Boimska}, {G{\'o}rski}, {Kazana},
  {Szleper}, {Zalewski}, {Bunkowski}, {Doroba}, {Kalinowski}, {Konecki},
  {Krolikowski}, {Araujo}, {Bargassa}, {Bastos}, {Boletti}, {Faccioli},
  {Gallinaro}, {Hollar}, {Leonardo}, {Niknejad}, {Pisano}, {Seixas}, {Varela},
  {Adzic}, {Dordevic}, {Milenovic}, {Milosevic}, {Aguilar-Benitez}, {Alcaraz
  Maestre}, {Barrio Luna}, {Bedoya}, {Cepeda}, {Cerrada}, {Colino}, {De La
  Cruz}, {Delgado Peris}, {Fern{\'a}ndez Del Val}, {Fern{\'a}ndez Ramos},
  {Flix}, {Fouz}, {Gonzalez Lopez}, {Lopez}, {Hernandez}, {Josa}, {Le{\'o}n
  Holgado}, {Moran}, {Perez Dengra}, {P{\'e}rez-Calero Yzquierdo}, {Puerta
  Pelayo}, {Redondo}, {Redondo Ferrero}, {Romero}, {S{\'a}nchez Navas},
  {Sastre}, {Urda G{\'o}mez}, {Vazquez Escobar}, {Willmott}, {de Troc{\'o}niz},
  {Alvarez Gonzalez}, {Cuevas}, {Fernandez Menendez}, {Folgueras}, {Gonzalez
  Caballero}, {Gonz{\'a}lez Fern{\'a}ndez}, {Palencia Cortezon}, {Ram{\'o}n
  {\'A}lvarez}, {Rodr{\'\i}guez Bouza}, {Soto Rodr{\'\i}guez}, {Trapote}, {Vico
  Villalba}, {Brochero Cifuentes}, {Cabrillo}, {Calderon}, {Duarte Campderros},
  {Fernandez}, {Fernandez Madrazo}, {Garc{\'\i}a Alonso}, {Gomez}, {Lasaosa
  Garc{\'\i}a}, {Martinez Rivero}, {Martinez Ruiz del Arbol}, {Matorras},
  {Matorras Cuevas}, {Piedra Gomez}, {Prieels}, {Scodellaro}, {Vila}, {Vizan
  Garcia}, {Jayananda}, {Kailasapathy}, {Sonnadara}, {Wickramarathna},
  {Dharmaratna}, {Liyanage}, {Perera}, {Wickramage}, {Abbaneo}, {Auffray},
  {Auzinger}, {Baechler}, {Baillon}, {Barney}, {Bendavid}, {Berm{\'u}dez
  Mart{\'\i}nez}, {Bianco}, {Bilin}, {Bin Anuar}, {Bocci}, {Brondolin},
  {Caillol}, {Camporesi}, {Cerminara}, {Chernyavskaya}, {Chhibra}, {Choudhury},
  {Cipriani}, {d'Enterria}, {Dabrowski}, {David}, {De Roeck}, {Defranchis},
  {Deile}, {Dobson}, {D{\"u}nser}, {Dupont}, {Fallavollita}, {Florent},
  {Forthomme}, {Franzoni}, {Funk}, {Ghosh}, {Giani}, {Gigi}, {Gill}, {Glege},
  {Gouskos}, {Govorkova}, {Haranko}, {Hegeman}, {Innocente}, {James}, {Janot},
  {Kaspar}, {Kieseler}, {Kratochwil}, {Laurila}, {Lecoq}, {Leutgeb},
  {Louren{\c{c}}o}, {Maier}, {Malgeri}, {Mannelli}, {Marini}, {Meijers},
  {Mersi}, {Meschi}, {Moortgat}, {Mulders}, {Orfanelli}, {Orsini}, {Pantaleo},
  {Perez}, {Peruzzi}, {Petrilli}, {Petrucciani}, {Pfeiffer}, {Pierini},
  {Piparo}, {Pitt}, {Qu}, {Quast}, {Rabady}, {Racz}, {Reales Guti{\'e}rrez},
  {Rovere}, {Sakulin}, {Salfeld-Nebgen}, {Scarfi}, {Selvaggi}, {Sharma},
  {Silva}, {Sphicas}, {Stahl Leiton}, {Summers}, {Tatar}, {Treille}, {Tropea},
  {Tsirou}, {Wanczyk}, {Wozniak}, {Zeuner}, {Caminada}, {Ebrahimi}, {Erdmann},
  {Horisberger}, {Ingram}, {Kaestli}, {Kotlinski}, {Lange}, {Missiroli},
  {Noehte}, {Rohe}, {Aarrestad}, {Androsov}, {Backhaus}, {Calandri}, {Datta},
  {De Cosa}, {Dissertori}, {Dittmar}, {Doneg{\`a}}, {Eble}, {Galli}, {Gedia},
  {Glessgen}, {G{\'o}mez Espinosa}, {Grab}, {Hits}, {Lustermann}, {Lyon},
  {Manzoni}, {Marchese}, {Martin Perez}, {Mascellani}, {Nessi-Tedaldi},
  {Niedziela}, {Pauss}, {Perovic}, {Pigazzini}, {Ratti}, {Reichmann},
  {Reissel}, {Reitenspiess}, {Ristic}, {Riti}, {Ruini}, {Sanz Becerra},
  {Seidita}, {Steggemann}, {Valsecchi}, {Wallny}, {Amsler}, {B{\"a}rtschi},
  {Botta}, {Brzhechko}, {Canelli}, {Cormier}, {De Wit}, {Del Burgo},
  {Heikkil{\"a}}, {Huwiler}, {Jin}, {Jofrehei}, {Kilminster}, {Leontsinis},
  {Liechti}, {Macchiolo}, {Meiring}, {Mikuni}, {Molinatti}, {Neutelings},
  {Reimers}, {Robmann}, {Cruz}, {Schweiger}, {Senger}, {Takahashi}, {Adloff},
  {Kuo}, {Lin}, {Rout}, {Tiwari}, {Yu}, {Ceard}, {Chao}, {Chen}, {Chen},
  {Cheng}, {Hou}, {Khurana}, {Kole}, {Li}, {Lu}, {Paganis}, {Psallidas},
  {Steen}, {Wu}, {Yazgan}, {Asawatangtrakuldee}, {Srimanobhas},
  {Wachirapusitanand}, {Agyel}, {Boran}, {Demiroglu}, {Dolek}, {Dumanoglu},
  {Eskut}, {Guler}, {Gurpinar Guler}, {Isik}, {Kara}, {Kayis Topaksu},
  {Kiminsu}, {Onengut}, {Ozdemir}, {Polatoz}, {Simsek}, {Tali}, {Tok},
  {Turkcapar}, {Uslan}, {Zorbakir}, {Karapinar}, {Ocalan}, {Yalvac}, {Akgun},
  {Atakisi}, {G{\"u}lmez}, {Kaya}, {Kaya}, {Tekten}, {Cakir}, {Cankocak},
  {Komurcu}, {Sen}, {Aydilek}, {Cerci}, {Hacisahinoglu}, {Hos}, {Isildak},
  {Kaynak}, {Ozkorucuklu}, {Simsek}, {Sunar Cerci}, {Grynyov}, {Levchuk},
  {Anthony}, {Brooke}, {Bundock}, {Clement}, {Cussans}, {Flacher}, {Glowacki},
  {Goldstein}, {Heath}, {Kreczko}, {Krikler}, {Paramesvaran}, {Seif El
  Nasr-Storey}, {Smith}, {Stylianou}, {Pass}, {White}, {Ball}, {Bell},
  {Belyaev}, {Brew}, {Brown}, {Cockerill}, {Cooke}, {Ellis}, {Harder},
  {Harper}, {Holmberg}, {Jain}, {Linacre}, {Manolopoulos}, {Newbold}, {Olaiya},
  {Petyt}, {Reis}, {Salvi}, {Schuh}, {Shepherd-Themistocleous}, {Tomalin},
  {Williams}, {Bainbridge}, {Bloch}, {Bonomally}, {Borg}, {Brown},
  {Buchmuller}, {Cacchio}, {Montoya}, {Cepaitis}, {Chahal}, {Colling}, {Dancu},
  {Dauncey}, {Davies}, {Davies}, {Della Negra}, {Fayer}, {Fedi}, {Hall},
  {Hassanshahi}, {Howard}, {Iles}, {Langford}, {Lyons}, {Magnan}, {Malik},
  {Martelli}, {Mieskolainen}, {Monk}, {Nash}, {Pesaresi}, {Radburn-Smith},
  {Raymond}, {Richards}, {Rose}, {Scott}, {Seez}, {Shukla}, {Tapper}, {Uchida},
  {Uttley}, {Vage}, {Virdee}, {Vojinovic}, {Wardle}, {Webb}, {Winterbottom},
  {Coldham}, {Cole}, {Khan}, {Kyberd}, {Reid}, {Abdullin}, {Brinkerhoff},
  {Caraway}, {Dittmann}, {Hatakeyama}, {Kanuganti}, {McMaster}, {Saunders},
  {Sawant}, {Sutantawibul}, {Toms}, {Wilson}, {Bartek}, {Dominguez},
  {Escamilla}, {Uniyal}, {Vargas Hernandez}, {Chudasama}, {Cooper}, {Croce},
  {Gleyzer}, {Perez}, {Rumerio}, {Usai}, {West}, {Akpinar}, {Albert}, {Arcaro},
  {Cosby}, {Demiragli}, {Erice}, {Fontanesi}, {Gastler}, {May}, {Rohlf},
  {Salyer}, {Sperka}, {Spitzbart}, {Suarez}, {Tsatsos}, {Yuan}, {Benelli},
  {Coubez}, {Cutts}, {Hadley}, {Heintz}, {Hogan}, {Kwon}, {Landsberg}, {Lau},
  {Li}, {Luo}, {Narain}, {Pervan}, {Sagir}, {Simpson}, {Wong}, {Yan}, {Yu},
  {Zhang}, {Abbott}, {Bonilla}, {Brainerd}, {Breedon}, {Calderon De La Barca
  Sanchez}, {Chertok}, {Conway}, {Cox}, {Erbacher}, {Haza}, {Jensen}, {Kukral},
  {Mocellin}, {Mulhearn}, {Pellett}, {Regnery}, {Yao}, {Zhang}, {Bachtis},
  {Cousins}, {Datta}, {Hauser}, {Ignatenko}, {Iqbal}, {Lam}, {Manca}, {Nash},
  {Saltzberg}, {Stone}, {Valuev}, {Clare}, {Gary}, {Gordon}, {Hanson}, {Long},
  {Manganelli}, {Si}, {Wimpenny}, {Branson}, {Cittolin}, {Cooperstein}, {Diaz},
  {Duarte}, {Gerosa}, {Giannini}, {Guiang}, {Kansal}, {Krutelyov}, {Lee},
  {Letts}, {Masciovecchio}, {Mokhtar}, {Pieri}, {Quinnan}, {Sathia Narayanan},
  {Sharma}, {Tadel}, {Vourliotis}, {W{\"u}rthwein}, {Xiang}, {Yagil},
  {Campagnari}, {Citron}, {Collura}, {Dorsett}, {Incandela}, {Kilpatrick},
  {Kim}, {Li}, {Masterson}, {Mei}, {Oshiro}, {Richman}, {Sarica}, {Schmitz},
  {Setti}, {Sheplock}, {Siddireddy}, {Stuart}, {Wang}, {Bornheim}, {Cerri},
  {Dutta}, {Latorre}, {Lawhorn}, {Mao}, {Newman}, {Nguyen}, {Spiropulu},
  {Vlimant}, {Wang}, {Xie}, {Zhu}, {Alison}, {An}, {Andrews}, {Bryant},
  {Dutta}, {Ferguson}, {Harilal}, {Liu}, {Mudholkar}, {Murthy}, {Paulini},
  {Roberts}, {Sanchez}, {Terrill}, {Cumalat}, {Ford}, {Hassani},
  {Karathanasis}, {MacDonald}, {Marini}, {Perloff}, {Savard}, {Schonbeck},
  {Stenson}, {Ulmer}, {Wagner}, {Zipper}, {Alexander}, {Bright-Thonney},
  {Chen}, {Cranshaw}, {Fan}, {Fan}, {Gadkari}, {Hogan}, {Monroy}, {Patterson},
  {Reichert}, {Reid}, {Ryd}, {Thom}, {Wittich}, {Zou}, {Albrow}, {Alyari},
  {Apollinari}, {Apresyan}, {Bauerdick}, {Berry}, {Berryhill}, {Bhat},
  {Burkett}, {Butler}, {Canepa}, {Cerati}, {Cheung}, {Chlebana}, {Di Petrillo},
  {Dickinson}, {Elvira}, {Feng}, {Freeman}, {Gandrakota}, {Gecse}, {Gray},
  {Green}, {Gr{\"u}nendahl}, {Guerrero}, {Gutsche}, {Harris}, {Heller},
  {Herwig}, {Hirschauer}, {Horyn}, {Jayatilaka}, {Jindariani}, {Johnson},
  {Joshi}, {Klijnsma}, {Klima}, {Kwok}, {Lammel}, {Lincoln}, {Lipton}, {Liu},
  {Madrid}, {Maeshima}, {Mantilla}, {Mason}, {McBride}, {Merkel}, {Mrenna},
  {Nahn}, {Ngadiuba}, {Noonan}, {Norberg}, {Papadimitriou}, {Pastika}, {Pedro},
  {Pena}, {Ravera}, {Hall}, {Ristori}, {Sexton-Kennedy}, {Smith}, {Soha},
  {Spiegel}, {Strait}, {Taylor}, {Tkaczyk}, {Tran}, {Uplegger}, {Vaandering},
  {Zoi}, {Avery}, {Bourilkov}, {Cadamuro}, {Chang}, {Cherepanov}, {Field},
  {Koenig}, {Kolosova}, {Konigsberg}, {Korytov}, {Kuznetsova}, {Lo}, {Matchev},
  {Menendez}, {Mitselmakher}, {Muthirakalayil Madhu}, {Rawal}, {Rosenzweig},
  {Rosenzweig}, {Shi}, {Wang}, {Wu}, {Adams}, {Askew}, {Bower}, {Habibullah},
  {Hagopian}, {Kolberg}, {Martinez}, {Prosper}, {Viazlo}, {Wulansatiti},
  {Yohay}, {Zhang}, {Baarmand}, {Butalla}, {Elkafrawy}, {Hohlmann}, {Kumar
  Verma}, {Rahmani}, {Yumiceva}, {Adams}, {Cavanaugh}, {Dittmer}, {Evdokimov},
  {Gerber}, {Hofman}, {Lemos}, {Merrit}, {Mills}, {Oh}, {Roy}, {Rudrabhatla},
  {Tonjes}, {Varelas}, {Wang}, {Ye}, {Yoo}, {Alhusseini}, {Dilsiz}, {Emediato},
  {Karaman}, {K{\"o}seyan}, {Merlo}, {Mestvirishvili}, {Nachtman}, {Neogi},
  {Ogul}, {Onel}, {Penzo}, {Snyder}, {Tiras}, {Amram}, {Blumenfeld},
  {Corcodilos}, {Davis}, {Gritsan}, {Kyriacou}, {Maksimovic}, {Roskes},
  {Sekhar}, {Swartz}, {V{\'a}mi}, {Abreu}, {Alcerro Alcerro}, {Anguiano},
  {Baringer}, {Bean}, {Flowers}, {King}, {Krintiras}, {Lazarovits}, {Le
  Mahieu}, {Lindsey}, {Marquez}, {Minafra}, {Murray}, {Nickel}, {Rogan},
  {Royon}, {Salvatico}, {Sanders}, {Smith}, {Wang}, {Wilson}, {Allmond},
  {Duric}, {Ivanov}, {Kaadze}, {Kalogeropoulos}, {Kim}, {Maravin}, {Mitchell},
  {Modak}, {Nam}, {Roy}, {Rebassoo}, {Wright}, {Adams}, {Baden}, {Baron},
  {Belloni}, {Bethani}, {Eno}, {Hadley}, {Jabeen}, {Kellogg}, {Koeth}, {Lai},
  {Lascio}, {Mignerey}, {Nabili}, {Palmer}, {Papageorgakis}, {Wang}, {Wong},
  {Busza}, {Cali}, {Chen}, {D'Alfonso}, {Eysermans}, {Freer}, {Gomez-Ceballos},
  {Goncharov}, {Harris}, {Kovalskyi}, {Krupa}, {Lee}, {Long}, {Mironov},
  {Paus}, {Rankin}, {Roland}, {Roland}, {Shi}, {Stephans}, {Wang}, {Wang},
  {Wyslouch}, {Yang}, {Chatterjee}, {Crossman}, {Hiltbrand}, {Joshi},
  {Kapsiak}, {Krohn}, {Kubota}, {Mahon}, {Mans}, {Revering}, {Rusack},
  {Saradhy}, {Schroeder}, {Strobbe}, {Wadud}, {Cremaldi}, {Bloom}, {Bryson},
  {Claes}, {Fangmeier}, {Finco}, {Golf}, {Joo}, {Kamalieddin}, {Kravchenko},
  {Reed}, {Siado}, {Snow}, {Tabb}, {Wightman}, {Yan}, {Zecchinelli}, {Agarwal},
  {Bandyopadhyay}, {Hay}, {Iashvili}, {Kharchilava}, {McLean}, {Morris},
  {Nguyen}, {Pekkanen}, {Rappoccio}, {Williams}, {Alverson}, {Barberis},
  {Haddad}, {Han}, {Krishna}, {Li}, {Lidrych}, {Madigan}, {Marzocchi}, {Morse},
  {Nguyen}, {Orimoto}, {Parker}, {Skinnari}, {Tishelman-Charny}, {Wamorkar},
  {Wang}, {Wisecarver}, {Wood}, {Bhattacharya}, {Bueghly}, {Chen}, {Gilbert},
  {Hahn}, {Liu}, {Odell}, {Schmitt}, {Velasco}, {Band}, {Bucci}, {Cremonesi},
  {Das}, {Goldouzian}, {Hildreth}, {Hurtado Anampa}, {Jessop}, {Lannon},
  {Lawrence}, {Loukas}, {Lutton}, {Mariano}, {Marinelli}, {Mcalister},
  {McCauley}, {Mcgrady}, {Mohrman}, {Moore}, {Musienko}, {Ruchti}, {Townsend},
  {Wayne}, {Yockey}, {Zarucki}, {Zygala}, {Bylsma}, {Carrigan}, {Durkin},
  {Hill}, {Joyce}, {Lesauvage}, {Nunez Ornelas}, {Wei}, {Winer}, {Yates},
  {Addesa}, {Das}, {Dezoort}, {Elmer}, {Frankenthal}, {Greenberg}, {Haubrich},
  {Higginbotham}, {Kopp}, {Kwan}, {Lange}, {Loeliger}, {Marlow}, {Ojalvo},
  {Olsen}, {Stickland}, {Tully}, {Malik}, {Bakshi}, {Barnes}, {Chawla}, {Das},
  {Gutay}, {Jones}, {Jung}, {Kondratyev}, {Koshy}, {Liu}, {Negro},
  {Neumeister}, {Paspalaki}, {Piperov}, {Purohit}, {Schulte}, {Stojanovic},
  {Thieman}, {Virdi}, {Wang}, {Xiao}, {Xie}, {Dolen}, {Parashar}, {Acosta},
  {Baty}, {Carnahan}, {Dildick}, {Ecklund}, {Fern{\'a}ndez Manteca}, {Freed},
  {Gardner}, {Geurts}, {Kumar}, {Li}, {Padley}, {Redjimi}, {Rotter}, {Yang},
  {Yigitbasi}, {Zhang}, {Bodek}, {de Barbaro}, {Demina}, {Dulemba}, {Fallon},
  {Garcia-Bellido}, {Hindrichs}, {Khukhunaishvili}, {Parygin}, {Popova},
  {Taus}, {Van Onsem}, {Goulianos}, {Chiarito}, {Chou}, {Gershtein},
  {Halkiadakis}, {Hart}, {Heindl}, {Jaroslawski}, {Karacheban}, {Laflotte},
  {Lath}, {Montalvo}, {Nash}, {Osherson}, {Routray}, {Salur}, {Schnetzer},
  {Somalwar}, {Stone}, {Thayil}, {Thomas}, {Wang}, {Acharya}, {Delannoy},
  {Fiorendi}, {Holmes}, {Nibigira}, {Spanier}, {Bouhali}, {Dalchenko},
  {Delgado}, {Eusebi}, {Gilmore}, {Huang}, {Kamon}, {Kim}, {Luo}, {Malhotra},
  {Mueller}, {Overton}, {Rathjens}, {Safonov}, {Akchurin}, {Damgov}, {Hegde},
  {Lamichhane}, {Lee}, {Mengke}, {Muthumuni}, {Peltola}, {Volobouev},
  {Whitbeck}, {Appelt}, {Greene}, {Gurrola}, {Johns}, {Melo}, {Romeo},
  {Sheldon}, {Tuo}, {Velkovska}, {Viinikainen}, {Cardwell}, {Cox}, {Cummings},
  {Hakala}, {Hirosky}, {Ledovskoy}, {Li}, {Neu}, {Perez Lara}, {Karchin},
  {Aravind}, {Banerjee}, {Black}, {Bose}, {Dasu}, {De Bruyn}, {Everaerts},
  {Galloni}, {He}, {Herndon}, {Herve}, {Koraka}, {Lanaro}, {Loveless},
  {Madhusudanan Sreekala}, {Mallampalli}, {Mohammadi}, {Mondal}, {Parida},
  {Pinna}, {Savin}, {Shang}, {Sharma}, {Smith}, {Teague}, {Tsoi}, {Vetens},
  {Warden}, {Afanasiev}, {Andreev}, {Andreev}, {Aushev}, {Azarkin}, {Babaev},
  {Belyaev}, {Blinov}, {Boos}, {Borshch}, {Budkouski}, {Bunichev}, {Chadeeva},
  {Chekhovsky}, {Danilov}, {Dermenev}, {Dimova}, {Dremin}, {Dubinin}, {Dudko},
  {Epshteyn}, {Ershov}, {Gavrilov}, {Gavrilov}, {Gninenko}, {Golovtcov},
  {Golubev}, {Golutvin}, {Gorbunov}, {Ivanov}, {Kachanov}, {Kardapoltsev},
  {Karjavine}, {Karneyeu}, {Kim}, {Kirakosyan}, {Kirpichnikov}, {Kirsanov},
  {Klyukhin}, {Kodolova}, {Konstantinov}, {Korenkov}, {Kozyrev}, {Krasnikov},
  {Lanev}, {Levchenko}, {Litomin}, {Lychkovskaya}, {Makarenko}, {Malakhov},
  {Matveev}, {Murzin}, {Nikitenko}, {Obraztsov}, {Oskin}, {Ovtin}, {Palichik},
  {Perelygin}, {Perfilov}, {Petrushanko}, {Popov}, {Radchenko}, {Rusinov},
  {Savina}, {Savrin}, {Shalaev}, {Shmatov}, {Shulha}, {Skovpen},
  {Slabospitskii}, {Smirnov}, {Sosnov}, {Sulimov}, {Tcherniaev}, {Terkulov},
  {Teryaev}, {Tlisova}, {Toropin}, {Uvarov}, {Uzunian}, {Vorobyev},
  {Voytishin}, {Yuldashev}, {Zarubin}, {Zhizhin}, {Zhokin}, \& {CMS
  Collaboration}}]{Tumasyan23}
{Tumasyan}, A., {Adam}, W., {Andrejkovic}, J.~W., {et~al.} 2023, European
  Physical Journal C, 83, 933, \dodoi{10.1140/epjc/s10052-023-11952-7}

\bibitem[{{van Dokkum} {et~al.}(2018){van Dokkum}, {Danieli}, {Cohen},
  {Merritt}, {Romanowsky}, {Abraham}, {Brodie}, {Conroy}, {Lokhorst}, {Mowla},
  {O'Sullivan}, \& {Zhang}}]{vanDokkum18}
{van Dokkum}, P., {Danieli}, S., {Cohen}, Y., {et~al.} 2018, \nat, 555, 629,
  \dodoi{10.1038/nature25767}

\bibitem[{{Wake} {et~al.}(2017){Wake}, {Bundy}, {Diamond-Stanic}, {Yan},
  {Blanton}, {Bershady}, {S{\'a}nchez-Gallego}, {Drory}, {Jones}, {Kauffmann},
  {Law}, {Li}, {MacDonald}, {Masters}, {Thomas}, {Tinker}, {Weijmans}, \&
  {Brownstein}}]{Wake17}
{Wake}, D.~A., {Bundy}, K., {Diamond-Stanic}, A.~M., {et~al.} 2017, \aj, 154,
  86, \dodoi{10.3847/1538-3881/aa7ecc}

\bibitem[{{Wang} {et~al.}(2020){Wang}, {Catinella}, {Saintonge}, {Pan},
  {Serra}, \& {Shao}}]{Wang20}
{Wang}, J., {Catinella}, B., {Saintonge}, A., {et~al.} 2020, \apj, 890, 63,
  \dodoi{10.3847/1538-4357/ab68dd}

\bibitem[{{Wang} {et~al.}(2016){Wang}, {Koribalski}, {Serra}, {van der Hulst},
  {Roychowdhury}, {Kamphuis}, \& {Chengalur}}]{Wang16}
{Wang}, J., {Koribalski}, B.~S., {Serra}, P., {et~al.} 2016, \mnras, 460, 2143,
  \dodoi{10.1093/mnras/stw1099}

\bibitem[{{Westfall} {et~al.}(2019){Westfall}, {Cappellari}, {Bershady},
  {Bundy}, {Belfiore}, {Ji}, {Law}, {Schaefer}, {Shetty}, {Tremonti}, {Yan},
  {Andrews}, {Brownstein}, {Cherinka}, {Coccato}, {Drory}, {Maraston},
  {Parikh}, {S{\'a}nchez-Gallego}, {Thomas}, {Weijmans}, {Barrera-Ballesteros},
  {Du}, {Goddard}, {Li}, {Masters}, {Ibarra Medel}, {S{\'a}nchez}, {Yang},
  {Zheng}, \& {Zhou}}]{Westfall19}
{Westfall}, K.~B., {Cappellari}, M., {Bershady}, M.~A., {et~al.} 2019, \aj,
  158, 231, \dodoi{10.3847/1538-3881/ab44a2}

\bibitem[{{Wylezalek} {et~al.}(2022){Wylezalek}, {Cicone}, {Belfiore},
  {Bertemes}, {Cazzoli}, {Wagg}, {Wang (王无忌)}, {Aravena}, {Maiolino},
  {Martin}, {Bothwell}, {Brownstein}, {Bundy}, \& {De Breuck}}]{Wylezalek22}
{Wylezalek}, D., {Cicone}, C., {Belfiore}, F., {et~al.} 2022, \mnras, 510,
  3119, \dodoi{10.1093/mnras/stab3356}

\bibitem[{{Yao} {et~al.}(2006){Yao}, {Amsler}, {Asner}, {Barnett}, {Beringer},
  {Burchat}, {Carone}, {Caso}, {Dahl}, {D'Ambrosio}, {De Gouvea}, {Doser},
  {Eidelman}, {Feng}, {Gherghetta}, {Goodman}, {Grab}, {Groom}, {Gurtu},
  {Hagiwara}, {Hayes}, {Hern{\'a}ndez-Rey}, {Hikasa}, {Jawahery}, {Kolda},
  {Kwon}, {Mangano}, {Manohar}, {Masoni}, {Miquel}, {M{\"o}nig}, {Murayama},
  {Nakamura}, {Navas}, {Olive}, {Pape}, {Patrignani}, {Piepke}, {Punzi},
  {Raffelt}, {Smith}, {Tanabashi}, {Terning}, {T{\"o}rnqvist}, {sTrippe},
  {Vogel}, {Watari}, {Wohl}, {Workman}, {Zyla}, {Armstrong}, {Harper},
  {Lugovsky}, {Schaffner}, {Artuso}, {Babu}, {Band}, {Barberio}, {Battaglia},
  {Bichsel}, {Biebel}, {Bloch}, {Blucher}, {Cahn}, {Casper}, {Cattai},
  {Ceccucci}, {Chakraborty}, {Chivukula}, {Cowan}, {Damour}, {DeGrand},
  {Desler}, {Dobbs}, {Drees}, {Edwards}, {Edwards}, {Elvira}, {Erler},
  {Ezhela}, {Fetscher}, {Fields}, {Foster}, {Froidevaux}, {Gaisser}, {Garren},
  {Gerber}, {Gerbier}, {Gibbons}, {Gilman}, {Giudice}, {Gritsan},
  {Gr{\"u}newald}, {Haber}, {Hagmann}, {Hinchliffe}, {H{\"o}cker},
  {Igo-Kemenes}, {JAckson}, {Johnson}, {Karlen}, {Kayser}, {Kirkby}, {Klein},
  {Kleinknecht}, {Knowles}, {Kowalewski}, {Kreitz}, {Kursche}, {Kuyanov},
  {Lahav}, {Langacker}, {Liddle}, {Ligeti}, {Liss}, {Littenberg}, {Liu},
  {Lugovsky}, {Lugovsky}, {Mannel}, {Manley}, {Marciano}, {Martin}, {Milstead},
  {Narain}, {Nason}, {Nir}, {Peacock}, {Prell}, {Quadt}, {Raby}, {Ratcliff},
  {Razuvaev}, {Renk}, {Richardson}, {Roesler}, {Rolandi}, {Ronan}, {Rosenberg},
  {Sachrajda}, {Sakai}, {Sarkar}, {Schmitt}, {Schneider}, {Scott},
  {Sj{\"o}strand}, {Smoot}, {Sokolsky}, {Spanier}, {Spieler}, {Stahl},
  {Stanev}, {Streitmatter}, {Sumiyoshi}, {Tkachenko}, {Trilling}, {Valencia},
  {van Bibber}, {Vincter}, {Ward}, {Webber}, {Wells}, {Whalley},
  {Wolfenstsein}, {Womersley}, {Woody}, {Yamamoto}, {Zenin}, {Zhang}, \&
  {Zhu}}]{Yao06}
{Yao}, W.~M., {Amsler}, C., {Asner}, D., {et~al.} 2006, Journal of Physics G
  Nuclear Physics, 33, 1, \dodoi{10.1088/0954-3899/33/1/001}

\bibitem[{{Yoon} {et~al.}(2021){Yoon}, {Park}, {Chung}, \& {Zhang}}]{Yoon21}
{Yoon}, Y., {Park}, C., {Chung}, H., \& {Zhang}, K. 2021, \apj, 922, 249,
  \dodoi{10.3847/1538-4357/ac2302}

\end{thebibliography}

\end{document}